\documentclass[twocolumn,tighten]{aastex63b}
\usepackage{graphicx}		
\definecolor{orcidlogocol}{HTML}{A6CE39}
\usepackage{url}
\usepackage{hyperref}
\usepackage{epstopdf}
\usepackage{color}
\usepackage{multirow}
\usepackage{ragged2e}
\usepackage{tabularx}
\usepackage{url}
\usepackage{soul}
\usepackage{textcomp}
\usepackage{gensymb}
\usepackage{amssymb,amsmath,amsfonts,amsthm}
\usepackage{mathrsfs}
\usepackage{needspace}
\usepackage[T1]{fontenc}
\usepackage[english]{babel}
\usepackage[utf8]{inputenc}
\usepackage{float}
\usepackage[encapsulated]{CJK}
\usepackage{ucs}
\newcommand{\cntext}[1]{\begin{CJK}{UTF8}{gbsn}#1\end{CJK}}

\usepackage{xspace}
\DeclareGraphicsExtensions{.pdf,.png,.jpeg}
\def\sgra{Sgr\,A$^{\ast}$}
\def\bhspin{a_*}
\def\m87{M87$^{\ast}$}
\def\lsim{\mathrel{\raise.3ex\hbox{$<$\kern-.75em\lower1ex\hbox{$\sim$}}}}
\def\gsim{\mathrel{\raise.3ex\hbox{$>$\kern-.75em\lower1ex\hbox{$\sim$}}}}
\def\gtwid{\mathrel{\raise.3ex\hbox{$>$\kern-.75em\lower1ex\hbox{$\sim$}}}}
\def\proptwid{\mathrel{\raise.3ex\hbox{$\propto$\kern-.75em\lower1ex\hbox{$\sim$}}}}

\def\apjl{ApJL}
\def\apjs{ApJS}
\def\ao{Appl.~Opt.}
\def\aap{A\&A}
\defcitealias{Paper1}{EHTC~I}
\defcitealias{Paper2}{EHTC~II}
\defcitealias{Paper3}{EHTC~III}
\defcitealias{Paper4}{EHTC~IV}
\defcitealias{Paper5}{EHTC~V}
\defcitealias{Paper6}{EHTC~VI}

\newcommand{\ed}[1]{{#1}}

\newcommand{\themis}{{\sc Themis}\xspace} 
\def\uas{\ensuremath{\mu}as\xspace} 
\newcommand{\foo}{Many Institutions}

\begin{document}
\shorttitle{Monitoring the Morphology of M87$^{\ast}$ in 2009--2017 with the EHT}
\title{Monitoring the Morphology of M87$^{\ast}$ in 2009--2017 with the Event Horizon Telescope}


\author[0000-0002-8635-4242]{Maciek Wielgus}
\affiliation{Black Hole Initiative at Harvard University, 20 Garden Street, Cambridge, MA 02138, USA}
\affiliation{Center for Astrophysics | Harvard \& Smithsonian, 60 Garden Street, Cambridge, MA 02138, USA}
\email{maciek.wielgus@gmail.com}

\author[0000-0002-9475-4254]{Kazunori Akiyama}
\affiliation{National Radio Astronomy Observatory, 520 Edgemont Rd, Charlottesville, VA 22903, USA}
\affiliation{Massachusetts Institute of Technology Haystack Observatory, 99 Millstone Road, Westford, MA 01886, USA}
\affiliation{National Astronomical Observatory of Japan, 2-21-1 Osawa, Mitaka, Tokyo 181-8588, Japan}
\affiliation{Black Hole Initiative at Harvard University, 20 Garden Street, Cambridge, MA 02138, USA}

\author[0000-0002-9030-642X]{Lindy Blackburn}
\affiliation{Black Hole Initiative at Harvard University, 20 Garden Street, Cambridge, MA 02138, USA}
\affiliation{Center for Astrophysics | Harvard \& Smithsonian, 60 Garden Street, Cambridge, MA 02138, USA}

\author[0000-0001-6337-6126]{Chi-kwan Chan}
\affiliation{Steward Observatory and Department of Astronomy, University of Arizona, 933 N. Cherry Ave., Tucson, AZ 85721, USA}
\affiliation{Data Science Institute, University of Arizona, 1230 N. Cherry Ave., Tucson, AZ 85721, USA}

\author[0000-0003-3903-0373]{Jason Dexter}
\affiliation{JILA and Department of Astrophysical and Planetary Sciences, University of Colorado, Boulder, CO 80309, USA}

\author[0000-0002-9031-0904]{Sheperd S. Doeleman}
\affiliation{Black Hole Initiative at Harvard University, 20 Garden Street, Cambridge, MA 02138, USA}
\affiliation{Center for Astrophysics | Harvard \& Smithsonian, 60 Garden Street, Cambridge, MA 02138, USA}

\author[0000-0002-7128-9345]{Vincent L. Fish}
\affiliation{Massachusetts Institute of Technology Haystack Observatory, 99 Millstone Road, Westford, MA 01886, USA}

\author[0000-0002-5297-921X]{Sara Issaoun}
\affiliation{Department of Astrophysics, Institute for Mathematics, Astrophysics and Particle Physics (IMAPP), Radboud University, P.O. Box 9010, 6500 GL Nijmegen, The Netherlands}

\author[0000-0002-4120-3029]{Michael D. Johnson}
\affiliation{Black Hole Initiative at Harvard University, 20 Garden Street, Cambridge, MA 02138, USA}
\affiliation{Center for Astrophysics | Harvard \& Smithsonian, 60 Garden Street, Cambridge, MA 02138, USA}

\author[0000-0002-4892-9586]{Thomas P. Krichbaum}
\affiliation{Max-Planck-Institut f\"ur Radioastronomie, Auf dem H\"ugel 69, D-53121 Bonn, Germany}

% ...



\begin{abstract}
The Event Horizon Telescope (EHT) has recently delivered the first resolved images of M87$^{\ast}$, the supermassive black hole in the center of the M87 galaxy. These images were produced using 230\,GHz observations performed in 2017 April. Additional observations are required to investigate the persistence of the primary image feature -- a ring with azimuthal brightness asymmetry -- and to quantify the image variability on event horizon scales. To address this need, we analyze 
M87$^{\ast}$ data collected with prototype EHT arrays
in 2009, 2011, 2012, and 2013. While these observations do not contain enough information to produce images, they are sufficient to constrain simple geometric models. 
We develop a~modeling approach based on the framework utilized for the 2017 EHT data analysis and validate our procedures using synthetic data. Applying the same approach to the observational data sets, we find the M87$^{\ast}$ morphology in 2009--2017 to be consistent with a~persistent asymmetric ring of ${\sim}40\,\mu$as diameter. The position angle of the peak intensity varies in time. In particular, we find a~significant difference between the position angle measured in 2013 and 2017. These variations are in broad agreement with predictions of a~subset of general relativistic magnetohydrodynamic simulations.
We show that quantifying the variability across multiple observational epochs has the potential to constrain the physical properties of the source, such as the accretion state or the black hole spin.
\end{abstract}


\section{Introduction}
\label{sec:intro}

\begin{figure}[t!]
\centering
\includegraphics[trim={1.5cm 0.8cm 0.2cm 0.cm},clip,width=1.0\linewidth]{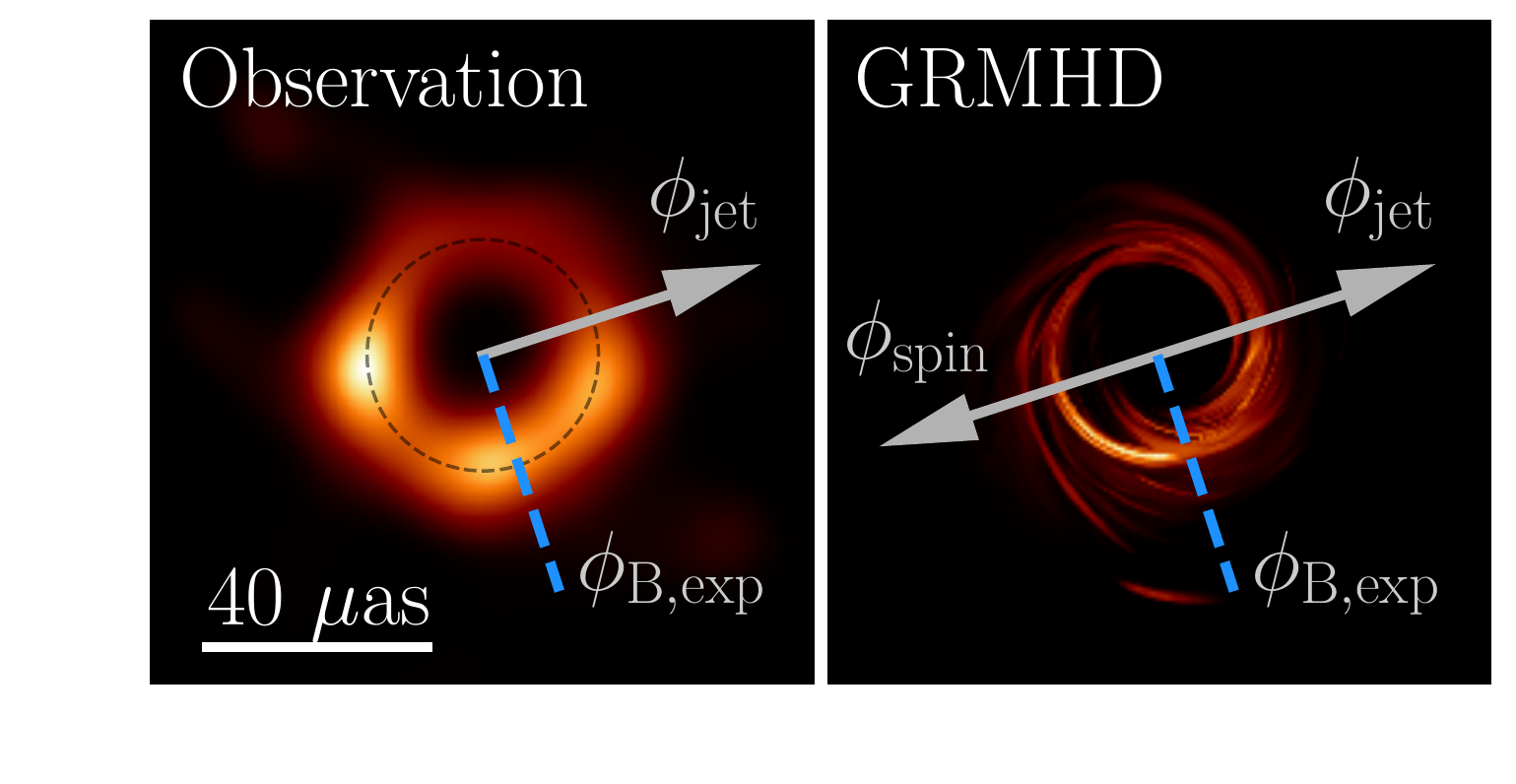}
\caption{\textit{Left panel:} one of the images of \m87 obtained in \citetalias{Paper4} (see \autoref{subsec:images} for details).
A~42\,$\mu$as circle is plotted with a dashed line for reference. The observed position angle of the approaching jet $\phi_{\rm jet}$ is 288$^\circ$ east of north \citep{Walker2018}. Under the assumed physical interpretation of the ring, we expect to find the bright side of the crescent on average approximately 90$^\circ$ clockwise from $\phi_{\rm jet}$ \citepalias{Paper5}. 
We assume a~convention $\phi_{\rm B,exp} = 198^\circ$, indicated with a~blue dashed line. \textit{Right panel:} a~random snapshot (note that this is not a fit to the EHT image) from a~GRMHD simulation adopting the expected properties of \m87 (\autoref{subsec:grmhd_snapshots}). The spin vector of the black hole is partially directed into the page, counteraligned with the approaching jet (and aligned with the deboosted receding jet); its projection onto the observer's screen is located at the position angle of $\phi_{\rm spin} = \phi_{\rm jet} - 180^\circ$. 
}
\label{fig:images_intro}
\end{figure}

The compact radio source in the center of the M87 galaxy, hereafter \m87, has been observed at 1.3 millimeter wavelength (230\,GHz frequency) using very long baseline interferometry (VLBI) since 2009. These observations, performed by early configurations of the Event Horizon Telescope \citep[EHT,][]{Doeleman2009} array, measured the size of the compact emission to be $\sim\,40\,\mu$as, with large systematic uncertainties related to the limited baseline coverage \citep{Shep2012,Kazu2015}. The addition of new sites and sensitivity improvements leading up to the April 2017 observations yielded the first resolved images of the source \citep[][hereafter EHTC~I-VI]{Paper1,Paper2,Paper3,Paper4,Paper5,Paper6}. These images revealed an asymmetric ring (a crescent) with a~diameter $d\,=\,42\pm3\,\mu$as and a~position angle of the bright side $\phi_{\rm B}$ between $150^\circ$ and $200^\circ$ east of north \citepalias[counterclockwise from north/up as seen on the sky,][]{Paper6}, see the left panel of \autoref{fig:images_intro}. The apparent size and appearance of the observed ring agree with theoretical expectations for a~$6.5 \times 10^9 M_\odot$ black hole driving a~magnetized accretion inflow/outflow system, inefficiently radiating via synchrotron emission \citepalias[\citealt{Yuan2014},][]{Paper5}. Trajectories of the emitted photons are subject to strong deflection in the vicinity of the event horizon, resulting in a~lensed ring-like feature seen by a distant observer -- the anticipated shadow of~a black hole \citep{Bardeen1973,Luminet1979,Falcke2000,Broderick2009}.

General relativistic magnetohydrodynamic (GRMHD) simulations of relativistic plasma in the accretion flow and jet-launching region close to the black hole \citepalias[][\citealt{Porth2019}]{Paper5} predict that the \m87 source structure will exhibit a~prominent asymmetric ring throughout multiple years of observations, with a~mean diameter $d$ primarily determined by the black hole mass-to-distance ratio and a position angle $\phi_{\rm B}$ primarily determined by the orientation of the black hole spin axis. The detailed appearance of \m87 may also be influenced by many poorly constrained effects, such as the black hole spin magnitude, magnetic field structure in the accretion flow \citepalias[\citealt{Narayan2012},][]{Paper5}, the electron heating mechanism \citep[e.g.,][]{Moscibrodzka2016,ChaelRowan2018}, nonthermal electrons \citep[e.g.,][]{Jordy2019}, and misalignment between the jet and the black hole spin \citep{White2020,Chatterjee2020}. Moreover, turbulence in the accretion flow, perhaps driven by the magnetorotational instability \citep{MRI}, is expected to cause stochastic variability in the image with correlation timescales of up to a~few weeks ($\sim$~dynamical time for \m87). The model uncertainties and expected time-dependent variability of these theoretical predictions strongly motivate the need for additional observations of \m87, especially on timescales long enough to yield uncorrelated snapshots of the turbulent flow.

To this end, we analyze archival EHT observations of \m87 from observing campaigns in 2009, 2011, 2012, and 2013. While these observations do not have enough baseline coverage to form images \citepalias{Paper4}, they are sufficient to constrain simple geometrical models, following procedures similar to those presented in \citetalias{Paper6}. We employ asymmetric ring models that are motivated by both results obtained with the mature 2017 array and the expectation from GRMHD simulations that the ring feature is persistent.

We begin, in \autoref{sec:observations}, by summarizing the details of these archival observations with the ``proto-EHT'' arrays. In \autoref{sec:modeling}, we describe our procedure for fitting simple geometrical models to these observations. In \autoref{sec:synthetic}, we test this procedure using synthetic proto-EHT observations of GRMHD snapshots and of the EHT images of \m87. We then use the same procedure to fit models to the archival observations of \m87 in \autoref{sec:ModelingRealData}. We discuss the implications of these results for our theoretical understanding of \m87 in \autoref{sec:discussion}, and briefly summarize our findings in \autoref{sec:summary}.


\section{Observations and data}
\label{sec:observations}

\begin{table*}[htp]
     \caption{\m87 data sets analyzed in this paper.
     }
    \begin{center}
    \tabcolsep=0.095cm
    \begin{tabularx}{1.0\linewidth}{cccccccccc}
    \hline
    \hline
    & &  & &\multicolumn{5}{c}{Detections on Nonredundant Baselines} &  \\
    {Year } &  Telescopes & Dates &  Baselines$^\text{a}$ & Zero & Short & Medium$^\text{b}$ & Long$^\text{c}$ & Total & CPs \\
    {} & & & & $<\!0.1$G$\lambda$ & $<\!1$G$\lambda$ & $<\!3.6$G$\lambda$ & $>\!3.6$G$\lambda$   & $>\!0.1$G$\lambda$  \\
    \hline
    2009 & CA, AZ, JC & Apr 5, 6 & 3/3/3 & --- & 12 & 16/5 &---& 28 & --- \\
    2011 & CA, AZ, JC, SM, CS & Mar 29, 31; Apr 1, 2, 4 & 10/6/3 & 52 & 33 & 21/6 &---& 54 & --- \\
    2012 & CA, AZ, SM & Mar 21 & 3/3/3 & 14 & 11 & 19/6 &---& 44 & 7\\
    2013 & CA, AZ, SM, JC, AP & Mar 21 -- 23, Mar 26 & 10/7/5 & 39 & 41 & 23/4 &19/1& 83 & --- \\
    2017 & AZ, SM, JC, AP, LM, PV, AA & Apr 6$^\text{d}$ & 21/21/10 & 24 & --- & 33/13 &92/16& 125 & 67 \\
    2017 & AZ, SM, JC, AP, LM, PV, AA & Apr 11$^\text{d}$ & 21/21/10 & 22  & --- & 28/9 &72/16& 100 & 54 \\
    \hline 
    \hline
    \end{tabularx}
    \label{tab:detections}
    \end{center}
$^\text{a}$ theoretically available\,/\,with detections\,/\,nonredundant, nonzero with detections, $^\text{b}$ all / SMT-Hawai'i, $^\text{c}$ all / SMT-Chile, $^\text{d}$~single-day data set
\end{table*}

\begin{figure*}[t!]
\centering
\includegraphics[width=0.99\linewidth]{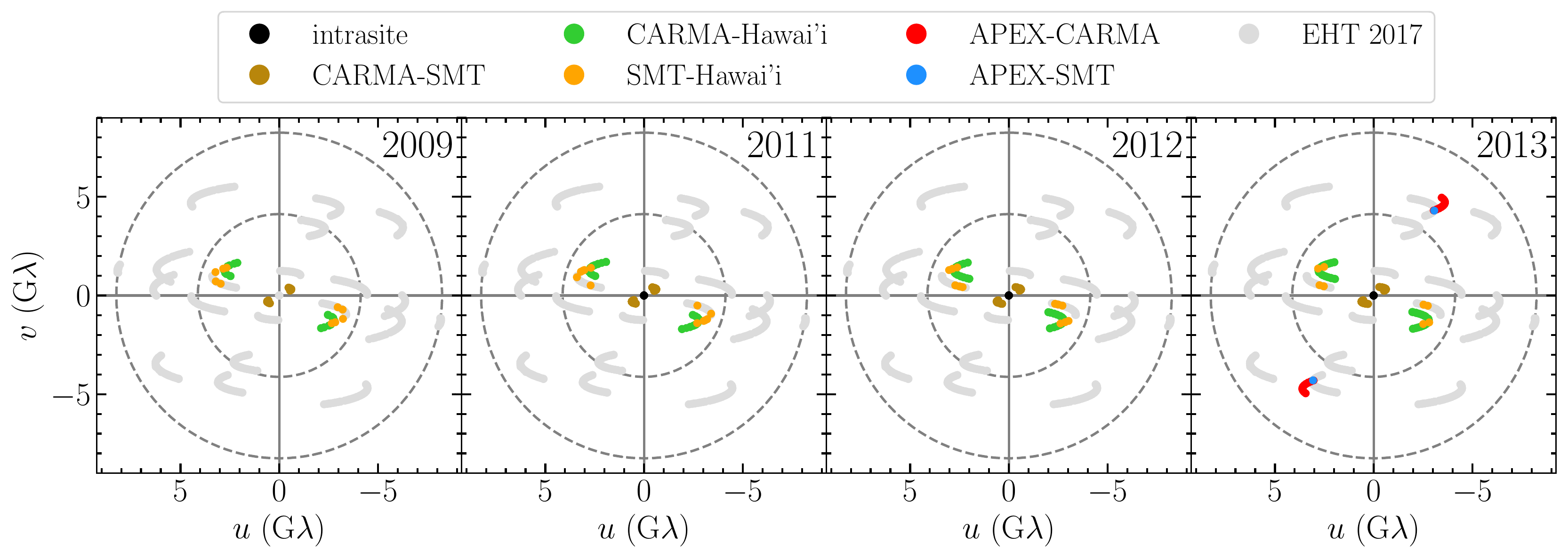}
\caption{$(u,v)$-coverage of the \m87 observations performed in 2009--2013 with various proto-EHT arrays. Gray lines indicate detections obtained during the 2017 observations with a~mature EHT array, including several new sites, but without the baselines to CARMA. Dashed circles correspond to angular scales of 50\,$\mu$as (inner) and 25\,$\mu$as (outer).}
\label{fig:coverage}
\end{figure*}

Our analysis covers five separate 1.3\,mm VLBI observing campaigns conducted in 2009, 2011, 2012, 2013, and 2017. The \m87 data from 2011 and 2013 have not been published previously. For all campaigns except 2012, \m87 was observed on multiple nights. For the proto-EHT data sets (2009--2013) we simultaneously utilize the entire data set from each year, fitting to data from multiple days with a~single source model, when available. This is motivated by the \m87 dynamical timescale argument, little visibility amplitude variation reported by \citetalias{Paper3} on a~one-week timescale, as well as by the limited amount of available data and lack of evidence for interday variability in the proto-EHT data sets. 
We use incoherent averaging to estimate visibility amplitudes on each scan ($\sim$ few minutes of continuous observation) and bispectral averaging to estimate closure phases \citep{Rogers1995,MDJ2015,fish2016}. The frequency setup in 2009--2013 consisted of two 480\,MHz bands, centered at 229.089 and 229.601\,GHz. Whenever both bands or both parallel-hand polarization components were available, we incoherently averaged all simultaneous visibility amplitudes.
The data sets are summarized in \autoref{tab:detections}, where the number of detections for nonredundant baselines of different projected baseline lengths is given, with the corresponding $(u,v)$-coverage shown in \autoref{fig:coverage}. Redundant baselines yield independent observations of the same visibility. In \autoref{tab:detections} we also indicate the number of available nonredundant closure phases (CPs, not counting redundant and intrasite baselines, minimal set, see \citealt{closures}). As is the case for non-phase-referenced VLBI observations \citep{TMS}, we do not have access to absolute visibility phases. All visibility amplitudes observed in 2009--2013 are presented in \autoref{fig:fluxes}.

A~more detailed summary of the observational setup of the proto-EHT array in 2009--2013 and the associated data reduction procedures can be found in \citet{fish2016}. All data sets discussed in this paper are publicly available\footnote{ \url{https://eventhorizontelescope.org/for-astronomers/data}}.

\subsection{2009--2012}

Prior to 2013, the proto-EHT array included telescopes at three geographical locations:
(1) the Combined Array for Research in Millimeter-wave Astronomy (CARMA, CA) in Cedar Flat, California, (2) the Submillimeter Telescope (SMT, AZ) on Mt. Graham in Arizona, and (3) the Submillimeter Array (SMA, SM), the James Clerk Maxwell Telescope (JCMT, JC), and the Caltech Submillimeter Observatory (CSO, CS) on Maunakea in Hawai'i. These arrays were strongly east-west oriented, and the longest projected baselines, between SMT and Hawai'i, reached about 3.5 G$\lambda$, corresponding to the instrument resolution (maximum fringe spacing) of $\sim$\,60\,$\mu$as.

The 2011 observations of \m87 have not been published but follow the data reduction procedures described in \citet{Lu2013}. The 2009 and 2012 observations and data processing of \m87 have been published in \citet{Shep2012} and \citet{Kazu2015}, respectively. However, our analysis uses a modified processing of the 2012 data because the original processing erroneously applied the same correction for atmospheric opacity at the SMT twice.\footnote{An opacity correction raises visibility amplitudes on SMT baselines by $\sim$10\% in nominal conditions; our visibility amplitudes on SMT baselines are, thus, slightly lower than those reported by \citet{Kazu2015}. However, the calibration error does not change the primary conclusions of \citet{Kazu2015}.} The SMT calibration procedures have been updated since then \citep{Issaoun2017b}.

Each observation included multiple subarrays of CARMA as well as simultaneous measurements of the total source flux density with CARMA acting as a~connected-element interferometer; these properties then allow the CARMA amplitude gains to be ``network calibrated'' \citepalias[\citealt{Fish2011, MDJ2015},][]{Paper3}. Of these three observing campaigns, only 2012 provides closure phase information for \m87, and all closure phases measured on the single, narrow triangle SMT--SMA--CARMA were consistent with zero to within $2\,\sigma$ \citep{Kazu2015}, see \autoref{fig:closures}. 

\subsection{2013}
The 2013 observing epoch did not include the CSO, but added the 
Atacama Pathfinder Experiment facility (APEX, AP) in the Atacama Desert in Chile. This additional site brought for the first time the long (${\approx 5-6\,\rm G}\lambda$) baselines CARMA--APEX and SMT--APEX, that are roughly orthogonal to the CARMA--Hawai'i and SMT--Hawai'i baselines, see \autoref{fig:coverage}. The addition of APEX increased the instrument resolution (maximum fringe spacing) to $\sim$\,35\,$\mu$as.
While the 2013 observations of \sgra\ were presented in several publications \citep{MDJ2015,fish2016,RuSen2018}, the \m87 observations obtained during the 2013 campaign have not been published previously.

The proto-EHT array observed \m87 on March 21st, 22nd, 23rd, and 26th 2013.  
CARMA--APEX detections were found
on March 22nd (11 detections) and 23rd (7 detections) with a~single SMT--APEX detection on March 23rd. March 23rd (MJD 36374) was the only day with detections on baselines to each of the four geographical sites. No detections between Hawai'i and APEX were found, and there were no simultaneous detections over a~closed triangle that would allow for the measurement of closure phase.

\subsection{2017}
In 2017, the EHT observed \m87 with five geographical sites \citepalias{Paper1,Paper2}, without CSO and CARMA, but with the addition of the Large Millimeter Telescope Alfonso Serrano (LMT, LM) on the Volc\'{a}n Sierra Negra in Mexico, the IRAM 30-m telescope (PV) on Pico Veleta in Spain, and the phased-up Atacama Large Millimeter/submillimeter Array \cite[ALMA, AA,][]{phasingALMA,Goddi2019}. 
The expansion of the array resulted in significant improvements in $(u,v)$-coverage, shown with gray lines in \autoref{fig:coverage}, and instrument resolution raised to $\sim$\,$25\,\mu$as. In addition to hardware setup developments \citepalias{Paper2}, the recorded bandwidth was increased from 2$\times$0.5\,GHz to 2$\times$2\,GHz (226-230\,GHz).
The 2017 data processing pipeline used ALMA as an anchor station \citepalias{Paper3}. Its high sensitivity greatly improved the signal phase stability \citepalias[\citealt{hops2019,casa},][]{Paper3}
and enabled data analysis based on robustly detected closure quantities \ed{obtained from coherently averaged visibilities} \citepalias[][\citealt{closures}]{Paper4} rather than on visibility amplitudes alone.
These improvements
allowed for 
an unambiguous analysis of the \m87 image by constraining the set of physical \citepalias{Paper5} and geometric \citepalias{Paper6} models representing the source morphology.

\begin{figure*}
\centering
\includegraphics[width=0.99\linewidth]{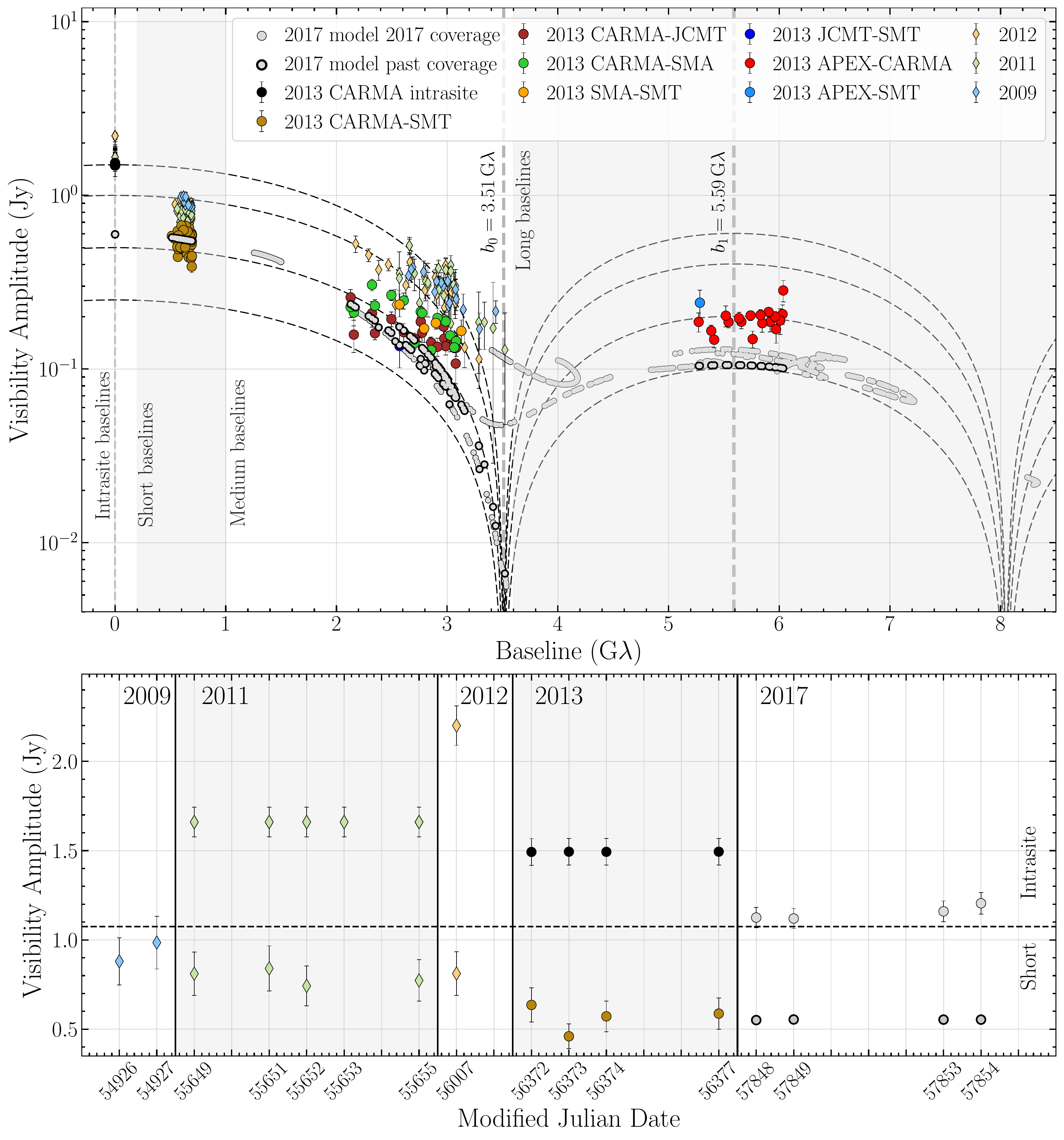}

\caption{\textit{Top:} Visibility amplitudes of \m87 detections in 2009--2013 as a function of projected baseline length $\sqrt{u^2 + v^2}$. The source model derived from the EHT 2017 observations is shown with gray dots. Gray dots with black borders show the predicted visibility amplitudes of the source model at the baselines of the prior observations in 2009--2013. Dashed black lines correspond to the family of Fourier transforms of a symmetric, infinitely thin ring of diameter $d_0 = 45.0\,\mu$as. \textit{Bottom:} total arcsecond-scale flux density (on intrasite baselines, network-calibrated), and compact emission flux density from the short CARMA--SMT baseline. In the case of the short baselines in 2017, predictions of the source model are given.}
\label{fig:fluxes}
\end{figure*}

\subsection{\m87 data properties}
\label{subs:salient_features}
VLBI observations sample the Fourier transform of the intensity distribution on the sky $I(x,y)$ via the van Cittert--Zernike theorem \citep{vancittert,zernike}
\begin{equation}
    V(u,v) = \iint I(x,y) e^{-2 \pi \text{i}(xu + vy) }  \text{d} x \text{d} y \ ,
\end{equation}
where the measured Fourier coefficients $V(u,v)$ are referred to as ``visibilities'' \citep{TMS}. \ed{When an array of $N$ telescopes observes a~source, $N(N-1)/2$ independent visibility measurements are obtained, provided detections on all baselines are found.} Certain properties of the geometry described by $I(x,y)$ can be inferred directly from inspecting the visibility data.

In the top panel of \autoref{fig:fluxes} we summarize all the \m87 detections obtained during 2009--2013 observations as a~function of projected baseline length $ \sqrt{u^2 + v^2}$.
Dashed lines represent $\widehat{R}$, the analytic Fourier transform of an infinitely thin ring with a total intensity $I_0$ and a~diameter $d_0$,
\begin{equation}
\widehat{R}\left( \sqrt{u^2 + v^2} \right) = I_0 J_0 \left(  \pi d_0 \sqrt{u^2 + v^2}\right) \ ,
\end{equation}
where $J_0$ is a~zeroth-order Bessel function of the first kind. This simple analytic model predicts the visibility null located at 
\begin{equation}
b_0 \approx 3.51 \left(\frac{d_0}{45\,\mu \text{as}} \right)^{-1} \text{G}\lambda \, ,
\label{eq:null}
\end{equation}
and a wide plateau around the first maximum, located at
\begin{equation}
b_1 \approx 5.59 \left(\frac{d_0}{45\,\mu \text{as}} \right)^{-1}  \text{G}\lambda \, ,
\label{eq:plateau}
\end{equation}
recovering about 40\% of the flux density seen on short baselines. In \autoref{fig:fluxes} we use $d_0\,=\,45.0\,\mu$as and show $\widehat{R}$ curves corresponding to $I_0\,=\,$0.25,\,0.5,\,1.0,\,1.5 to guide the eye. The behaviors of the visibility amplitudes, particularly the fall-off rate seen on medium-length baselines (1.0--3.6\,G$\lambda$) in all data sets and the flux density recovery on long baselines to APEX in 2013, are roughly consistent with that of a~simple ring model. Moreover, all detections on baselines to APEX have a~similar flux density of $\sim\,0.2$\,Jy, while the projected baseline length varies between 5.2 and 6.1\,G$\lambda$. In the analytic thin ring model framework, this can be readily understood, because baselines to APEX sample the wide plateau around the maximum located at $b_1$, \autoref{eq:plateau}.

The gray dots in \autoref{fig:fluxes} correspond to the source model constructed based on the 2017 EHT observations -- the mean of the four images reconstructed for April 5th, 6th, 10th and 11th 2017 with the \texttt{eht-imaging} pipeline \citepalias[\citealt{Chael2016},][]{Paper4}. In the 2017 model, east--west baselines, such as SMT--Hawai'i, probe a deep visibility null located around $b_0$ (\autoref{eq:null}), where sampled amplitudes drop below 0.01 Jy. North--south baselines do not show a similar feature, which indicates source asymmetry. Irrespective of the orientation, visibility amplitudes flatten out around $b_1$. Gray dots with black envelopes represent the 2017 source model sampled at the $(u,v)$-coordinates of the past observations, for which all medium-length baselines were oriented in the east--west direction. 

One can immediately notice interesting discrepancies. The visibility amplitudes measured on long baselines to APEX (projected baseline length $\sim\,b_1$) in 2013 were about a~factor of 2 larger than the corresponding 2017 source model predictions. At the same time, the flux density on the~short CARMA--SMT baseline is consistent between the 2013 measurements and the 2017 model predictions. This shows that the image on the sky has changed between 2013 and 2017 in a~structural way, which cannot be explained with a~simple total intensity scaling. 
We also notice that several detections obtained in 2009--2011, corresponding to projected $(u,v)$-distances of 3.2-3.5 G$\lambda$ on Hawai'i--USA baselines, record flux density above 0.1\,Jy. At the same time, the 2017 model predicts that these baselines sample a~visibility null region around $b_0$, with a flux density lower by an order of magnitude. However, the compact flux (on short baselines) did not change by more than a~factor of two, remaining between 0.5 and 1.0\,Jy throughout the 2009--2017 observations, see the bottom panel of \autoref{fig:fluxes}. This suggests that the null location in the past (if present) was different than that observed in 2017, which may correspond to 
a~fluctuation of the crescent position angle or a~changing degree of source symmetry.

\begin{figure}
\centering
\includegraphics[width=0.99\linewidth]{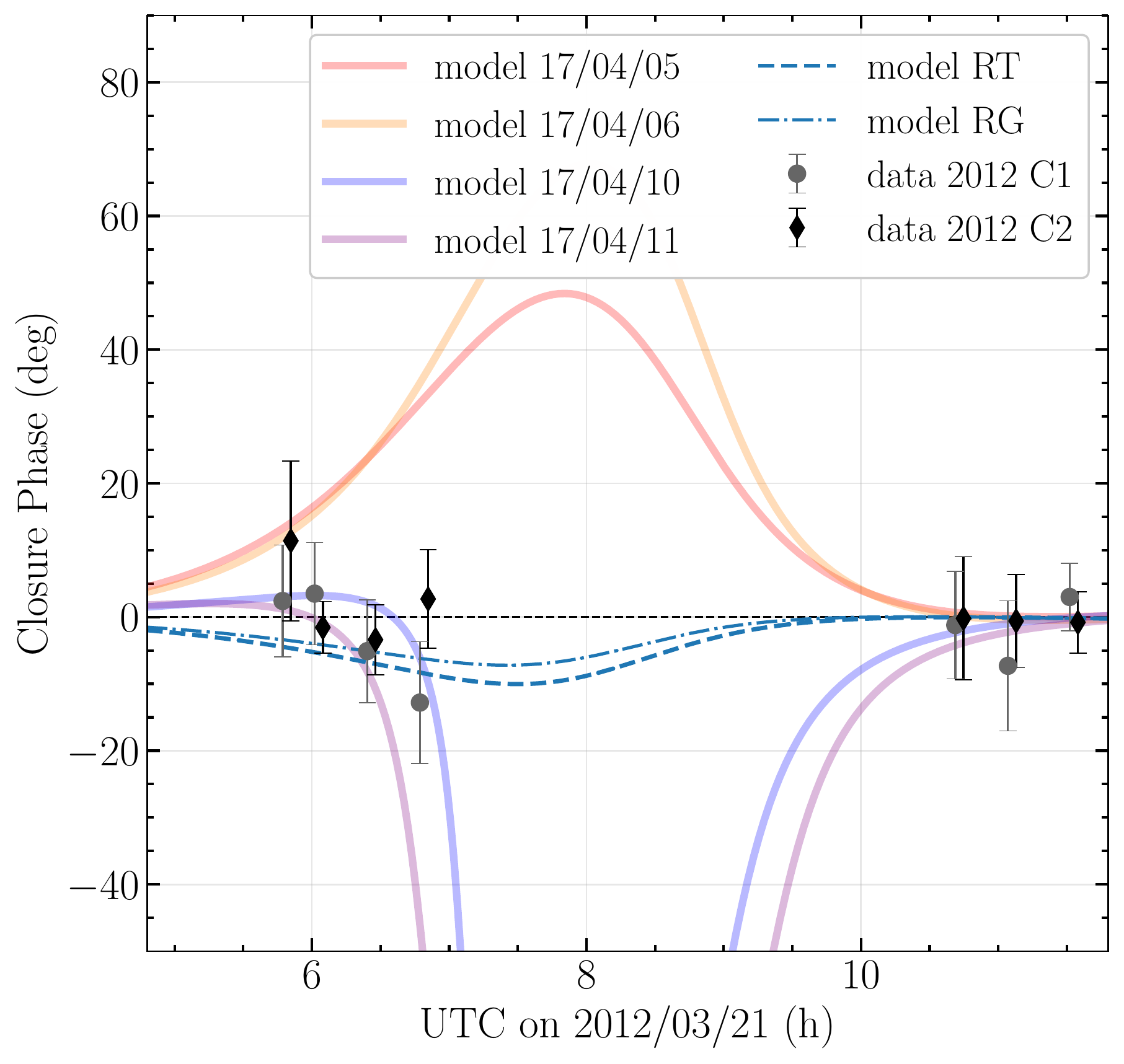}
\caption{ Consistency of the closure phases on the SMT--SMA--CARMA triangle between the values observed in 2012 \citep{Kazu2015} and numerically resampled source models constructed based on the 2017 observations. The predictions of the asymmetric ring models RT and RG fitted to 2012 observations are also given, see \autoref{sec:modeling} and \autoref{sec:ModelingRealData}. Data corresponding to two CARMA subarrays, C1 and C2, are shown separately.
}
\label{fig:closures}
\end{figure}

Apart from the visibility amplitude data, a~limited number of closure phases from the narrow triangle SMT--SMA--CARMA has been obtained from the 2012 data set \citep{Kazu2015}. All of these closure phases are measured to be consistent with zero, which suggests a~high degree of east--west symmetry in the geometry of the source observed in 2012. While the closure phases on this triangle were not observed in 2017 (CARMA was not part of the 2017 array), we can numerically resample the 2017 images \citepalias[\texttt{eht-imaging} reconstructions,][]{Paper4} to verify the consistency. In \autoref{fig:closures} we show the closure phases obtained in 2012, averaged between bands, the two CARMA subarrays shown separately. Near-zero closure phases observed in 2012 are roughly consistent with at least some models from 2017. Unfortunately, technical difficulties that occurred during the 2012 campaign precluded obtaining measurements between UTC 7.5 and 10.5, where nonzero closure phases are predicted by all 2017 models.

Altogether, we see strong suggestions that the 2009--2013 data sets describe a similar 
geometry to the 2017 results, but there are also substantial hints that the detailed properties of the source structure evolved between observations. These differences can be quantified with geometric modeling of the source morphology.



\section{Modeling Approach}
\label{sec:modeling}

\begin{figure}[t]
\centering
\includegraphics[width=1.0\columnwidth]{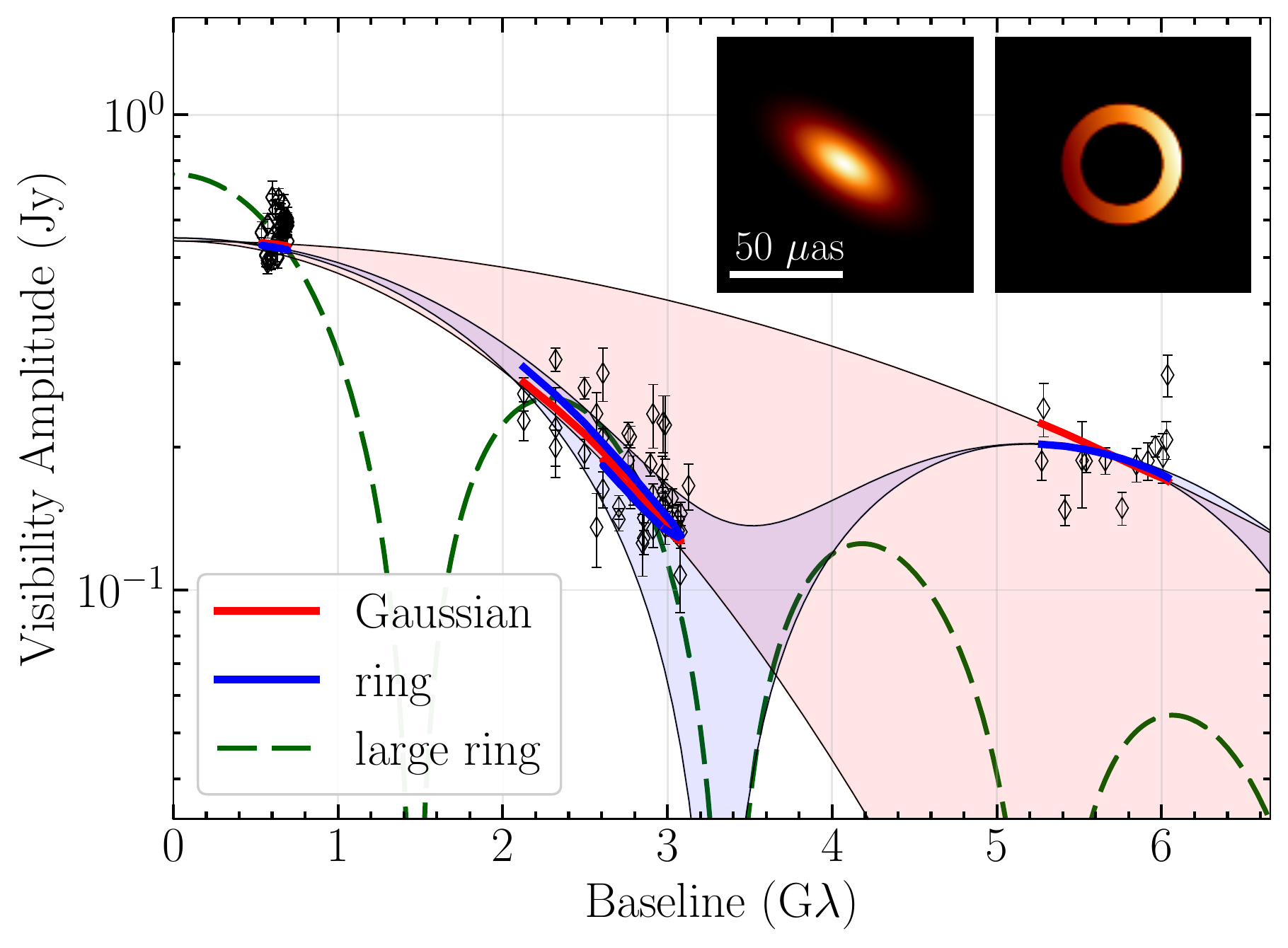}
\caption{Comparison of maximum likelihood (ML) asymmetric Gaussian and asymmetric ring models fitted to the 2013 data. Data are shown as points with error bars corresponding to the thermal errors. The shaded regions cover all amplitudes for a given model. The red and blue lines represent models evaluated at the $(u,v)$-coordinates of the observations. Both models offer a very similar fit quality. ML estimators are shown as inset figures. The model of a~ring with roughly double the diameter (dashed curve) fits the intermediate baselines, but is excluded by long-baseline amplitudes.
}
\label{fig:ring_or_gauss}
\end{figure}

The sparse nature of the pre-2017 data sets precludes reconstructing images in the manner employed for the 2017 data \citepalias{Paper4}.  However, the earlier data are still capable of providing interesting constraints on more strictly parameterized classes of models.  \autoref{fig:ring_or_gauss} shows the 2013 data set overplotted with a best-fit ring model\footnote{This is the maximum likelihood estimator for the slashed thick ring model (RT), as discussed in \autoref{sec:ModelSpecification}.} (in blue; 5 degrees of freedom) and asymmetric Gaussian model (in red; 4 degrees of freedom). Both models attain similar fit qualities, as determined by Bayesian and Akaike information criteria \citep[see, e.g.,][]{Liddle2007}.
In the absence of prior information, we would be unable to confidently select a~preferred model. 
However, the robust image morphology reconstructed from the 2017 data provides a natural and strong prior for selecting an appropriate parameterization. The ``generalized crescent'' (GC) geometric models considered in \citetalias{Paper6} yielded fit qualities comparable to those of image reconstructions for the 2017 data, and in this paper, we apply two variants of the GC model to the pre-2017 data sets. Owing to data sparsity, we restrict the parameter space of the models to a~subset of that considered in \citetalias{Paper6} containing only a handful of parameters of interest.  

Throughout this paper, we use perceptually uniform color maps from the \texttt{ehtplot} library\footnote{\url{https://github.com/liamedeiros/ehtplot}} to display the images. In some of the figures we present models blurred to a~resolution of 15\,$\mu$as, adopted in this paper as the effective resolution of the EHT. The EHT instrument resolution measured as the maximum fringe spacing in 2017 is about 25\,$\mu$as, however, for the image reconstruction methods employed in \citetalias{Paper4} a~moderate effect of superresolution can be expected \citep{Mareki2014,Chael2016}. The effect may be much more prominent for the geometric models, which are not fundamentally limited by the resolution.

\subsection{Model specification} \label{sec:ModelSpecification}

Given the image morphology inferred from the 2017 data, the primary parameters of interest we would like to constrain using the earlier data sets are the size of the source, the orientation of any asymmetry, and the presence or absence of a central flux depression.  The analyses presented in this paper use two simple ring-like models -- similar to those presented in  \citet{Kamruddin2013} and \citet{Benkevitc2016} -- both of which are subsets of the GC models from \citetalias{Paper6}.

The first model we consider is a concentric ``slashed'' ring, where the ring intensity is modulated by a~linear gradient, hereafter denoted as RT.
In this model, the flux is contained within a circular annulus with the inner and outer radii $R_{\rm in}$ and $R_{\rm out}$, respectively.  The model is described by five parameters:

\begin{enumerate}
    \item the mean diameter of the ring $d = R_{\rm in} + R_{\rm out}$,
    \item the position angle of the bright side of the ring $0 \le \phi_{\rm B} < 2 \pi$,
    \item the fractional thickness of the ring $0< \psi = 1 - R_{\rm in}/R_{\rm out} < 1$,
    \item the total intensity $ 0 < I_0 < 2$\,Jy, and
    \item $\beta$, an intensity gradient (``slash'') across the ring in the direction given by $\phi_{\rm B}$, corresponding to the ratio between the dimmest and brightest points on the ring, $0 < \beta < 1$.  A ring of uniform brightness has $\beta=1$, while a ring with vanishing flux at the dimmest part has $\beta = 0$.
\end{enumerate}

\noindent This model reduces to a slashed disk for $\psi \rightarrow 1$. The assumed definition of mean diameter is consistent with the one used in \citetalias{Paper6}, allowing for direct comparisons. Except where otherwise specified, we use this first model for the analyses discussed in this paper.

As a check against model-specific biases, we consider a second model consisting of an infinitesimally thin slashed ring, blurred with a Gaussian kernel \citepalias{Paper4}.  The equivalent five parameters for this model are:

\begin{enumerate}
    \item the mean diameter of the ring $d = 2 R_{\rm in} = 2 R_{\rm out}$,
    \item the position angle of the bright side of the ring $0 \le \phi_{\rm B} < 2 \pi$,
    \item the width of the Gaussian blurring kernel $0 < \sigma < 40\,\mu$as,
    \item the total intensity $ 0 < I_0 < 2$\,Jy, and
    \item the slash $0 < \beta < 1$.
\end{enumerate}

\noindent This second model, hereafter referred to as RG, reduces to a~circular Gaussian for $d \ll \sigma$.

Both the RT and RG models provide a~measure of the source diameter ($d$), the orientation of the brightness asymmetry ($\phi_{\text{B}}$), and the presence of a central flux depression.  We quantify the latter property using the following general measure of relative ring thickness \citepalias[from][]{Paper6}
\begin{equation}
    f_{\rm w} = \frac{R_{\rm out} - R_{\rm in} + 2\sigma \sqrt{2 \ln 2}}{d} \ ,
\label{eq:fw}
\end{equation}
where $R_{\rm out} = R_{\rm in}$ for the RG model and $\sigma = 0$ for the RT model.

All data sets except 2009 contain observations from intrasite baselines (``zero baselines''), see \autoref{tab:detections}. For \m87, these baselines are sensitive to the flux from the extended jet emission on $\sim$\,arcsecond scales \citepalias[][see also the bottom panel of \autoref{fig:fluxes}]{Paper4} and do not directly inform us about the compact source structure on scales of $\sim$\,tens of microarcseconds. However, the intrasite baselines still provide useful constraints on station gain parameters (see \autoref{sec:FittingProcedure}), and hence, we do not flag them.  Rather, we parameterize this large-scale flux using a~large symmetric Gaussian component consisting of two parameters, flux and size. This component is entirely resolved out on intersite baselines and thus has no direct impact on the compact source geometry. Ultimately, the models that we use have 5 geometric parameters for the 2009 data set and 5+2=7 geometric parameters in all other cases.

\subsection{Fitting procedure and priors} \label{sec:FittingProcedure}

We perform the parameter estimation for this paper using \themis, an analysis framework developed by \cite{broderick20} for the specific requirements of EHT data analysis. \themis operates within a Bayesian formalism, employing a differential evolution Markov chain Monte Carlo (MCMC) sampler to explore the posterior space.  Prior to model fitting, the data products are prepared in a manner similar to that described in \citetalias{Paper6}.  Descriptions of the likelihood constructions for different classes of data products are given in \cite{broderick20}.

One important difference between the 2017 and pre-2017 data sets is that the latter contain almost exclusively visibility amplitude information, rather than having access to the robust closure quantities in both phase and amplitude \citep{TMS,closures} that aided interpretation of the 2017 data. In addition to thermal noise, visibility amplitudes suffer from uncertainties in the absolute flux calibration, including potential systematic effects such as losses related to telescope pointing imperfections \citepalias{Paper3}. These uncertainties are parameterized within \themis using station-based amplitude gain factors $g_i$, representing the scaling between the geometric model amplitudes $|\bar{V}_{ij}|$ and the gains-corrected model amplitudes $|\hat{V}_{ij}|$,
\begin{equation}
\label{eq:gains}
|\hat{V}_{ij}| = (1 + g_i)(1+ g_j) |\bar{V}_{ij}| \, .
\end{equation}
Model amplitudes $|\hat{V}_{ij}|$ are then compared with the measured visibility amplitudes $|V_{ij}|$.
Within \themis, the number of amplitude gain parameters $N_{\rm g}$ is equal to the number of (station, scan) pairs, i.e., the gains are assumed to be 
constant across a~single scan
but uncorrelated from one scan to another. By explicitly modeling station-based gains, we correctly account for the otherwise covariant algebraic structure of the visibility calibration errors \citep{closures}.
At each MCMC step, \themis marginalizes over the gain amplitude parameters (subject to Gaussian priors) using a quadratic expansion of the log-likelihood around its maximum given the current parameter vector; see \cite{broderick20} for details. For the analysis of the 2009--2013 data sets presented in this paper, we have adopted rather conservative 15\% amplitude gain uncertainties for each station, represented by symmetric Gaussian priors with a~mean value of 0.0 and standard deviation of 0.15. The width of these priors reflects our confidence in the flux density calibration rather than the statistical variation in the visibility data.

The RT model is parameterized within \themis in terms of $R_{\text{out}}$, $\phi_{\rm B}$, $\psi$, $I_0$, and $\beta$.  Uniform priors are used for each of these parameters, with ranges of $[0,200]$\,\uas for $R_{\text{out}}$, $[0,2\pi]$ for $\phi_{\rm B}$, $[0,1]$ for $\psi$, $[0,2]$\,Jy for $I_0$, and $[0,1]$ for $\beta$.  We achieve the ``infinitesimally thin'' ring of the RG model within \themis by imposing a strict prior on $\psi$ of $[10^{-7},10^{-6}]$, and the prior on $\sigma$ is uniform in the range $[0,40]$\,\uas.  Because $d$ and $f_{\rm w}$ are derived parameters, we do not impose their priors directly but rather infer them from appropriate transformations of the priors on $R_{\text{out}}$, $\psi$, and $\sigma$.  The effective prior on $d = R_{\text{out}} (2 - \psi)$ is given by
\begin{equation}
\pi(d) = \begin{cases}
\frac{\ln(2)}{200} , & 0\,\text{\uas} \leq d < 200\,\text{\uas} \\
\frac{1}{200} \ln\left( \frac{400\,\text{\uas}}{d} \right) , & 200\,\text{\uas} \leq d < 400\,\text{\uas} \\
0 , & \text{otherwise}
\end{cases} ,
\end{equation}

\noindent which is uniform within the range $[0,200]$\,\uas.  For the RT model, the effective prior on $f_{\rm w} = \psi / (2 - \psi)$ is given by

\begin{equation}
\pi_{\text{RT}}(f_{\rm w}) = \begin{cases}
\frac{2}{(1 + f_{\rm w})^2} , & 0 \leq f_{\rm w} \leq 1 \\
0 , & \text{otherwise}
\end{cases} ,
\end{equation}

\noindent which is not uniform but rather increases toward smaller values.  For the RG model, the effective prior on $f_{\rm w} = \sigma \sqrt{2 \ln(2)} / R_{\text{out}}$ is given by

\begin{equation}
\pi_{\text{RG}}(f_{\rm {\rm w}}) = \begin{cases}
\frac{1}{2 \alpha} , & 0 \leq f_{\rm w} \leq \alpha \\
\frac{\alpha}{2 f_{\rm w}^2} , & f_{\rm w} > \alpha \\
0 , & \text{otherwise}
\end{cases} ,
\end{equation}

\noindent where $\alpha = \sqrt{2 \ln(2)} / 5 \approx 0.235$ for our specified priors on $\sigma$ and $R_{\text{out}}$; this prior is uniform within the range $[0,\alpha]$.

\subsection{Degeneracies and limitations}
\label{sub:degenercies}

\begin{figure*}[ht]
\centering
\includegraphics[trim={0cm 0.3cm 0 0.1cm},clip,width=0.98\linewidth]{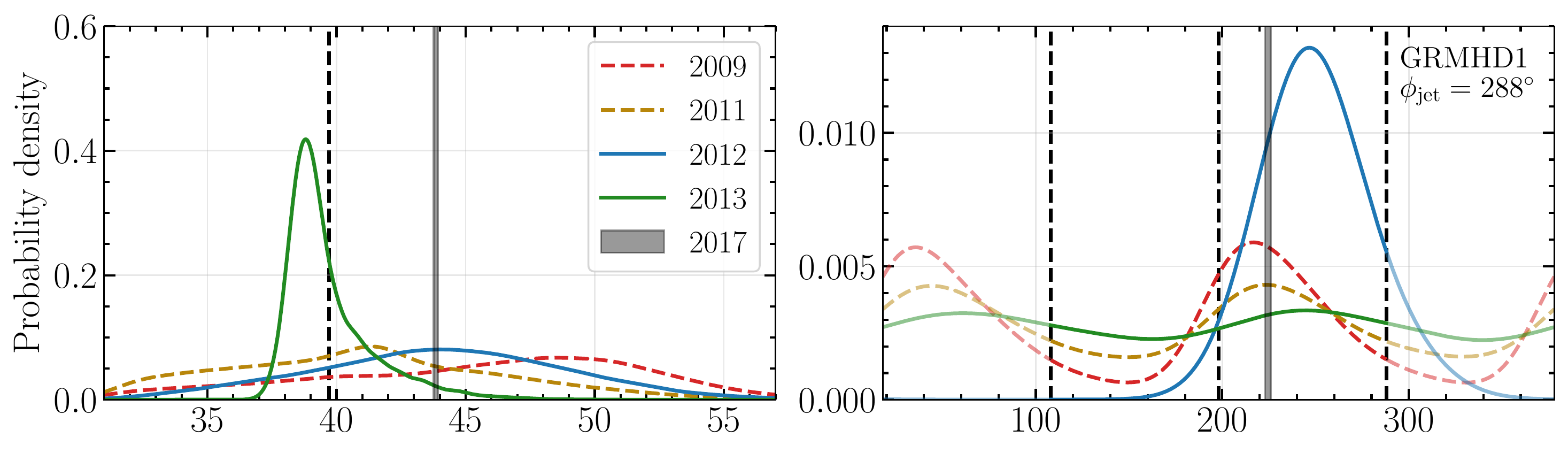}
\includegraphics[trim={0cm 0cm 0 0.2cm},clip,width=0.98\linewidth]{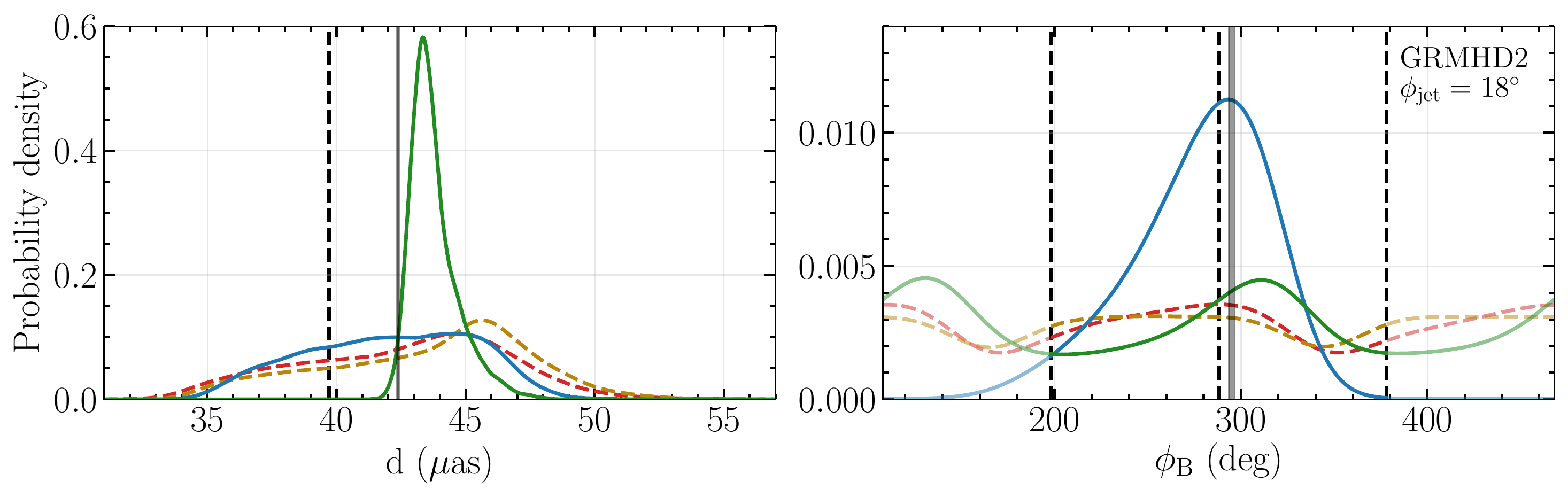}
\includegraphics[trim={0cm 0.3cm 0 0.1cm},clip,width=0.98\linewidth]{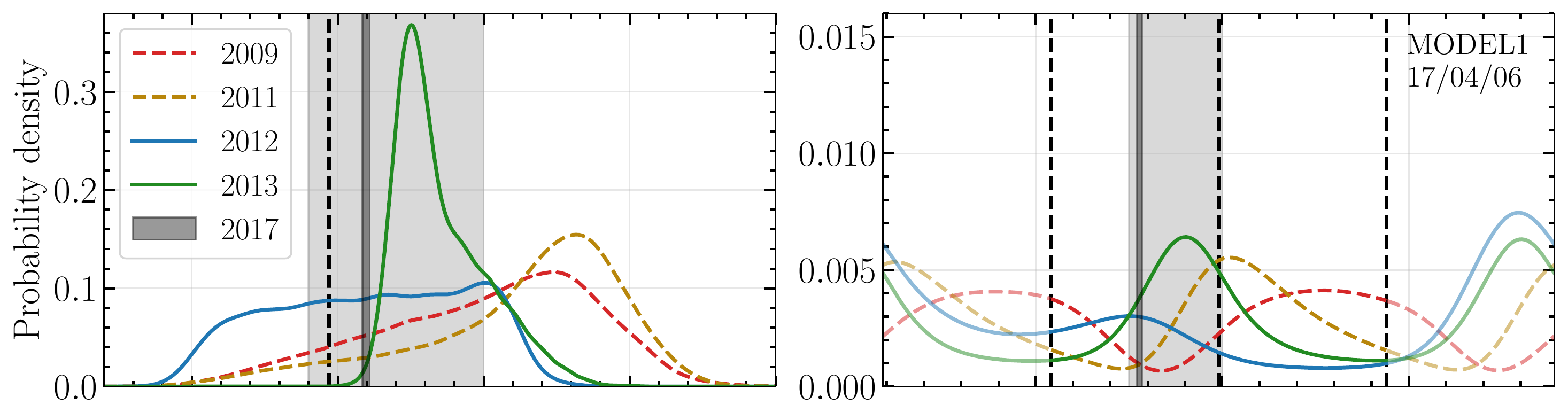}
\includegraphics[trim={0cm 0cm 0 0.2cm},clip,width=0.98\linewidth]{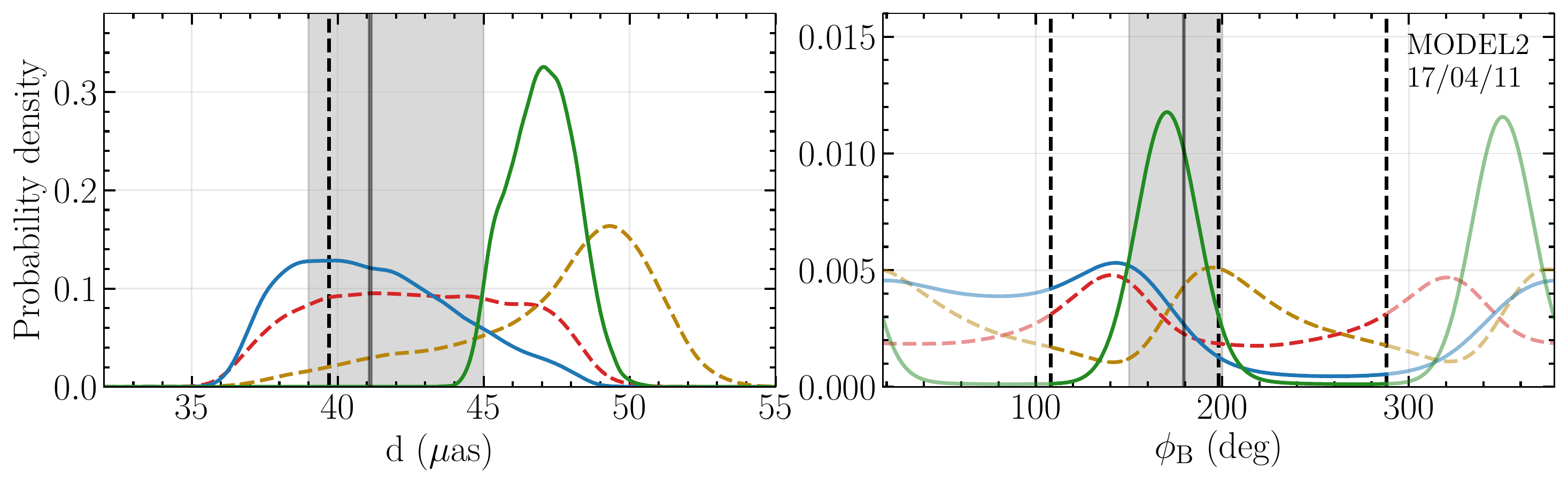}
\caption{\textit{Top two rows:} marginalized distributions of the mean diameter $d$ and brightness maximum position angle $\phi_{\mathrm{B}}$ for the RT model fits to GRMHD simulation snapshots GRMHD1 (first row) and GRMHD2 (second row). The 2017 posteriors are contained within the dark gray bands. The dashed vertical line in the left panels denotes diameter of $2\sqrt{27}M/D$. The vertical dashed lines in the right panels denote the convention angle $\phi_{\rm B, exp}$, $\phi_{\rm B, exp} - 90^\circ$, and the approaching jet position angle $\phi_{\rm{jet}} = \phi_{\rm B, exp} + 90^\circ$. The range of $(\phi_{\rm{jet}}\,-\,180^\circ,\phi_{\rm{jet}})$ is highlighted. \textit{Two bottom rows: } similar as above, but for the RT model fitted to MODEL1 and MODEL2. Lightly shaded areas correspond to values reported in \citetalias{Paper1}, diameter $d = 42 \pm 3\,\mu$as and position angle  $150^\circ < \phi_{\mathrm{B}} < 200^\circ$. 
}
\label{fig:grmhd_model_fits}
\end{figure*}

Modeling tests revealed the presence of a~large-diameter secondary ring mode in the posterior distributions for the 2009--2012 data sets, corresponding to the dashed green line in 
\autoref{fig:ring_or_gauss}.
This mode is excluded by the detections on long baselines (APEX baselines in 2013, multiple baselines in 2017) and detections on medium-length (${\sim}1.5$\,G$\lambda$) baselines (LMT--SMT in 2017).  Excising this secondary mode, as we do for the posteriors presented in \autoref{fig:grmhd_model_fits},
effectively limits the diameter $d$ to be less than $\sim$\,80\,\uas. In all cases, the prior range is sufficient to capture the entire volume of the primary posterior mode, corresponding to an emission region of radius $\sim20\,\mu$as. We have verified numerically that this procedure produces the same results as restricting the priors on $R_{\rm out}$ to $[0,45]\, \mu$as for the analysis of the 2009--2012 data sets.

As a consequence of the Fourier symmetry of a~real domain input signal, we have $V^*(u,v) = V(-u,-v)$. Hence, visibility amplitude data alone cannot break the degeneracy between the orientation of $\phi_{\rm B}$ and $\phi'_{\rm B} = \phi_{\rm B} + 180^\circ$, and effectively, we only constrain the axis of the crescent asymmetry. This is how the reported uncertainties should be interpreted. Having that in mind, for the 2009, 2011, and 2013, consisting exclusively of the visibility amplitude data, we choose the reported $\phi_{\rm B}$ using the prior information about the position angle of the jet $\phi_{\rm jet}$ to select the $\phi_{\rm B}$ such that $\phi_{\rm jet} - 180^\circ < \phi_{\rm B} < \phi_{\rm jet}$, where $\phi_{\rm jet} = 288^\circ$ \citep{Walker2018}. In other words, between the orientations $\phi_{\rm B}$ and $\phi'_{\rm B}$, we choose the one that is closer to the expected bright side position $\phi_{\rm B, exp} = 198^\circ$. This is motivated by the theoretical interpretation of the asymmetric ring feature \citepalias{Paper5}. In the case of the 2012 data set, for which a~very limited number of closure phases is available, we report the orientation $\phi_{\rm B}$ of the maximum likelihood (ML) estimators, noting the bimodal character of the posterior distributions and the aforementioned $180^\circ$ degeneracy. These caveats do not apply to the 2017 data set, for which substantial closure phase information is available and breaks the degeneracy.

It is important to recognize that the parameters of a~geometric model have no direct relation to the physical parameters of the source, unlike direct fitting using GRMHD simulation snapshots \citepalias[\citealt{Dexter2010,Kim2016,GENA2019},][]{Paper5} or ray-traced geometric source models \citep{Broderick2009,Broderick2016,Vincent2020} 
to the data. The crescent model is a~phenomenological description of the source morphology in the observer's plane. If physical parameters (such as black hole mass) are to be extracted, additional calibration, in general affected by the details of the assumed theoretical model and the $(u,v)$-coverage, needs to be performed \citepalias{Paper6}.


\begin{figure*}[ht]
\centering
\includegraphics[width=0.81\linewidth]{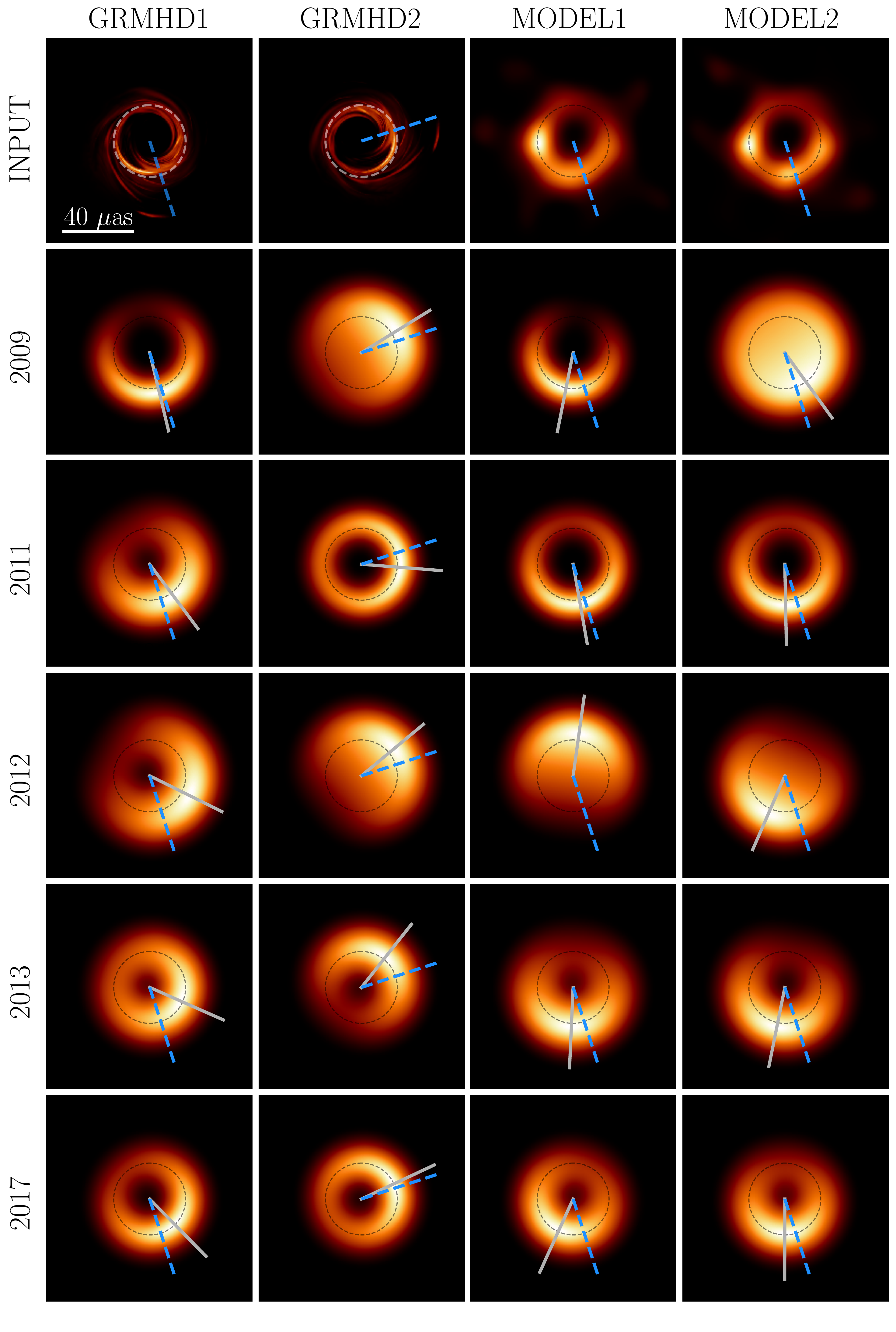}
\caption{ML estimators corresponding to the fits to four synthetic images, shown in the first row (no blurring). Estimators were obtained through synthetic VLBI observations with the $(u,v)$-coverage and uncertainties identical to those of the real observations performed in 2009--2017. The thick ring model (RT) was used, and the presented images of ML estimators were blurred to $15\,\mu$as resolution. Blue dashed lines indicate the convention for the expected position angle of the bright component $\phi_{\rm B, exp}$. The gray bar represents the ML estimate of $\phi_{\rm B}$. For the 2009, 2011, and 2013 data sets, the orientation is determined assuming that $|\phi_{\rm B, exp} - \phi_{\rm B}| < 90^\circ$. The dashed circles correspond to a~diameter of 42\,$\mu$as.}

\label{fig:synthetic_images}
\end{figure*}

\section{Modeling synthetic data}
\label{sec:synthetic}
In order to verify whether the $(u,v)$-coverage and signal-to-noise ratio of the 2009--2013 observations are sufficient to constrain source geometric properties with simple asymmetric ring models, we have designed tests using synthetic VLBI observations. The synthetic observations are generated with the \texttt{eht-imaging} software \citep{Chael2016,Chael2018} by sampling four emission models (GRMHD1, GRMHD2, MODEL1, MODEL2) with the $(u,v)$-coverage and thermal error budget reported for past observations. Additionally, corruption from time-dependent station-based gain errors has been folded into the synthetic observations. The ground-truth images that we use correspond to ray-traced snapshots of a~GRMHD simulation and published images of \m87 \citepalias{Paper4}, reconstructed based on the 2017 observations.

\subsection{GRMHD snapshots} 
\label{subsec:grmhd_snapshots}
For the first two synthetic data tests, we use a~random snapshot from a~GRMHD simulation of a low magnetic flux standard and normal evolution (SANE) 
accretion disk \citep{Narayan2012,Sadowski2013}  around a~black hole with spin $\bhspin\equiv Jc/GM^2=0.5$, shown in \autoref{fig:images_intro} (second panel), and in \autoref{fig:synthetic_images} (first panel). The GRMHD simulation was performed with the {\tt{iharm}} code \citep{Gammie2003}, and the ray tracing was done with {\tt{ipole}} \citep{moscibrodzka2018}.  
Following \citet{Moscibrodzka2016} and \citetalias{Paper5}, we assume a~thermal electron energy distribution function and relate the local ratio of ion ($T_{\rm i}$) and electron temperature ($T_{\rm e}$) to the plasma parameter $\beta_{\rm p}$ representing the ratio of gas to magnetic pressure
\begin{equation}
    \frac{T_{\rm i}}{T_{\rm e}} = R_{\rm high} \frac{\beta_{\rm p}^2}{1+\beta_{\rm p}^2} + R_{\rm low} \frac{1}{1+\beta_{\rm p}^2} \, ,
\label{eq:R}
\end{equation}
with $R_{\rm high} = 40$ and $R_{\rm low} = 1$ for the considered snapshot.
The prescription given by \autoref{eq:R} parameterizes complex plasma microphysics, allowing us to efficiently survey different models of electron heating, resulting in a different geometry of the radiating region. As an example, for the SANE models with large $R_{\rm high}$ the emission originates predominantly in the strongly magnetized jet base region, while for a~small $R_{\rm high}$ disk emission dominates \citepalias{Paper5}.

The image considered here is a~higher resolution version (1280$\times$1280 pixels) of one of the images generated for the Image Library of \citetalias{Paper5} and corresponds to a~$6.5 \times 10^9 M_{\odot}$ black hole at a~distance $D$ = 16.9 Mpc.
This choice results in an $M/D$ ratio\footnote{Hereafter, we use the natural units in which $G=c=1$.} of 3.80\,$\mu$as and an observed black hole shadow that is nearly circular with an angular diameter not substantially different from the Schwarzschild case, which is $2\sqrt{27}M/D = 39.45\,\mu$as \citep{Bardeen1973}. For reference, the dashed circles plotted in \autoref{fig:images_intro} and \autoref{fig:synthetic_images} have a~diameter of 42.0\,$\mu$as.
These parameters were chosen to be consistent with the ones inferred from the EHT 2017 observations \citepalias{Paper1}. The camera is oriented with an inclination angle of $22^\circ$.
The viewing angle was chosen to agree with the expected inclination of the \m87 jet \citep{Walker2018}. The choices of spin $\bhspin$, electron temperature parameter $R_{\rm high}$, and the SANE accretion state are arbitrary. The choice of $R_{\rm low}$ follows the assumptions made in \citetalias{Paper5}. We also assume that the accretion disk plane is perpendicular to the black hole jet (the disk is not tilted). The first image, GRMHD1, has been rotated in such a way that the projection of the simulated black hole spin axis counteraligns with the observed position angle of the approaching \m87 jet, $\phi_{\rm jet}\,=\,288^\circ$ \citep{Walker2018, kim2018}. The GRMHD2 test corresponds to the same snapshot, but rotated counterclockwise by 90$^\circ$ to $\phi_{\rm jet}\,=\,18^\circ$, hence displaying a brightness asymmetry in the east-west rather than in the north-south direction, see the second panel of \autoref{fig:synthetic_images}. Because the $(u,v)$-coverage in 2009--2013 was highly anisotropic, a dependence of the fidelity of the results on the image orientation may be expected.

\subsection{\m87 images}
\label{subsec:images}
For additional synthetic data tests, we consider images of \m87 generated based on the 2017 EHT observations, published in \citetalias{Paper4}. We consider two days of the 2017 observations with good coverage and reported structural source differences \citepalias[][\citealt{Arras2020}]{Paper3,Paper4} - 2017 April 6th (MODEL1) and 2017 April 11th (MODEL2), see the first row of \autoref{fig:synthetic_images}. MODEL2 was also shown in the first panel of \autoref{fig:images_intro}. While these models are constructed based on the observational data, we resample them numerically to obtain synthetic data sets considered in this section. Synthetic closure phases on the SMT--SMA--CARMA triangle computed from these models were shown in \autoref{fig:closures}. Note that there is a~subtle difference between resampling a~model constructed based on the 2017 data with a~numerical model of the 2017 array and direct modeling of the actual 2017 data, considered in \autoref{sec:ModelingRealData}. 
The sampled images were generated utilizing the \texttt{eht-imaging} pipeline through the~procedure outlined in \citetalias{Paper4}, with a~resolution of 64$\times$64 pixels. This test can be viewed as an attempt to evaluate what the outcome of the modeling efforts would have been had the 2017 EHT observations been carried out with one of the proto-EHT 2009--2013 arrays rather than with the mature 2017 array. 

\subsection{Results for the synthetic data sets}
\label{subsec:synthetic_results}

\autoref{fig:synthetic_images} shows a~summary of the maximum likelihood (ML) RT model fits to the synthetic data sets; each column shows the fits for a~single ground-truth image, and each row shows the fits for a~single array configuration.  Though our simple ring models cannot fully reproduce the properties of the abundant and high signal-to-noise data sampled with the 2017 array (i.e., fits to these data sets are characterized by poor reduced-$\chi^2$ values of $\chi^2_n \sim 5$), they nevertheless recover diameter and orientation values that are reasonably consistent with those reported in \citetalias{Paper6} for the 2017 observations, including the counterclockwise shift of the brightness position angle between 2017 April 6th (MODEL1) and 11th (MODEL2). We note that the underfitting of the data set sampled with the 2017 array results in an artificial narrowing of the parameter posteriors (the full posterior is captured in \citetalias{Paper6} by considering a~more complicated GC model).
ML estimators for 2009--2013 data sets, on the other hand, typically fit the data much closer than the thermal error budget. Because we model time-dependent station gains in a small array (often only two to three telescopes observing at the same time in 2009--2013 with missing detections on some baselines), the number of model parameters may be formally larger than the number of data points, and this complicates our estimation of the number of effective degrees of freedom \ed{(see also \autoref{subs:fit_quality})}. As a~consequence, we cannot generally utilize a~$\chi^2_n$ goodness-of-fit statistic as was done in \citetalias{Paper4} and \citetalias{Paper6}.\footnote{See, e.g., \citet{Andrae2010} for further comments about the problems with the $\chi^2_n$ metric and counting the degrees of freedom.}
 
The relevant parameter estimates and uncertainties from the RT model fits are listed in \autoref{tab:results_sim}. In \autoref{fig:grmhd_model_fits} we show the marginalized posteriors for the diameter and position angle parameters. For each synthetic data set we also indicate the image domain position angle $\eta$, defined as 
\begin{equation}
    \eta = \text{Arg} \left( \frac{\sum_{k} I_{k} e^{i\phi_k} }{\sum_{k} I_{k}} \right) \, ,
\label{eq:PA}
\end{equation}
where $I_k$ is the intensity and $\phi_k$ is the~position angle of the~$k$th pixel in the image. A~similar image domain position angle estimator was considered in \citetalias{Paper4}.
We notice that the image- and model-based estimators may occasionally display significant differences (e.g., GRMHD1). However, they are both sensitive to global properties of the brightness distribution, unlike some other estimators that could be considered, such as, e.g., the location of the brightest pixel. For the diameter $d$ estimates reported in \autoref{tab:results_sim} we list both the median and ML values, with 68\% and 95\% confidence intervals, respectively. For the orientation angle $\phi_{\text{B}}$ we list the ML values with 68\% confidence intervals, and for the fractional thickness $f_{\rm w}$ we list the 95th distribution percentile. Values of $\phi_B$ contained in parentheses indicate that the 68\% confidence interval exceeds 100$^{\circ}$, in which case we have concluded that the orientation is effectively unconstrained. We find that the diameter is well constrained in general, with the GRMHD data sets recovering a typical value of ${\sim}44$\,\uas and 95\% confidence intervals that never exceed ${\pm}$12\,\uas from this value; the analogous measurement for the MODEL data sets is $44\,\pm\,9\,\mu$as. Biases related to the array orientation can be seen - particularly with the 2013 coverage, the GRMHD2 test estimates an appreciably larger diameter than GRMHD1, inconsistent within the 68\% confidence interval.

\begin{table}[t!]
    \caption{Parameter estimates from fitting the RT model to the synthetic data sets. }
    \begin{center}
    \tabcolsep=0.09cm
    \begin{tabularx}{1.01\columnwidth}{cccccc}
    \hline
    \hline
     & Coverage & \multicolumn{2}{c}{$d\,(\mu$as)} & $\phi_{\rm B}$ (deg) & \multicolumn{1}{c}{$f_{\rm w}$} \\
    {Estimator} & & Median & ML &   ML  &  At Most \\
    {Confidence} & & 68\% & 95\% & 68\% &  95\% \\
    \hline
     GRMHD1 & 2009 & $46.4^{+5.3}_{-7.9}$ & $51.1^{+4.9}_{-18.5}$ & $194^{+56}_{-5}$ & 0.88 \\
      & 2011 & $40.9^{+5.6}_{-5.5}$ & $46.8^{+5.1}_{-13.9}$ & $217^{+55}_{-35}$ & 0.93 \\
      $\eta^\text{a}= 170^\circ$ & 2012 & $44.0^{+4.9}_{-5.2}$ & $49.3^{+4.3}_{-15.1}$ & $244^{+25}_{-24}$ & 0.90 \\
        & 2013 & $39.2^{+2.0}_{-0.8}$ & $41.3^{+2.1}_{-4.0}$ & $(246)$ & 0.56 \\
        & 2017$^\text{b}$ & $43.8^{+0.1}_{-0.1}$ & $43.8^{+0.1}_{-0.1}$ & $225^{+1}_{-1}$ & 0.33 \\
     \hline
   GRMHD2 & 2009 & $43.1^{+3.4}_{-5.0}$ & $37.2^{+11.8}_{-2.7}$ & $(302)$ & 0.95 \\
      & 2011 & $44.3^{+2.9}_{-5.4}$ & $47.4^{+2.2}_{-12.4}$ & $(265)$ & 0.94 \\
       $\eta^\text{a}= 260^\circ$ & 2012 & $42.2^{+3.3}_{-3.8}$ & $36.3^{+10.0}_{-0.5}$ & $310^{+6}_{-55}$ & 0.96 \\
        & 2013 & $43.6^{+1.1}_{-0.6}$ & $43.5^{+2.4}_{-1.2}$ & $322^{+38}_{-55}$ & 0.50 \\
        & 2017$^\text{b}$ & $42.4^{+0.1}_{-0.1}$ & $42.4^{+0.1}_{-0.1}$ & $295^{+1}_{-1}$ & 0.33 \\
     \hline
     MODEL1 & 2009 & $45.7^{+3.0}_{-4.7}$ & $46.5^{+4.6}_{-9.1}$ &$169^{+10}_{-80}$ & 0.95 \\
      & 2011 & $47.1^{+2.7}_{-4.6}$ & $49.6^{+2.2}_{-11.4}$ & $190^{+65}_{-5}$ & 0.90 \\
      $\eta^\text{a}= 155^\circ$ & 2012 & $41.3^{+3.6}_{-4.0}$ & $36.4^{+9.9}_{-1.2}$ & $352^{+38}_{-47}$ & 0.96 \\
        & 2013 & $43.1^{+2.1}_{-1.0}$ & $45.1^{+1.8}_{-3.9}$ & $177^{+36}_{-27}$ & 0.63 \\
        & 2017$^\text{b}$ & $41.0^{+0.1}_{-0.1}$ & $41.0^{+0.1}_{-0.1}$ & $156^{+1}_{-1}$ & 0.47 \\
     \hline
     MODEL2 & 2009 & $42.7^{+3.8}_{-3.7}$ & $36.0^{+11.9}_{-1.0}$ & $169^{+30}_{-69}$ & 0.97 \\
      & 2011 & $48.3^{+2.1}_{-4.7}$ & $50.4^{+2.0}_{-10.7}$ & $180^{+72}_{-5}$ & 0.91 \\
      $\eta^\text{a}= 172^\circ$ & 2012 & $41.0^{+3.3}_{-2.7}$ & $37.0^{+9.7}_{-0.3}$ & $(179)$ & 0.98 \\
        & 2013 & $47.0^{+1.1}_{-1.3}$ & $47.3^{+1.8}_{-2.4}$ & $174^{+2}_{-10}$ & 0.53 \\
        & 2017$^\text{b}$ & $41.2^{+0.1}_{-0.1}$ & $41.2^{+0.1}_{-0.1}$ & $180^{+1}_{-1}$ & 0.50 \\
     \hline
    \hline 
    \end{tabularx}
    \label{tab:results_sim}
    \end{center}
     $^\text{a}$ $\eta$ calculated with \autoref{eq:PA},
    $^\text{b}$ using the $(u,v)$-coverage of 2017 April 6th
\end{table}

The orientation $\phi_{\rm B}$ is poorly constrained, with posterior distributions that depend strongly on the details of the $(u,v)$-coverage. Nevertheless, the 2009--2013 ML estimates provide orientations of the axis of asymmetry that are consistent within ${\pm}35^\circ$ with the results obtained using the 2017 synthetic coverage.  We note that in 3 out of 4 synthetic data sets, the limited number of closure phases provided by the simulated data sets with the 2012 coverage is enough to correctly break the degeneracy in the position angle $\phi_{\rm B}$, discussed in \autoref{sub:degenercies}. For the synthetic GRMHD data sets, the preference for the correct brightness position angle is very strong (see \autoref{fig:grmhd_model_fits}). For the MODEL data sets, the effect of closure phases is much less prominent, the distributions remain bimodal, and in the case of the MODEL1 data set, the ML estimator points at the wrong orientation, suggesting brightness located in the north.

We also consider the fractional thickness $f_{\rm w}$ of the ring, as defined in \autoref{eq:fw}. The fractional thickness provides a~measure of whether the data support the presence of a central flux depression, a~signature feature of the black hole shadow, or if it is consistent with a~disk-like morphology (i.e., $f_{\rm w} \approx 1$ for the RT model). We find that $f_{\rm w}$ is less well constrained than the diameter $d$, consistently with the conclusions of \citetalias{Paper6}. In some cases the ML estimator corresponds to a~limit of a~disk-like source morphology without a~central depression (see \autoref{fig:synthetic_images}). Only the 2013 and 2017 synthetic data sets allow us to confidently establish the presence of a~central flux depression, with posterior distributions excluding $f_{\rm w} > 0.7$ for all synthetic data sets (see \autoref{tab:results_sim}). For the 2009--2012 coverage synthetic data sets, $f_{\rm w}$ is not constrained sufficiently well to permit similar statements.
We find that the RG model produces results that are typically consistent with those of the RT model (see \autoref{app:imagesRG}, \autoref{fig:synthetic_images_blur}).



\section{Modeling real data} \label{sec:ModelingRealData}

\begin{figure*}[h!]
\centering
\includegraphics[trim={0.5cm 0.9cm 0.5cm 0.0cm},clip,width=1.0\linewidth]{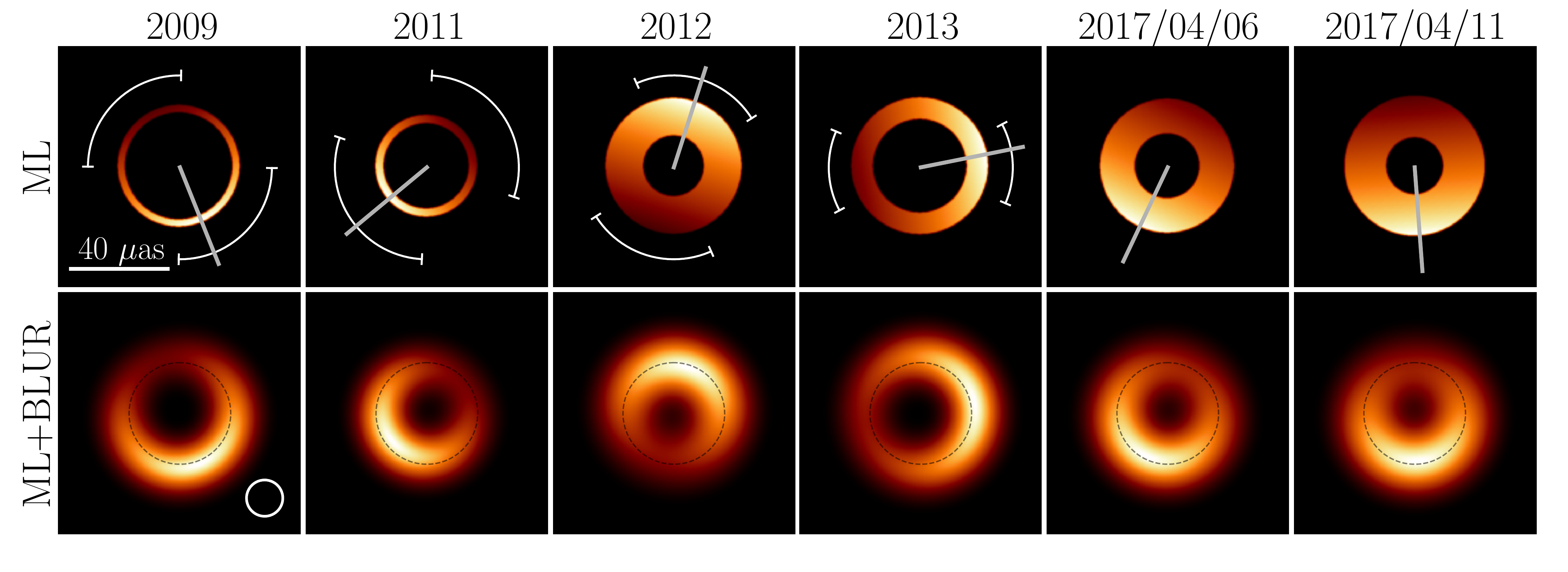}
\includegraphics[trim={0.5cm 0.9cm 0.5cm 0.5cm},clip,width=1.0\linewidth]{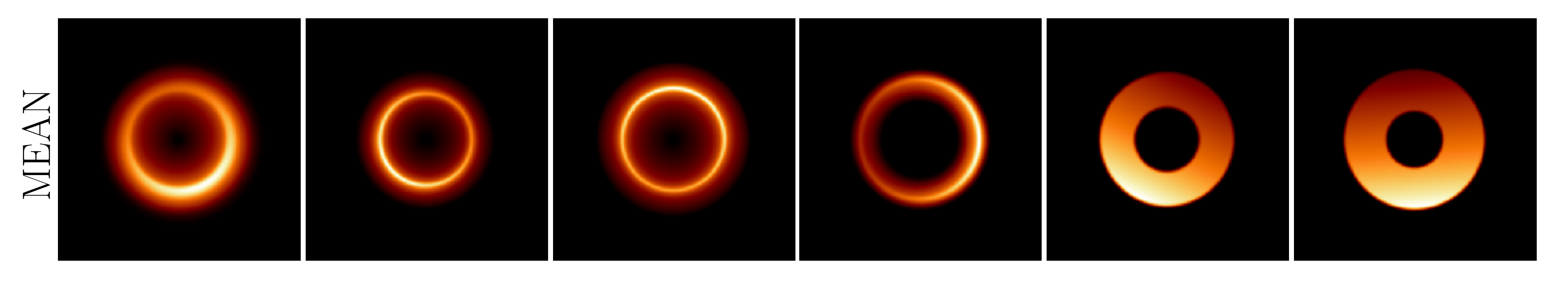}
\caption{\textit{First row:} ML estimators obtained from fitting the RT model to the 2009--2017 observations. The position angle $\phi_{\rm B}$ is indicated with a~bar. For the 2009, 2011, and 2013 data sets the orientation is determined assuming that $\phi_{\rm jet} - 180^\circ < \phi_{\rm B} < \phi_{\rm jet}$, where $\phi_{\rm jet} = 288^\circ$. Position angle 68\% confidence intervals are shown for the 2009--2013 data sets. \textit{Second row:} RT models from the first row blurred to a 15~$\mu$as resolution, indicated with a beam circle in the bottom-right corner of the first panel. The dashed circle of 42~$\mu$as diameter is plotted for reference. 
\textit{Third row:} mean of the 2$\times 10^4$ images drawn from the posterior of the RT model fits.  
}
\label{fig:oldrings}
\centering
\includegraphics[trim={0.5cm 0.9cm 0.5cm 0.0cm},clip,width=1.0\linewidth]{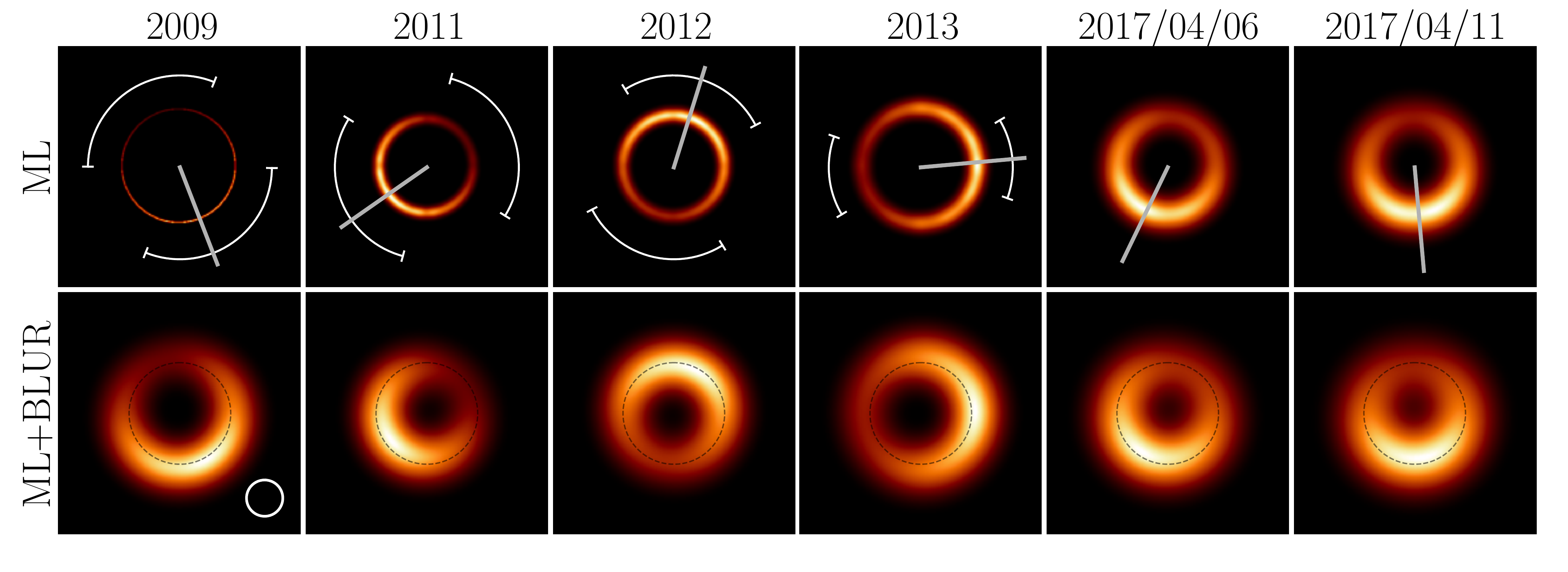}
\includegraphics[trim={0.5cm 0.9cm 0.5cm 0.5cm},clip,width=1.0\linewidth]{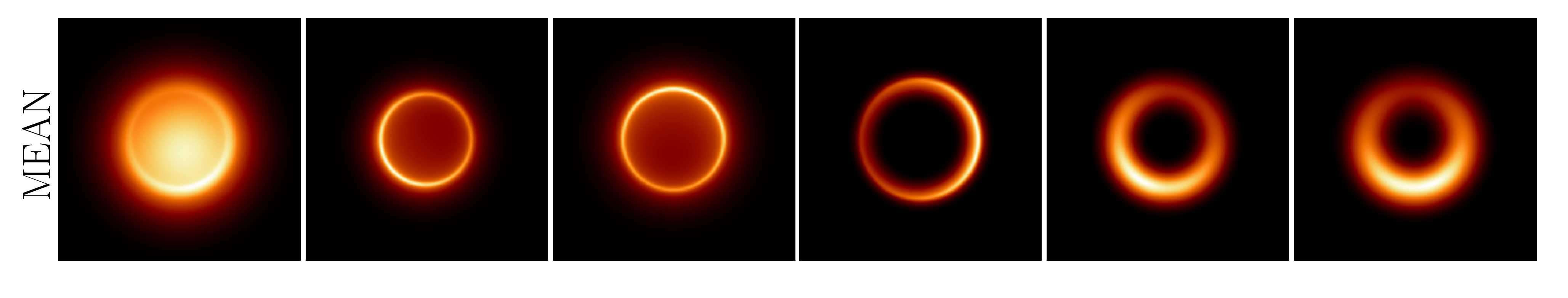}
\caption{Same as \autoref{fig:oldrings}, but for the RG model. The 2009--2012 posterior distributions contain a~Gaussian mode, manifesting as a~bright ring interior in the mean images. This is related to the lower spatial resolution of the 2009--2012 observations.}
\label{fig:oldrings_blur}
\end{figure*}

\begin{figure*}[t!]
\centering
\includegraphics[trim={0cm 0.3cm 0 0.1cm},clip,width=0.98\linewidth]{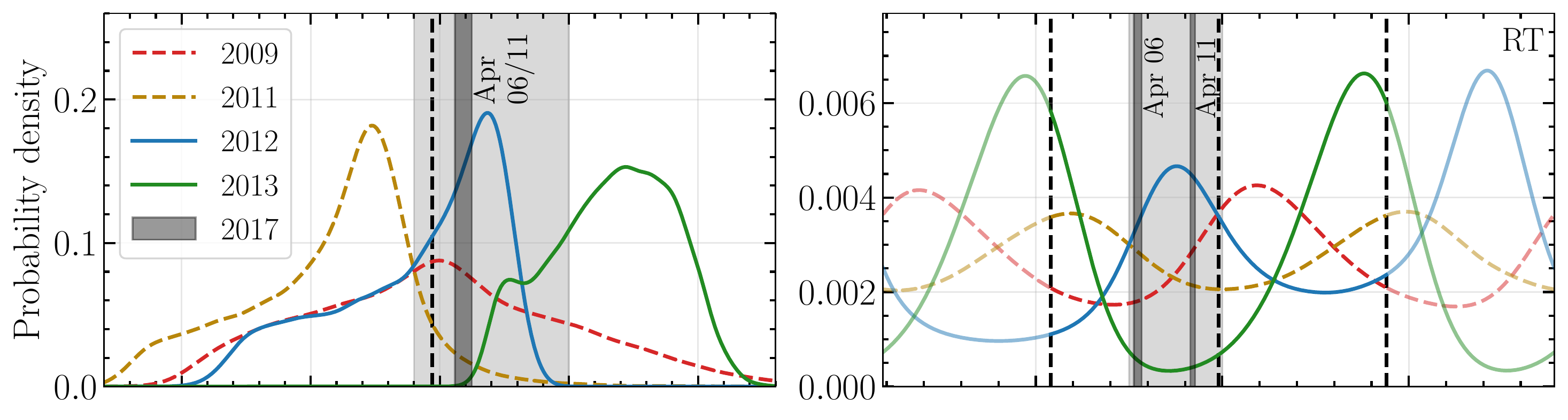}
\includegraphics[trim={0cm 0cm 0 0.2cm},clip,width=0.98\linewidth]{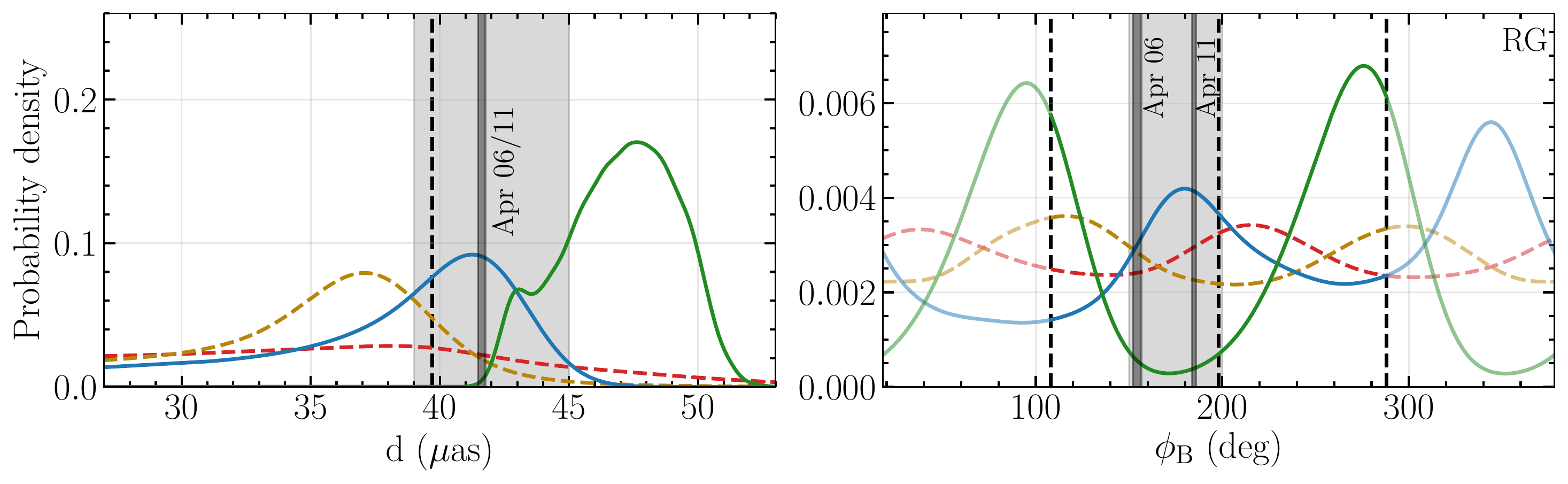}
\caption{ \textit{Top:} marginalized distributions of mean diameter $d$ and brightness maximum position angle $\phi_{\mathrm{B}}$ 
for the RT model fits to 2009--2017 observational data sets. Lightly shaded areas correspond to values reported in \citetalias{Paper1}, diameter $d = 42\,\pm\,3\,\mu$as and position angle  $150 ^\circ < \phi_{\mathrm{B}} < 200^\circ$. The 2017 posteriors are contained within the dark gray bands. The dashed vertical line in the left panels denotes the diameter of $2\sqrt{27}M/D$.
The vertical dashed lines in the right panels denote the convention angle $\phi_{\rm B, exp}= 198^\circ$, $\phi_{\rm B, exp} - 90^\circ$, and the approaching jet position angle $\phi_{\rm{jet}} = 288^\circ$. The range of $(\phi_{\rm{jet}}\,-\,180^\circ,\phi_{\rm{jet}})$ is highlighted. \textit{Bottom:} same, but for the RG model.}
\label{fig:results}
\end{figure*}

Encouraged by the results of the tests on synthetic data sets, we performed the same analysis on the 2009--2013 proto-EHT \m87 observations. We also present the analysis of the 2017 observations with the RT and RG models. In the latter case only lower band data (2\,GHz bandwidth centered at 227\,GHz) were used.

\subsection{Source geometry estimators}

In the first row of \autoref{fig:oldrings} we show the ML estimators obtained by fitting the RT model to each observational data set; in \autoref{fig:oldrings_blur} we show the same for the RG model. For the 2009, 2011, and 2013 data sets, which only constrain the axis of the crescent asymmetry, the orientation of the brightness peak was selected with a~prior derived from the approaching jet orientation on the sky, $\phi_{\rm B} \in (\phi_{\rm jet}-180^\circ, \phi_{\rm jet})$, \autoref{sub:degenercies}. The 2012 data set, for which some closure phases are available, indicates a~weak preference toward the brightness position located in the north rather than in the south (\autoref{fig:results}, and Figures \ref{fig:cornerRT2009-2013}-\ref{fig:cornerRG2009-2013} in the \autoref{app:cornerplots}). However, the posterior distribution remains bimodal and 47\% of its volume remains consistent with the jet orientation based prior. Hence, the distinction is not very significant -- it is entirely dependent on the sign of closure phases shown in \autoref{fig:closures}, which are all consistent with zero to within 2\,$\sigma$. Closure phases predicted by the ML estimators for the RT and RG models are indicated in \autoref{fig:closures}. The 2017 data sets are consistent with the orientation imposed by the jet position prior. The second rows of Figures \ref{fig:oldrings} and \ref{fig:oldrings_blur} show the ML estimators blurred to a~resolution of $15\,\mu$as. In the third rows of Figures \ref{fig:oldrings} and \ref{fig:oldrings_blur} we present ``mean images'' for each data set, obtained by sampling 2\,$\times\,10^{4}$ sets of model parameters from the MCMC chains and averaging the corresponding images. The mean images highlight structure that is ``typical'' of a~random draw from the posterior distribution, though we note that a~mean image itself does not necessarily provide a good fit to the data.  Because of the rotational degeneracy, the orientation is always assumed to be the one closer to the orientation given by the ML estimate of $\phi_{\rm B}$ for the construction of these images.

\subsection{Estimated parameters}

The marginalized posteriors for the mean diameter $d$ and position angle $\phi_{\rm B}$ for the observational data sets are shown in \autoref{fig:results} for both the RT and RG models, and tabulated values of the relevant estimates for the RT model are given in \autoref{tab:results2}. The posterior distributions for the 2009--2012 data sets have complex shapes, not all parameters are well constrained, and ML estimators do not necessarily coincide with the marginalized posteriors maxima of the individual model parameters, which can be seen in the corner plots (Figures \ref{fig:cornerRT2009-2013}-\ref{fig:cornerRG2009-2013} in the \autoref{app:cornerplots}). The behavior of the posterior distributions is much improved for the 2013 data set, and becomes exemplary in the case of the 2017 data sets (Figures \ref{fig:corner2017}-\ref{fig:corner2017RG} in the \autoref{app:cornerplots}).

\begin{table}[b]
     \caption{Parameter estimates from fitting the RT model to the observational data sets. }
    \begin{center}
    \tabcolsep=0.17cm
    \begin{tabularx}{0.92\columnwidth}{ccccc}
    \hline
    \hline
       & \multicolumn{2}{c}{$d\,(\mu$as)} & \multicolumn{1}{c}{$\phi_{\rm B}$(deg)} & \multicolumn{1}{c}{$f_{\rm w}$} \\
     Estimator & Median & ML & ML & At Most  \\
      Confidence & 68\% & 95\% &  68\% &  95\%  \\
    \hline
       2009 & $39.8^{+5.5}_{-5.2}$ & $47.3^{+3.1}_{-16.0}$ &  $202^{+68}_{-23}$&  0.93 \\
       2011 & $36.5^{+2.0}_{-4.3}$ & $38.6^{+2.8}_{-9.9}$ &  $(130)$ &  0.91 \\
        2012$^\text{a}$ & $40.1^{+2.1}_{-4.8}$ & $40.6^{+2.5}_{-8.3}$ &  $342^{+42}_{-40}$ &  0.92  \\
         2013 & $46.8^{+2.3}_{-2.9}$ & $47.6^{+3.0}_{-5.5}$ &  $281^{+17}_{-34}$  &  0.28 \\
         2017$^\text{b}$ & $41.1^{+0.1}_{-0.1}$ & $41.1^{+0.2}_{-0.2}$ &  $155^{+1}_{-1}$ &  0.37  \\
         2017$^\text{c}$ & $40.7^{+0.1}_{-0.1}$ &  $40.7^{+0.1}_{-0.1}$ &  $184^{+1}_{-1}$ &  0.44  \\
     \hline
    \hline 
    \end{tabularx}
    \label{tab:results2}
    \end{center}
    $^\text{a}$ secondary mode present at $\phi_{\rm B}- 180^\circ$ (see the text), $^\text{b}$ 2017 April 6th, $^\text{c}$ 2017 April 11th
\end{table}

Similar to the case of the synthetic data sets, we find that the diameter $d$ is well constrained; the RT model 95\% confidence intervals across all observational data sets always fall within $\pm$12\,\uas from $d = 40$\,\uas. 
The 2013 proto-EHT observations provide meaningful constraints on $\phi_{\rm B}$, indicating that the source asymmetry in 2013 was in the east-west direction, rather than in the north-south direction, as in the case of the 2017 data set. 
The 2009 and 2011 data sets do not constrain the orientation well.

All ML estimators and mean images from the RT model fits show a~clear shadow feature, indicating that a~disk-like, filled-in structure is disfavored by all of the observations (however, for 2009--2012 it cannot be excluded with high confidence based on the relative thickness parameter $f_{\rm w}$ distribution, \autoref{tab:results2} and \autoref{app:cornerplots}). This is contrary to the synthetic data results shown in \autoref{fig:synthetic_images}, where some of the ML estimators correspond to a~disk-like morphology. On the other hand, the mean images for the 2009--2012 RG model fits show a~significant flux density interior to the ring, indicating that these data sets are consistent with a~symmetric Gaussian source model, having no central flux depression. This is a~consequence of the resolution being limited by the lack of long baselines prior to 2013. Short and medium-length baselines alone provide insufficient information to fully exclude the Gaussian mode allowed by the RG model, or the~disk-like mode allowed by the RT model. For the same reason we see flattened posterior distributions of the RG diameter for 2009--2012 in \autoref{fig:results} -- these indicate consistency with a~small, strongly blurred ring with $f_{\rm w} > 1$, becoming a Gaussian in the limit of $\sigma \gg d$.

The slash parameter $\beta$ can be measured to be $0.3 \pm 0.1$ for the 2013 data set, which is consistent with the fits to the 2017 data sets that give $\beta \sim 0.20$ (RT) or $\beta \sim 0.35$ (RG). Fits to the 2012 data set indicate preference toward more symmetric brightness distribution, 2009 and 2011 do not provide meaningful constraints on $\beta$.

\subsection{Quality of the visibility amplitude fits}
\label{subs:fit_quality}
\begin{figure*}[t!]
\centering
\includegraphics[trim={0cm 0.0cm 0 0.0cm},clip,width=0.75\linewidth]{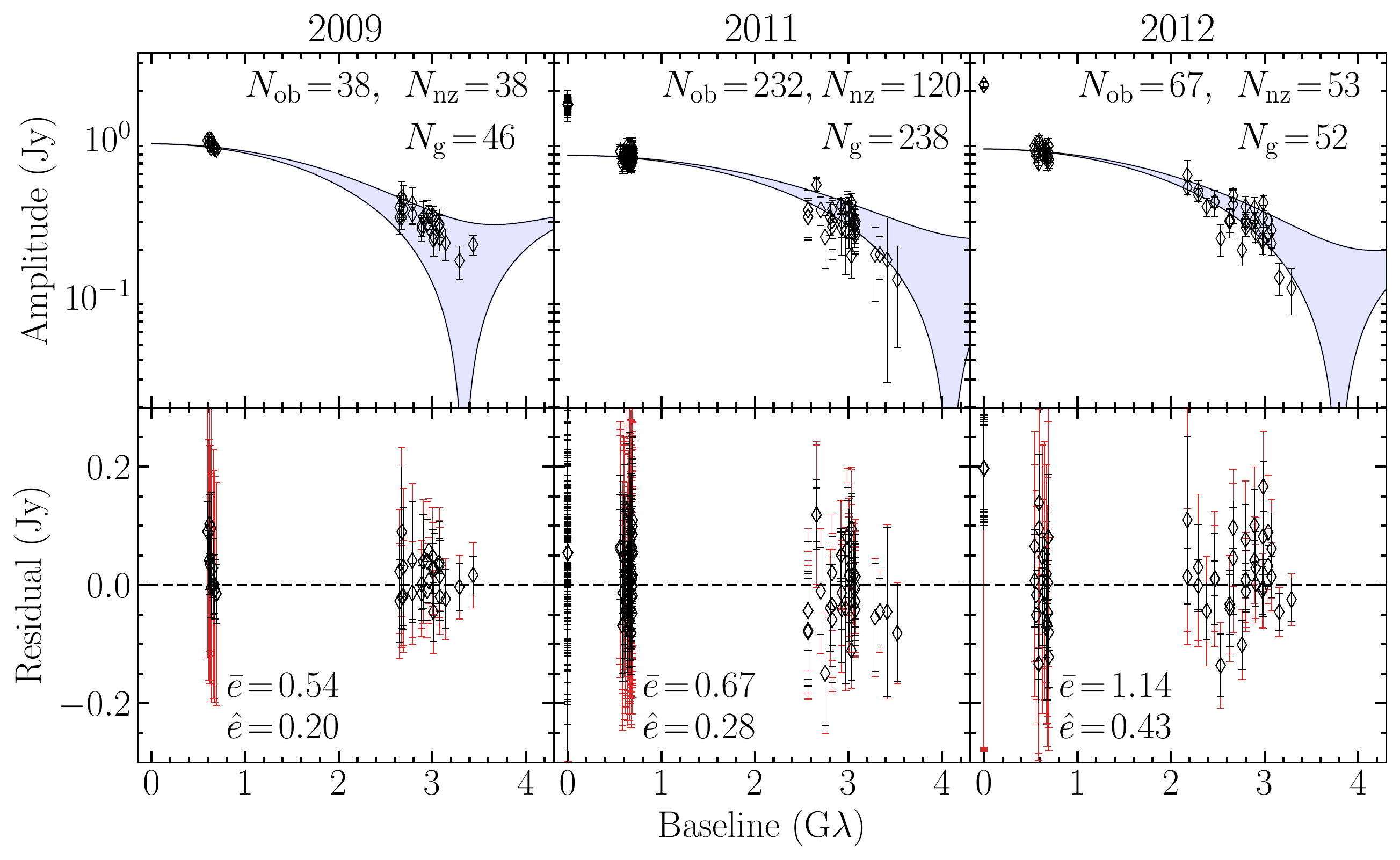}
\caption{\ed{\textit{Top row:} visibility amplitudes measured in 2009--2012 are shown as black diamonds with error bars corresponding to thermal uncertainties. Blue shaded regions correspond to the range of visibility amplitudes of the asymmetric ring RT model ML estimators, shown in the first row of \autoref{fig:oldrings_blur}. The total number of observed visibility amplitudes $N_{\rm ob}$ is given, along with the number of nonzero baseline visibility amplitudes $N_{\rm nz}$, and the number of modeled gains $N_{\rm g}$. \textit{Bottom row:} differences between measured amplitudes $|V_i|$ and the geometric model amplitudes $|\bar{V}_i|$. Black error bars correspond to thermal uncertainties, while red ones correspond to error budget inflated by adding systematics approximately capturing the gains uncertainties (\autoref{eq:inflated_errors}). Two fit quality metrics, defined with Equations \ref{eq:ebar}-\ref{eq:ehat}, are provided for each data set.}}
\label{fig:fit_quality}
\end{figure*}

\begin{figure*}[t!]
\centering
\includegraphics[trim={0cm 0cm 0 0.0cm},clip,width=0.75\linewidth]{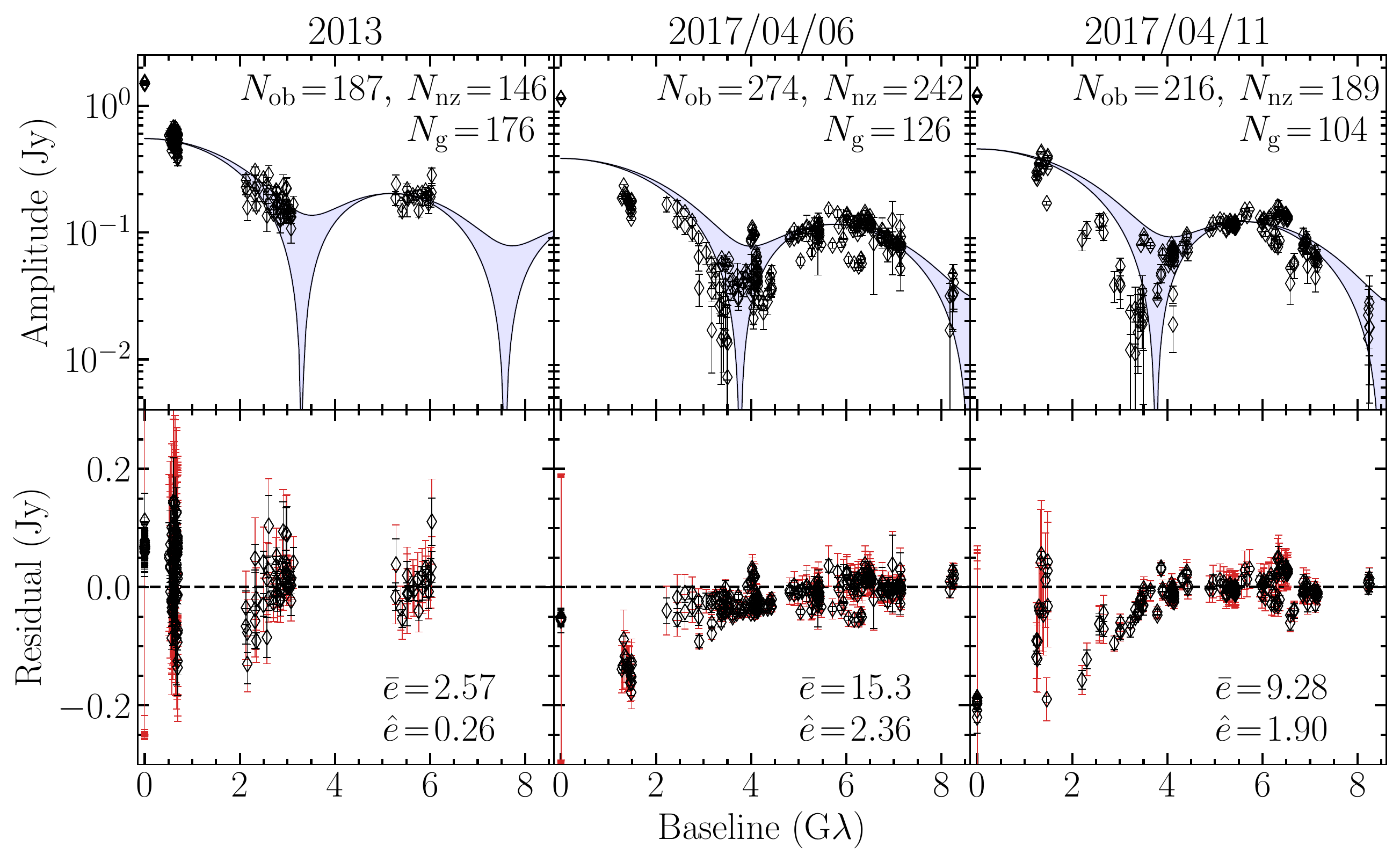}
\caption{\ed{Same as \autoref{fig:fit_quality}, but for the 2013--2017 data sets.}}
\label{fig:fit_quality2}
\end{figure*}

\ed{
The quality of fits and their behavior in terms of $\chi^2_n$ are similar to the synthetic data sets (see the comments in \autoref{subsec:synthetic_results} and \autoref{tab:results_2017}). In Figures \ref{fig:fit_quality}-\ref{fig:fit_quality2} we explicitly give the number of independent visibility amplitude observations for each data set $N_{\rm ob}$ and the number of independent visibilities on nonzero (intersite) baselines $N_{\rm nz}$. Note that the latter is larger than the number of detections on nonzero, nonredundant baselines given in \autoref{tab:detections}, as some detections are independent but redundant. We also provide the number of explicitly modeled amplitude gains $N_{\rm g}$ for each data set (see \autoref{sec:FittingProcedure} and \autoref{subsec:synthetic_results}). Given the pathologies in the $\chi^2_n$ metric described in \autoref{subsec:synthetic_results}, we characterize the quality of the ML estimator fits to data using the two following metrics
}
\begin{align}
    \bar{e} &= \frac{1}{N_{\rm nz}}\sum_{i=1}^{N_{\rm nz}} \left| \frac{|V_i| - |\bar{V}_{i}|}{\sigma_i} \right| \ , \label{eq:ebar} \\
     \hat{e} &= \frac{1}{N_{\rm nz}}\sum_{i=1}^{N_{\rm nz}} \left| \frac{|V_i| - |\hat{V}_{i}|}{\sigma_i} \right| \ .
\label{eq:ehat}
\end{align}
\ed{In Equations \ref{eq:ebar}-\ref{eq:ehat} we follow the notation of \autoref{eq:gains}, that is, $\bar{V}_i$ represents the visibilities of the geometric model while $\hat{V}_i$ corresponds to the model modified by applying the estimated gains, representing the final fit to observations $V_i$. Uncertainties $\sigma_i$ correspond to the thermal error budget. We only account for nonzero baselines, which describe the compact source properties. In the bottom rows of Figures \ref{fig:fit_quality}-\ref{fig:fit_quality2} we indicate two error bars. Black error bars correspond to the thermal uncertainties $\sigma_i$, while the red ones correspond to inflated uncertainties
\begin{equation}
    s_i = \sqrt{\sigma_i^2 + 2\cdot(0.15 V_i)^2} \ ,
\label{eq:inflated_errors}
\end{equation}
approximately capturing the uncertainty related to the amplitude gains. For all 2009--2013 data sets, the flexibility of the full model is sufficient to fit the sparse data to within the thermal uncertainty level with a~best-fit ML estimator.
}

\subsection{Consistency with the prior analysis}

In order to assess model-related biases and verify to what extent our simplified models recover geometric parameters consistent with the ones reported by EHT, i.e., image domain results given in \citetalias{Paper4} Tables 5 and 7, and geometric modeling results given in \citetalias{Paper6} Tables 2 and 3, we gather these results in \autoref{tab:results_2017}. For the details of the methods and algorithms, see explanations and references in \citetalias{Paper4} and \citetalias{Paper6}. We notice that (1) differences between methods may be as large as 30$^\circ$ for the same data set, (2) models considered in this work measure diameter and position angle consistently with more complex crescent models (GC) and with the image domain methods within the expected intermodel variation, (3) RT and RG models are too simplistic to fully capture the properties of the 2017 data sets, resulting in underfitting, indicated by higher values of $\chi^2_n$, and (4) posteriors of the RT and RG models are narrower for the 2017 data sets than the GC posteriors as an effect of the underfitting. 

\begin{table}[h!]
     \caption{Comparison between parameters extraction results reported in This Paper, \citetalias{Paper4}, and \citetalias{Paper6}.}
    \begin{center}
    \tabcolsep=0.10cm
    \begin{tabularx}{0.99\columnwidth}{ccccc}
    \hline
    \hline
     Source & Method & $d\,(\mu$as) & $\phi_{\rm B}$ or $\eta$ (deg)$^\text{a}$ & $\chi^2_n$ \\
    \hline
    \multicolumn{4}{c}{\, \, \, \, \, \,  2017 April 6th}  \\
    \hline
       this work  & RT & $41.1^{+0.09}_{-0.08}$ & $154.8 \pm 1.1$ & $2.99$ \\
        & RG & $41.6 \pm 0.07$ &  $154.3^{+1.2}_{-1.1}$ & $3.04$ \\
       \hline
        \citetalias{Paper6} & \themis & $43.5 \pm 0.14$ &  $153.0^{+2.0}_{-2.4}$ & $1.32$   \\
         (GC) & \texttt{dynesty} & $43.4^{+0.27}_{-0.26}$ & $148.5^{+1.4}_{-1.2}$ & $1.29$ \\
    \hline
       \citetalias{Paper4}   & \texttt{DIFMAP} & $40.1 \pm 7.4$  & $162.1 \pm 9.7$ & 2.10\\
         (image & \texttt{eht-imaging} & $39.6 \pm 1.8$ & $151.1 \pm 8.6$  & 1.28 \\
         domain) & \texttt{SMILI} & $40.9 \pm 2.4$ & $151.7 \pm 8.2$ & $1.34$  \\
    \hline  
    \multicolumn{4}{c}{\, \, \, \, \, \, 2017 April 11th}  \\
    \hline
     this work & RT & $40.7\pm 0.05$ & $184.2^{+0.6}_{-0.7}$ & $7.26$\\
        & RG & $41.6^{+0.05}_{-0.04}$ &  $185.0\pm 0.6$ & $7.49$ \\
      \hline
       \citetalias{Paper6}  & \themis & $42.2^{+0.43}_{-0.41}$ &  $201.1^{+2.5}_{-2.3}$ & $1.07$ \\
        (GC) & \texttt{dynesty} & $41.6^{+0.51}_{-0.46}$ & $175.9^{+2.1}_{-2.0}$ & $0.89$ \\
    \hline
       \citetalias{Paper4}   & \texttt{DIFMAP} & $40.7 \pm 2.6$ & $173.3 \pm 4.8$ & $2.19$ \\
         (image & \texttt{eht-imaging} & $41.0 \pm 1.4$ & $168.0 \pm 6.9$ & $0.97$\\
        domain)  & \texttt{SMILI} & $42.3 \pm 1.6$ & $167.6 \pm 2.8$ & $1.08$ \\
     \hline
    \hline 
    \end{tabularx}
    \label{tab:results_2017}
    \end{center}
     $^\text{a}$ results from this work and \citetalias{Paper6} correspond to visibility domain-based estimator $\phi_{\rm B}$, while results from \citetalias{Paper4} correspond to the image domain estimator $\eta$, similar to the one given by \autoref{eq:PA}
\end{table}

\begin{figure*}[t!]
\centering
\includegraphics[trim={0.1cm 0.0cm 0 0.0cm},clip,width=1.0\linewidth]{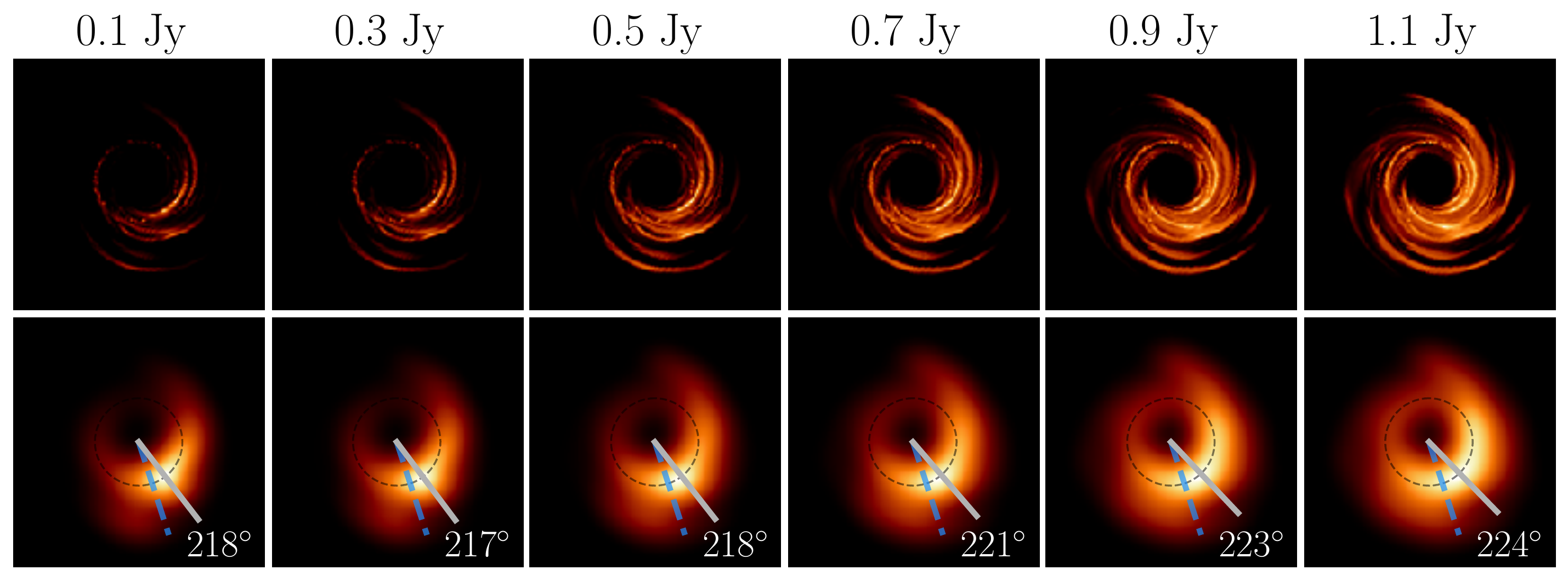}
\caption{\textit{Top row:} a single GRMHD simulation snapshot, ray-traced with total compact flux density adjusted to values between 0.1\,Jy and 1.1\,Jy. The brightness distribution in each panel has been normalized by the brightness maximum. EHT observations of \m87 found a total compact flux density of 0.8-0.9\,Jy in 2009--2012 and 0.5-0.6\,Jy in 2013 and 2017, see the bottom panel of \autoref{fig:fluxes}. \textit{Bottom row:} same snapshots blurred to the resolution of 15\,$\mu$as. The 42\,$\mu$as diameter ring is indicated with a~dashed circle. The position angle convention $\phi_{\rm B, exp}$ is shown with a~dashed blue bar. The gray bar indicates the image domain position angle $\eta$ calculated using \autoref{eq:PA}, and values of $\eta$ obtained are given in the bottom-right corner of each panel. }
\label{fig:scaling_flux}
\end{figure*}

\section{Discussion and conclusions}
\label{sec:discussion}

Processing five independent observations of \m87 in a unified framework offers important insights into the source morphology and stability. While the constraining power 
of the 2009--2013 data sets is rather weak in comparison to the 2017 observations, we find evidence for a~persistent ring structure that shows modest structural variability. Both the persistence and variability offer important constraints for models of \m87. We will now discuss our evidence and theoretical implications for the presence of the shadow feature in 2009--2017 (\autoref{subsec:ringlike_morphology}), the persistence of the source geometry as an argument for its theoretical interpretation (\autoref{sec:geometry_persistence}), and the variability of the source geometry (\autoref{subsec:variability}). We also summarize the limitations and caveats of the theoretical interpretation within the GRMHD framework (\autoref{sub:limitations}).

\subsection{Presence of the shadow feature}
\label{subsec:ringlike_morphology}

In \autoref{sec:ModelingRealData} we have discussed the fits of the asymmetric ring models to the \m87 observations, indicating that all data sets are consistent with such a~geometry. Within the framework of a~ring model, a~key question for the archival observations is whether we unambiguously detect the inner flux depression seen in the 2017 results, the expected feature of a~black hole shadow. While maximum likelihood estimators clearly indicate this feature in all cases  (\autoref{fig:oldrings}), a~detailed inspection of the posterior distributions shown in \autoref{app:cornerplots} allows us to conclude that only the 2013 archival data set provides a~robust detection of the central flux depression, constraining the relative thickness of the ring $f_{\rm w}$ to less than 0.5 (see \autoref{tab:results2}). For the 2009--2012 data sets, the RT model posteriors indicate some preference toward the presence of a flux depression; however, a~disk-like filled-in geometry cannot be excluded with a~high degree of confidence (\autoref{app:cornerplots}). This is not entirely surprising, as the 2009--2012 observations lack baselines with projected lengths $> 3.6$\,G$\lambda$, probing spatial frequencies higher than the visibility null $b_0$ (\autoref{subs:salient_features}), which are more sensitive to differences between an empty ring, a~filled-in disk, and a~Gaussian. 

However, the presence of the shadow feature is sensitive to changes in the optical depth. The total compact flux density in 2009--2012 was measured to be $0.8-0.9$\,Jy, significantly higher than the $0.5-0.6$\,Jy observed in 2013 and 2017 (\autoref{fig:fluxes}, bottom panel). 
Mildly elevated levels of X-ray
emission from the nucleus of M87 before 2016 were also reported \citep{Sun2018}. These measurements suggest a~higher mass accretion rate in 2009--2012 and hence a~larger density scale, in turn increasing the opacities and the optical depth, possibly changing the appearance of the black hole shadow \citep{Moscibrodzka2012,CK2015}. Is it possible that the shadow feature had been obscured by a~more optically thick medium in 2009--2012 than in the case of the more recent 2013 and 2017 observations? While the fully general answer to that question would require extensive testing of a~variety of GRMHD models, we address this concern by analyzing a~representative example from the library of simulated \m87 images. We consider a~random snapshot from a~SANE simulation with spin $a_*\,=\,0.5$, and electron temperature parameter $R_{\rm high} = 20$. The $R_{\rm low}$ parameter is equal to 1, as it is throughout this paper. The snapshot is then repeatedly ray-traced with its density scale (and hence opacities and emissivities) adjusted to give a total compact flux density equal to 0.1, 0.3, 0.5, 0.7, 0.9, and 1.1\,Jy. The resulting images, normalized by the brightness maximum, are shown in \autoref{fig:scaling_flux}. These findings indicate that variation in the total compact flux density between 0.1 and 1.1\,Jy does not eliminate the central brightness depression. Judging from the similarity between blurred images, the flux density scaling is also not expected to influence the estimated diameter or the position angle appreciably. However, it is likely to influence measures of asymmetry, such as the slash parameter $\beta$ in the RT/RG models, discussed in \autoref{sec:ModelSpecification}. Interestingly, we see a~preference toward more symmetric source geometry (larger $\beta$) in the 2012 posteriors (\autoref{app:cornerplots}). We conclude that the central flux depression was most likely present throughout the 2009--2017 observations and that a~lack of a~high-confidence detection of this feature in 2009--2012 is presumably a~consequence of the very limited $(u,v)$-coverage.

None of the EHT observations so far took place during an unambiguous flaring activity period of the M87 nucleus, such as the events discussed in \cite{Abramowski2012}.
We note that such an event could potentially influence the image morphology more strongly than the moderate increase of total brightness considered in \autoref{fig:scaling_flux}. Future simultaneous multiwavelength observational campaigns will shed light on the structural changes in the \m87 compact radio image in relation to the enhanced activity in different parts of the spectrum, allowing the site of particle acceleration to be
localized.

\subsection{Black hole shadow or a~transient feature }
\label{sec:geometry_persistence}
\begin{figure}[t]
\centering
\includegraphics[trim={2.5cm 0.5cm 0 0.0cm},clip,width=1.03\linewidth]{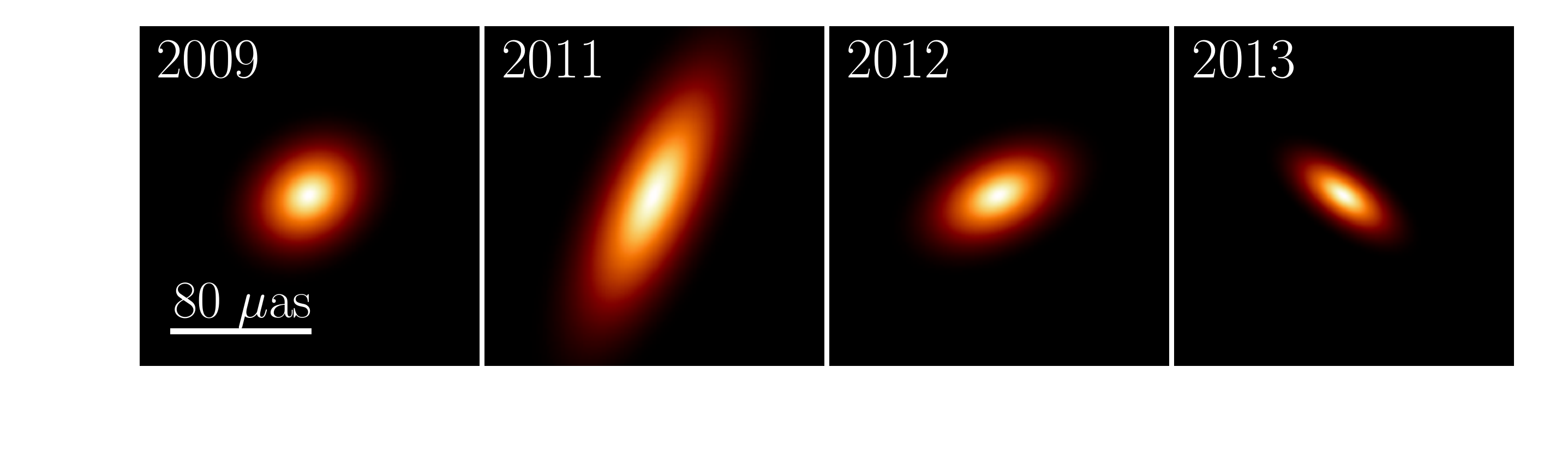}
\caption{Year to year consistency of the best-fitting (ML) asymmetric Gaussian model to the \m87 data sets.}
\label{fig:oldgauss}
\end{figure}

As discussed in \autoref{sec:modeling}, the 2009--2013 data sets can be successfully modeled with both an asymmetric Gaussian and a~ring model, with each giving a similar fit quality. However, the maximum likelihood Gaussian models are very inconsistent in size, shape, and orientation across different years, see \autoref{fig:oldgauss}. In contrast, the best-fitting ring models, as seen in \autoref{fig:oldrings}, are similar in size over all epochs under the priors described in Sections \ref{sec:FittingProcedure} and \ref{sub:degenercies}.
Even when considered separately from the 2017 results, this consistency supports our choice to interpret the 2009--2013 morphology as a~ring and to draw conclusions from the results of fitting asymmetric ring models to the data. The 95\% confidence intervals of all 2009--2017 diameter posteriors lie within
40$\pm$12\,$\mu$as,
while ML estimators of the diameter from all years lie within
43\,$\pm$5\,$\mu$as, see \autoref{tab:results2}.
These values are also consistent with the \m87 diameter measurement of 42$\pm$3\,$\mu$as reported by \citetalias{Paper6} and with the expected size of an observed black hole of mass/distance corresponding to the stellar dynamics measurement \citep{Blakeslee2009,Gebhardt2011}.
All 2009--2017 diameter measurements are inconsistent with the gas dynamics mass measurement \citep{walsh13}, which predicts a~diameter roughly half as large. 
The consistency in the diameter across multiple observational epochs supports the interpretation of the ring-like feature as emission from the immediate surroundings of the supermassive black hole.

The four EHT observations of \m87 in 2017 spanned about a week, corresponding to a~timescale of $\sim 15\,M$. With such a~short time span, we cannot exclude a~transient origin for the source morphology using 2017 data alone. However, such a~feature would need to attain the geometry and apparent size expected of a~shadow of a~black hole (of independently measured mass-to-distance ratio) through an unusual coincidence. Moreover, transient features are unlikely to persist with a~similar geometry throughout multiple years of observations, corresponding to $\sim 10^3\!-\!10^4\,M$ timescales (the total span of our analyzed observations is $7900 \,M$). For example, a~lensed background source would need a~low transverse velocity of $v \lesssim 40 \, \rm km \, \rm s^{-1}$ to travel $\lesssim 1 \,M$ between 2009 and 2017. This is much smaller than the gas velocities seen in the nucleus of M87 \citep[e.g.,][]{macchetto1997}. A~bright feature moving with the jet ($v \sim c$) should travel $\gtrsim 1$ pc on that timescale, a~factor of roughly $10^3$ larger than the physical size of the ring itself.  A~bright knot or other stationary jet feature would need to persist with a~similar location, flux density, and ring morphology to remain consistent with these results. The 8 yr span of the 2009--2017 monitoring is also much longer than the typical variability timescale of the M87 nucleus observed at 230\,GHz, which is $\sim\!50$ days \citep{Bower2015}. While features remaining stationary for many years in otherwise rapidly flowing jets have been reported and interpreted as standing recollimation shocks \citep{mojave2009}, such a~configuration would constitute one more unusual coincidence. Thus, we conclude that with multiple years of observations remaining consistent with a~$\sim 40\,\mu$as ring model, it is highly unlikely that the origin of the observed geometry could be a~transient feature.

\subsection{Time variability of the source geometry}
\label{subsec:variability}

In addition to conclusions from the persistence of the ring structure, we can also draw inferences from the variability observed in the ring structure across the 2009--2017 data sets. In particular, the spread of the diameter and brightness position angle estimates (\autoref{tab:results2}) are significantly larger than the spread for corresponding static synthetic data sets (\autoref{tab:results_sim}). As a specific example, the circular standard deviation of the ML position angle estimators given in \autoref{tab:results_sim} is equal to 19$^\circ$, 19$^\circ$, 11$^\circ$, 4$^\circ$ for GRMHD1, GRMHD2, MODEL1, and MODEL2, respectively. For the observational data (\autoref{tab:results2}) the circular standard deviation is equal to 48$^\circ$.  This larger spread suggests that we are detecting intrinsic structural variability despite the large uncertainties in the parameters estimated with pre-2017 observations. Moreover, unambiguous signatures of intrinsic variability on a~timescale of years can be seen directly in the visibility data (\autoref{subs:salient_features}).

\begin{figure}[t]
\centering
\includegraphics[trim={0cm 0.cm 0 -0.0cm},clip,width=1.0\linewidth]{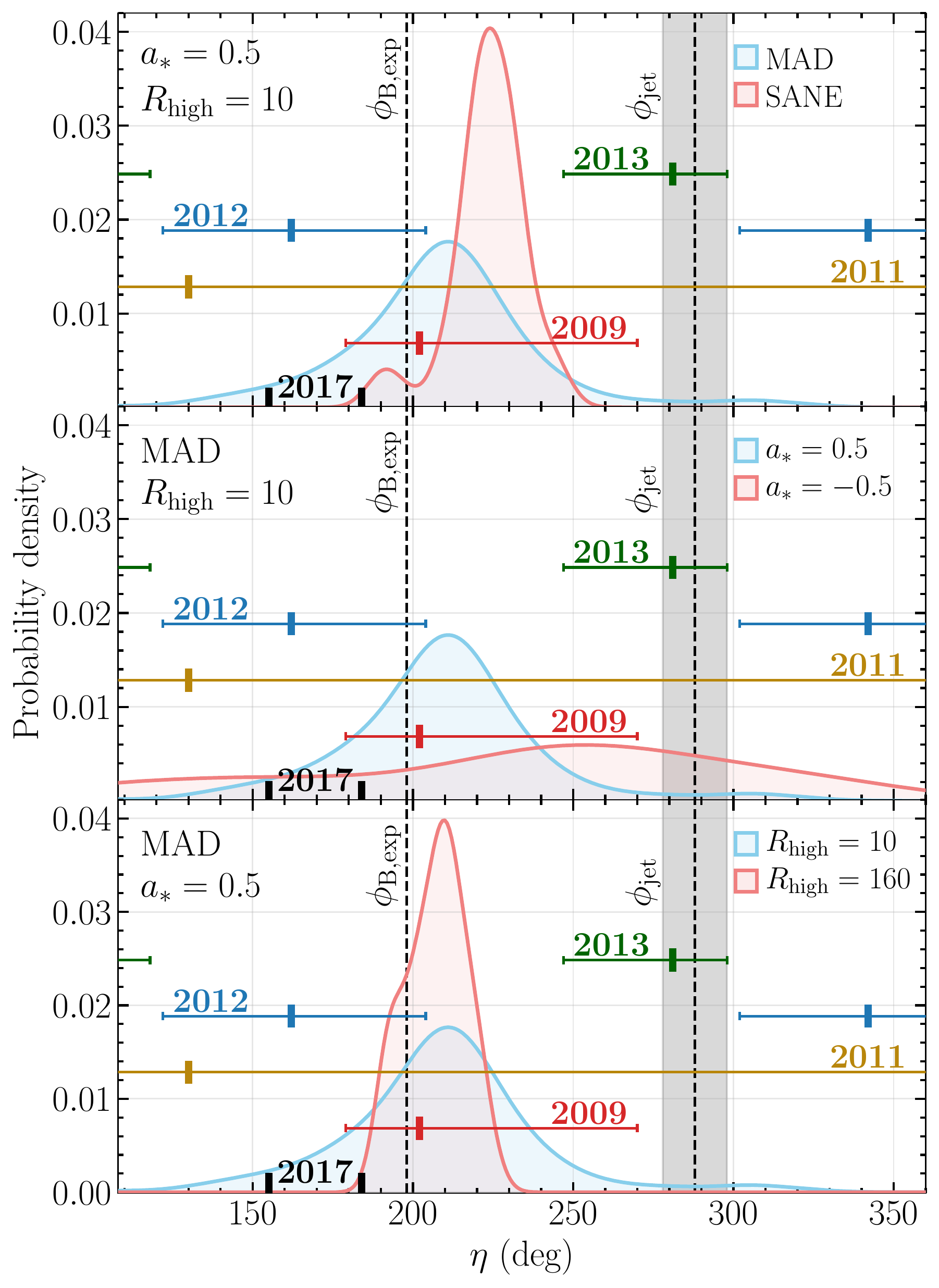}
\caption{Histograms of the brightness position angle $\eta$, measured in the image domain for 500 ray-traced snapshots of GRMHD simulations from the EHT \m87 Simulation Library \citepalias{Paper5}. In each panel, the blue histogram represents the same fixed model S1: MAD, $a_*\!=\!0.5$, $R_{\rm high}\!=\!10$, and the red histogram represents a~model in which a~single parameter has been altered with respect to the fixed model (S2-S4). Orientations measured in the 2009--2017 observational data sets with an~ML estimator are indicated, with 68\% confidence intervals (\autoref{tab:results2}, two results from 2017 are shown without their very narrow error bars). The gray area around $\phi_{\rm jet}$ corresponds to the observed jet position angle variation \citep{Walker2018}.
}
\label{fig:sim_PA}
\end{figure}

Because GRMHD simulations naturally model both the source structure and its variability, they provide an important pathway for drawing conclusions from the observed variability. 
As a preliminary demonstration for a~comprehensive study that will be published separately, we consider a~small subset of the EHT Image Library \citepalias{Paper5}. The simulations are parameterized with the black hole spin $a_*$, the electron temperature parameter $R_{\rm high}$ (see \autoref{subsec:grmhd_snapshots}) and the accretion state -- strongly magnetized magnetically arrested disk \citep[MAD,][]{Narayan2003}, or standard and normal evolution (SANE) flow, such as those considered in Sections~\ref{subsec:grmhd_snapshots} and \ref{subsec:ringlike_morphology}. Other parameters (e.g., total compact flux density of 0.5\,Jy, inclination, jet position angle) are adjusted to match the observed properties of \m87. For our exploratory study, we utilize the following four simulations: (S1) MAD, $a_*=0.5$, $R_{\rm high}=10$ ; (S2) SANE, $a_*=0.5$, $R_{\rm high}=10$ ; (S3) MAD, $a_*=-0.5$, $R_{\rm high}=10$; (S4) MAD, $a_*=0.5$, $R_{\rm high}=160$.
For each simulation, we take 500 ray-traced snapshots with $5\,M$ separation in time. For each snapshot we calculate the position angle of the bright component $\eta$ using \autoref{eq:PA}.
We then construct histograms of $\eta$ for each of the four simulations, shown in \autoref{fig:sim_PA}. 

Each of the simulation parameters influences the distribution of $\eta$, both in terms of its mean and spread. Some of these differences can be readily understood, for instance, in the case of prograde accretion onto a~spinning black hole, the radiation is boosted both with Doppler and with frame-dragging effects. The position angle of the bright component is thus expected to be relatively more influenced by the geometry (assumed to be fixed) and not by the stochastic component than the retrograde case, in which Doppler effect and frame-dragging counteract and the geometry becomes relatively less important. Irrespective of the mechanism, some variants of simulations have great difficulties explaining either source orientation on 2017 April 6th (bright component too far clockwise), or in 2013 (too far counterclockwise), as indicated in \autoref{fig:sim_PA}. Thus, continued EHT observations, with tight constraints on $\eta$ spaced over multiple years, will constrain these types of models on the basis of variability in $\eta$. 

\subsection{Limitations of the current approach}
\label{sub:limitations}

There are caveats to the simple analysis outlined above that should be addressed in more focused future studies. The simulations that we consider do not capture the full extent of the physics relevant for \m87. In particular, the electron temperature could be calculated in a~more self-consistent fashion than via a~temperature ratio prescription of \autoref{eq:R}, by separately evolving the energy of ions and electrons \citep{Sadowski2017}. Then, one could evolve a~population of nonthermal electrons \citep{Chael2017}, determine the electrons heating with a~well-motivated subgrid prescription \citep{ChaelRowan2018,Jordy2019}, or even employ nonideal MHD to model nonthermal emission caused by the particles accelerated through magnetic reconnection in a~more self-consistent manner \citep{Bart2019}. In the current analysis we also make an assumption of no tilt between the plane of accretion and the black hole spin. If the tilt is present, an additional degree of freedom (``camera longitude'') corresponding to a~position angle of the black hole spin (misaligned with the jet position angle $\phi_{\rm jet}$) in the image plane will influence the observed crescent orientation \citep{Chatterjee2020}. In that case, analysis of the position angle distributions could place joint constraints on the tilt magnitude, the longitude, and other parameters of the simulation. Large-scale parameter surveys with these extensions to the GRMHD setup are currently precluded by the immense computational costs.

A separate concern is whether the orientation of crescent models fitted to the VLBI data is consistent with the image domain $\eta$ (\autoref{eq:PA}). In the case of the synthetic data sets considered in \autoref{sec:synthetic}, the two GRMHD data sets exhibited quite large biases while the two MODEL data sets showed a high level of consistency, so this issue requires further study. Characterizing GRMHD simulations in terms of VLBI observables is the subject of continued work.

\section{Summary}
\label{sec:summary}

We have performed geometric modeling of the 2009--2017 EHT observations of \m87. Motivated by EHT imaging and modeling results using the 2017 observations and the stability of fits across the archival observations, we have used a simple asymmetric ring model. We found that the fitted ring diameter is stable throughout these observations, which strongly argues in favor of its association with the shadow of a supermassive black hole. We observe indications of modest intrinsic variability in the total flux density of the ring and in its position angle. 

Specifically, we find the brightness asymmetry along the east-west direction in the previously unpublished 2013 observations, while all other data sets are consistent with the north-south asymmetry direction seen in the 2017 EHT images. This degree of position angle variation is seen in some GRMHD simulations of \m87, while others do not show position angle variations as broad as those observed between 2013 and 2017. Thus, the source variability over these observations provides new constraints on the simulation parameters, including the black hole spin, accretion flow magnetization, and electron heating model. As an example, the GRMHD MAD model with spin $a_*\!=\!0.5$ and $R_{\rm high}\!=\!160$ (last panel of \autoref{fig:sim_PA}), which was determined to be viable by \citetalias{Paper5}, is inconsistent with the presented position angle measurements. We expect that unmodeled physical effects such as black hole and accretion flow spin misalignment may also be important in interpreting this variability. 

Our results extend the temporal span of EHT constraints on the ring morphology by nearly three orders of magnitude, from ${\sim}15\,M$ over the 2017 observations to ${\sim}7900\,M$ between the 2009 and 2017 campaigns. Because the correlation timescale for \m87 is expected to be at least a few tens of $M$, the longer span is critical for decoupling stable image features such as the black hole shadow from transient features associated with the turbulent accretion flow. As continued EHT observations become available, the variation of the estimated position angle should allow us to discriminate between viable GRMHD models, providing constraints on the physical parameters of \m87 and opening an exciting new avenue for quantitative time-domain studies of structural variability in \m87.

\section*{acknowledgements}
The authors of the present paper thank the following organizations and programs: 
the Academy of Finland (projects 274477, 284495, 312496);
the Advanced European Network of E-infrastructures for Astronomy with the SKA (AENEAS) project, supported by the European Commission Framework Programme Horizon 2020 Research and Innovation action under grant agreement 731016;
the Alexander von Humboldt Stiftung; 
the Black Hole Initiative at Harvard University, through a grant (60477) from the John Templeton Foundation; 
the China Scholarship Council;
Comisi{\'o}n Nacional de Investigaci{\'o}n Cient\'{\i}fica y Tecnol{\'o}gica (CONICYT, Chile, via PIA ACT172033, Fondecyt projects 1171506 and 3190878, BASAL AFB-170002, ALMA-conicyt 31140007);
Consejo Nacional de Ciencia y Tecnolog{\'i}a (CONACYT, Mexico, projects 104497, 275201, 279006, 281692);
the Delaney Family via the Delaney Family John A.\ Wheeler Chair at Perimeter Institute; 
Direcci{\'o}n General de Asuntos del Personal Acad{\'e}mico-—Universidad Nacional Aut{\'o}noma de M{\'e}xico (DGAPA-—UNAM, project IN112417); 
the European Research Council Synergy Grant "BlackHoleCam: Imaging the Event Horizon of Black Holes" (grant 610058); 
the Generalitat Valenciana postdoctoral grant APOSTD/2018/177 and GenT Program (project CIDEGENT/2018/021); 
the Gordon and Betty Moore Foundation (grants GBMF-3561, GBMF-5278); 
the Istituto Nazionale di Fisica Nucleare (INFN) sezione di Napoli, iniziative specifiche TEONGRAV;
the International Max Planck Research School for Astronomy and Astrophysics at the Universities of Bonn and Cologne; 
the Jansky Fellowship program of the National Radio Astronomy Observatory (NRAO);
the Japanese Government (Monbukagakusho: MEXT) Scholarship; 
the Japan Society for the Promotion of Science (JSPS) Grant-in-Aid for JSPS Research Fellowship (JP17J08829);
the Key Research Program of Frontier Sciences, Chinese Academy of Sciences (CAS, grants QYZDJ-SSW-SLH057, QYZDJ-SSW-SYS008, ZDBS-LY-SLH011);
the Leverhulme Trust Early Career Research Fellowship;
the Max-Planck-Gesellschaft (MPG);
the Max Planck Partner Group of the MPG and the CAS;
the MEXT/JSPS KAKENHI (grants 18KK0090, JP18K13594, JP18K03656, JP18H03721, 18K03709, 18H01245, 25120007);
the MIT International Science and Technology Initiatives (MISTI) Funds; 
the Ministry of Science and Technology (MOST) of Taiwan (105-2112-M-001-025-MY3, 106-2112-M-001-011, 106-2119-M-001-027, 107-2119-M-001-017, 107-2119-M-001-020, and 107-2119-M-110-005);
the National Aeronautics and Space Administration (NASA, Fermi Guest Investigator grant 80NSSC17K0649 and Hubble Fellowship grant HST-HF2-51431.001-A awarded by the Space Telescope Science Institute, which is operated by the Association of Universities for Research in Astronomy, Inc., for NASA, under contract NAS5-26555); 
the National Institute of Natural Sciences (NINS) of Japan;
the National Key Research and Development Program of China (grant 2016YFA0400704, 2016YFA0400702); 
the National Science Foundation (NSF, grants AST-0096454, AST-0352953, AST-0521233, AST-0705062, AST-0905844, AST-0922984, AST-1126433, AST-1140030, DGE-1144085, AST-1207704, AST-1207730, AST-1207752, MRI-1228509, OPP-1248097, AST-1310896, AST-1312651, AST-1337663, AST-1440254, AST-1555365, AST-1715061, AST-1615796, AST-1716327, OISE-1743747, AST-1816420); 
the Natural Science Foundation of China (grants 11573051, 11633006, 11650110427, 10625314, 11721303, 11725312, 11933007); 
the Natural Sciences and Engineering Research Council of Canada (NSERC, including a Discovery Grant and the NSERC Alexander Graham Bell Canada Graduate Scholarships-Doctoral Program);
the National Youth Thousand Talents Program of China;
the National Research Foundation of Korea (the Global PhD Fellowship Grant: grants NRF-2015H1A2A1033752, 2015-R1D1A1A01056807, the Korea Research Fellowship Program: NRF-2015H1D3A1066561); 
the Netherlands Organization for Scientific Research (NWO) VICI award (grant 639.043.513) and Spinoza Prize SPI 78-409; the New Scientific Frontiers with Precision Radio Interferometry Fellowship awarded by the South African Radio Astronomy Observatory (SARAO), which is a facility of the National Research Foundation (NRF), an agency of the Department of Science and Technology (DST) of South Africa;
the Onsala Space Observatory (OSO) national infrastructure, for the provisioning of its facilities/observational support (OSO receives funding through the Swedish Research Council under grant 2017-00648)
the Perimeter Institute for Theoretical Physics (research at Perimeter Institute is supported by the Government of Canada through the Department of Innovation, Science and Economic Development and by the Province of Ontario through the Ministry of Research, Innovation and Science);
the Russian Science Foundation (grant 17-12-01029); 
the Spanish Ministerio de Econom\'{\i}a y Competitividad (grants PGC2018-098915-B-C21,  AYA2016-80889-P); 
the State Agency for Research of the Spanish MCIU through the "Center of Excellence Severo Ochoa" award for the Instituto de Astrof\'{\i}sica de Andaluc\'{\i}a (SEV-2017-0709); 
the Toray Science Foundation; 
the US Department of Energy (USDOE) through the Los Alamos National Laboratory (operated by Triad National Security, LLC, for the National Nuclear Security Administration of the USDOE (Contract 89233218CNA000001));
the Italian Ministero dell'Istruzione Universit\`{a} e Ricerca through the grant Progetti Premiali 2012-iALMA (CUP C52I13000140001);
the European Union’s Horizon 2020 research and innovation programme under grant agreement No 730562 RadioNet;
ALMA North America Development Fund;
the Academia Sinica;
Chandra TM6-17006X;
the GenT Program (Generalitat Valenciana) Project CIDEGENT/2018/021.
This work used the Extreme Science and Engineering Discovery Environment (XSEDE), supported by NSF grant ACI-1548562, and CyVerse, supported by NSF grants DBI-0735191, DBI-1265383, and DBI-1743442.  XSEDE Stampede2 resource at TACC was allocated through TG-AST170024 and TG-AST080026N.  XSEDE JetStream resource at PTI and TACC was allocated through AST170028.  The simulations were performed in part on the SuperMUC cluster at the LRZ in Garching, on the LOEWE cluster in CSC in Frankfurt, and on the HazelHen cluster at the HLRS in Stuttgart.  This research was enabled in part by support provided by Compute Ontario (http://computeontario.ca), Calcul Quebec (http://www.calculquebec.ca) and Compute Canada (http://www.computecanada.ca).
We thank the staff at the participating observatories, correlation centers, and institutions for their enthusiastic support.
This paper makes use of the following ALMA data: 
%
ADS/JAO.ALMA\#2016.1.01154.V. 
ALMA is a partnership of the European Southern Observatory (ESO; Europe, representing its member states), NSF, and National Institutes of Natural Sciences of Japan, together with National Research Council (Canada), Ministry of Science and Technology (MOST; Taiwan), Academia Sinica Institute of Astronomy and Astrophysics (ASIAA; Taiwan), and Korea Astronomy and Space Science Institute (KASI; Republic of Korea), in cooperation with the Republic of Chile. The Joint ALMA Observatory is operated by ESO, Associated Universities, Inc.\  (AUI)/NRAO, and the National Astronomical Observatory of Japan (NAOJ).  The NRAO is a facility of the NSF operated under cooperative agreement by AUI. APEX is a collaboration between the Max-Planck-Institut f\"{u}r Radioastronomie (Germany), ESO, and the Onsala Space Observatory (Sweden). The SMA is a joint project between the SAO and ASIAA and is funded by the Smithsonian Institution and the Academia Sinica. The JCMT is operated by the East Asian Observatory on behalf of the NAOJ, ASIAA, and KASI, as well as the Ministry of Finance of China, Chinese Academy of Sciences, and the National Key R\&D Program (No.\ 2017YFA0402700) of China. Additional funding support for the JCMT is provided by the Science and Technologies Facility Council (UK) and participating universities in the UK and Canada. The LMT is a project operated by the Instituto Nacional de Astr\'{o}fisica, \'{O}ptica, y Electr\'{o}nica (Mexico) and the University of Massachusetts at Amherst (USA). The IRAM~30-m telescope on Pico Veleta, Spain is operated by IRAM and supported by CNRS (Centre National de la Recherche Scientifique, France), MPG (Max-Planck-Gesellschaft, Germany) and IGN (Instituto Geogr\'afico Nacional, Spain).
The SMT is operated by the Arizona Radio Observatory, a part of the Steward Observatory of the University of Arizona, with financial support of operations from the State of Arizona and financial support for instrumentation development from the NSF. The SPT is supported by the National Science Foundation through grant PLR- 1248097. Partial support is also provided by the NSF Physics Frontier Center grant PHY-1125897 to the Kavli Institute of Cosmological Physics at the University of Chicago, the Kavli Foundation and the Gordon and Betty Moore Foundation grant GBMF 947. The SPT hydrogen maser was provided on loan from the GLT, courtesy of ASIAA.
The EHTC has received generous donations of FPGA chips from Xilinx Inc., under the Xilinx University Program. The EHTC has benefited from technology shared under open-source license by the Collaboration for Astronomy Signal Processing and Electronics Research (CASPER). The EHT project is grateful to T4Science and Microsemi for their assistance with Hydrogen Masers. This research has made use of NASA's Astrophysics Data System.  We gratefully acknowledge the support provided by the extended staff of the ALMA, both from the inception of the ALMA Phasing Project through the observational campaigns of 2017 and 2018. We would like to thank A. Deller and W. Brisken for EHT-specific support with the use of DiFX.  We acknowledge the significance that Maunakea, where the SMA and JCMT EHT stations are located, has for the indigenous Hawaiian people.
%
%
%

\bibliography{eht2013_bibl.bib}

\appendix
\section{ML estimators for the RG model}
\label{app:imagesRG}
\begin{figure*}[h!]
\centering
\includegraphics[trim={0cm 0.5cm 0 0.1cm},clip,width=0.64\linewidth]{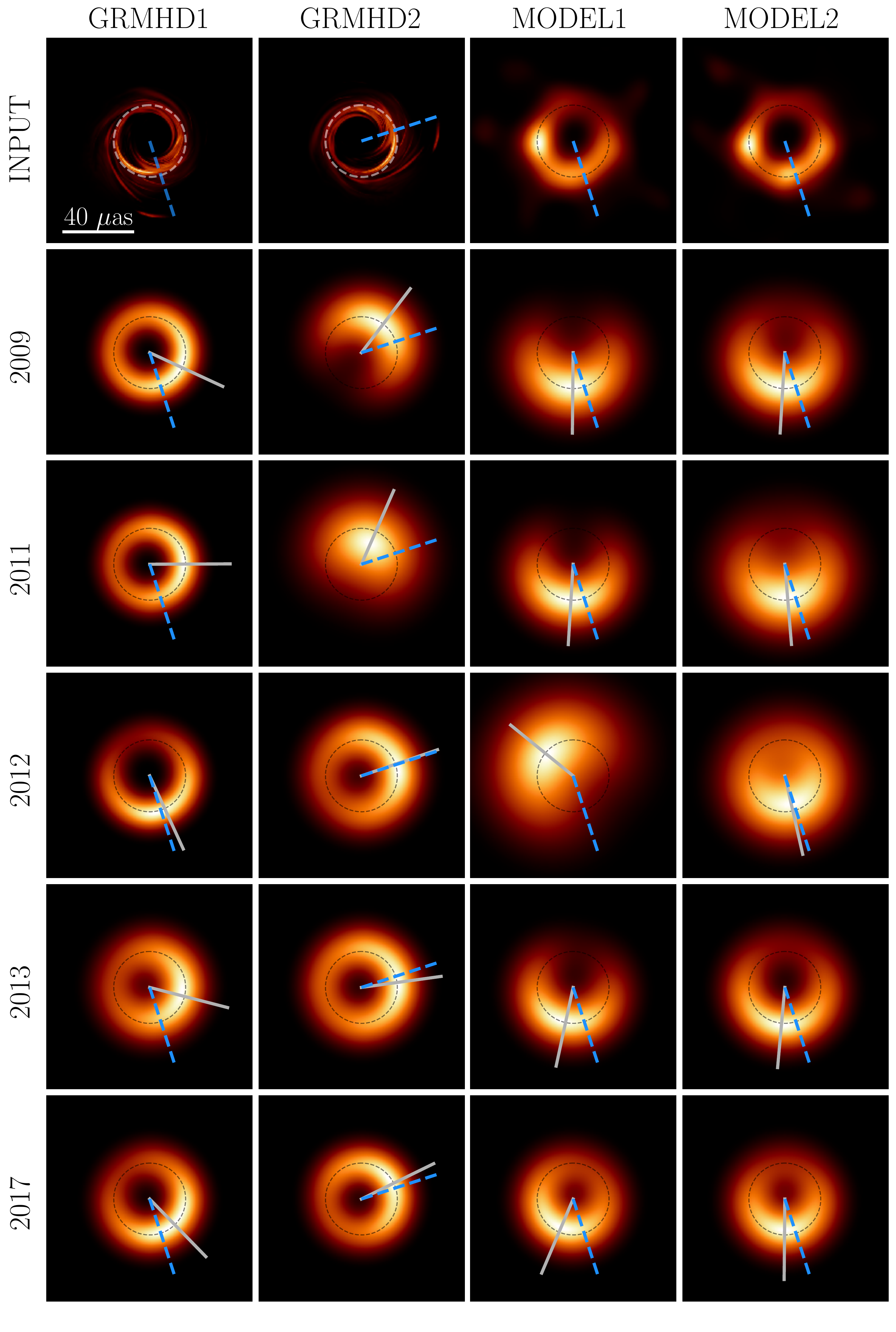}
\caption{Maximum likelihood estimators corresponding to the fits to four synthetic images, shown in the first row (no blurring). Estimators were obtained through synthetic VLBI observations with the $(u,v)$-coverage and uncertainties identical to those of the real observations performed in 2009--2017. The blurred ring model (RG) was used, and the presented images of ML estimators were blurred to a~$15\,\mu$as resolution. Blue dashed lines indicate the convention for the expected position angle of the bright component $\phi_{\rm B, exp}$. The gray bar represents the ML estimate of $\phi_{\rm B}$. The dashed circles correspond to a~diameter of 42\,$\mu$as. }
\label{fig:synthetic_images_blur}
\end{figure*}

Figure \ref{fig:synthetic_images_blur} presents RG models best-fitting the synthetic data sets, following the procedures described in Section \ref{sec:synthetic}.

\newpage
\section{Corner plots for the RT and the RG models}
\label{app:cornerplots}

\begin{figure*}[h!]
\centering
\includegraphics[width=0.495\columnwidth]{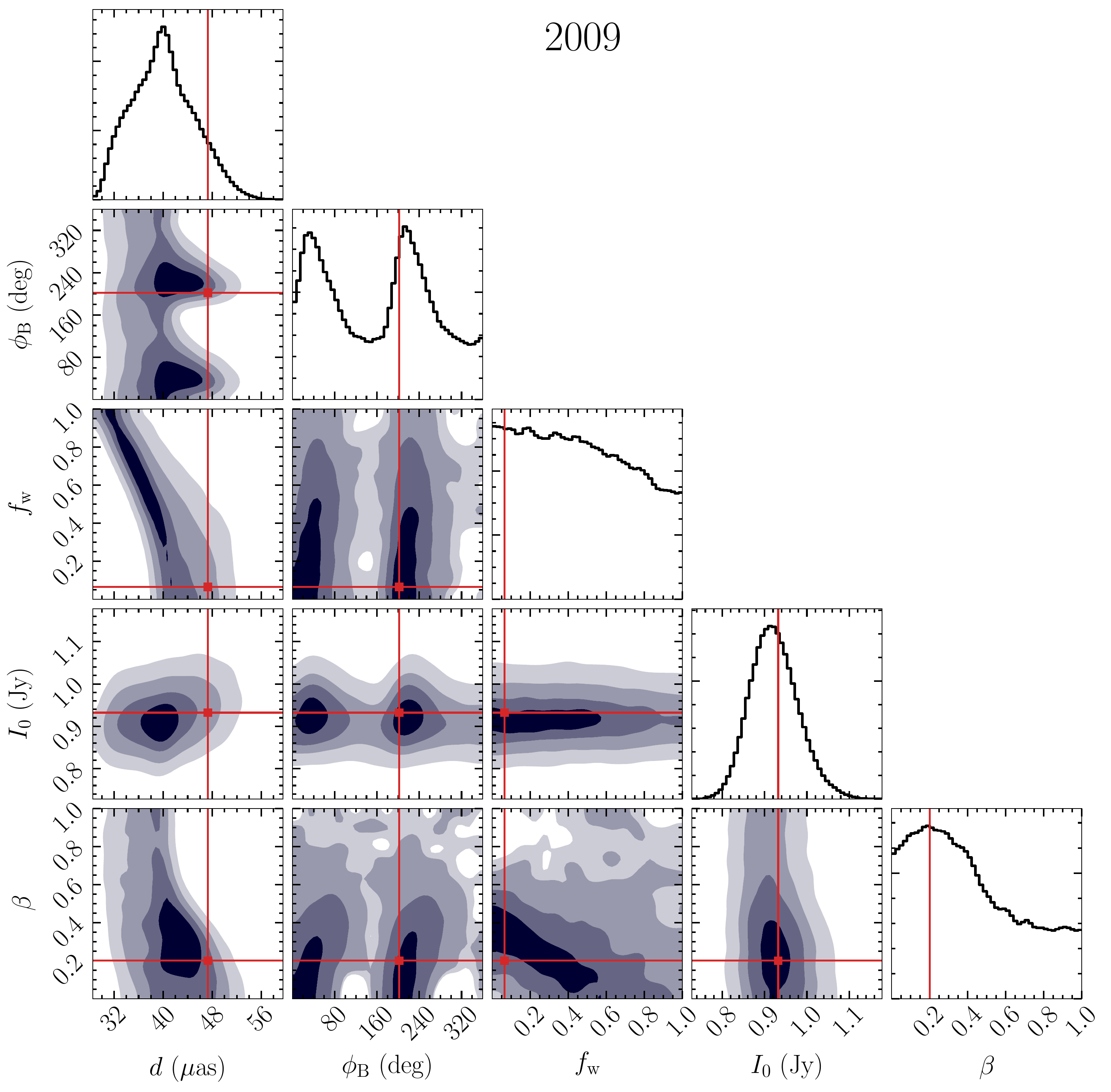}
\includegraphics[width=0.495\columnwidth]{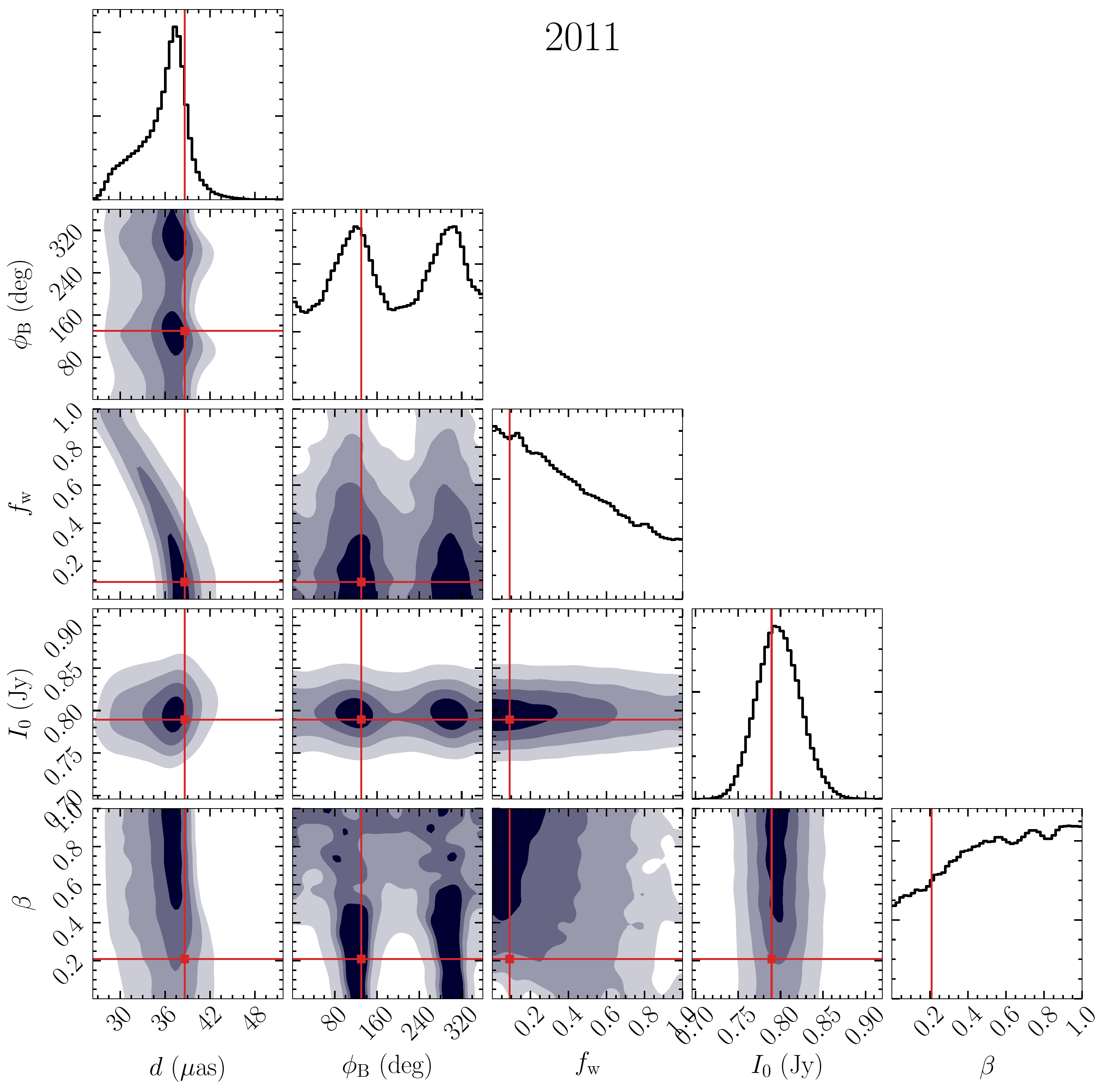}
\includegraphics[width=0.495\columnwidth]{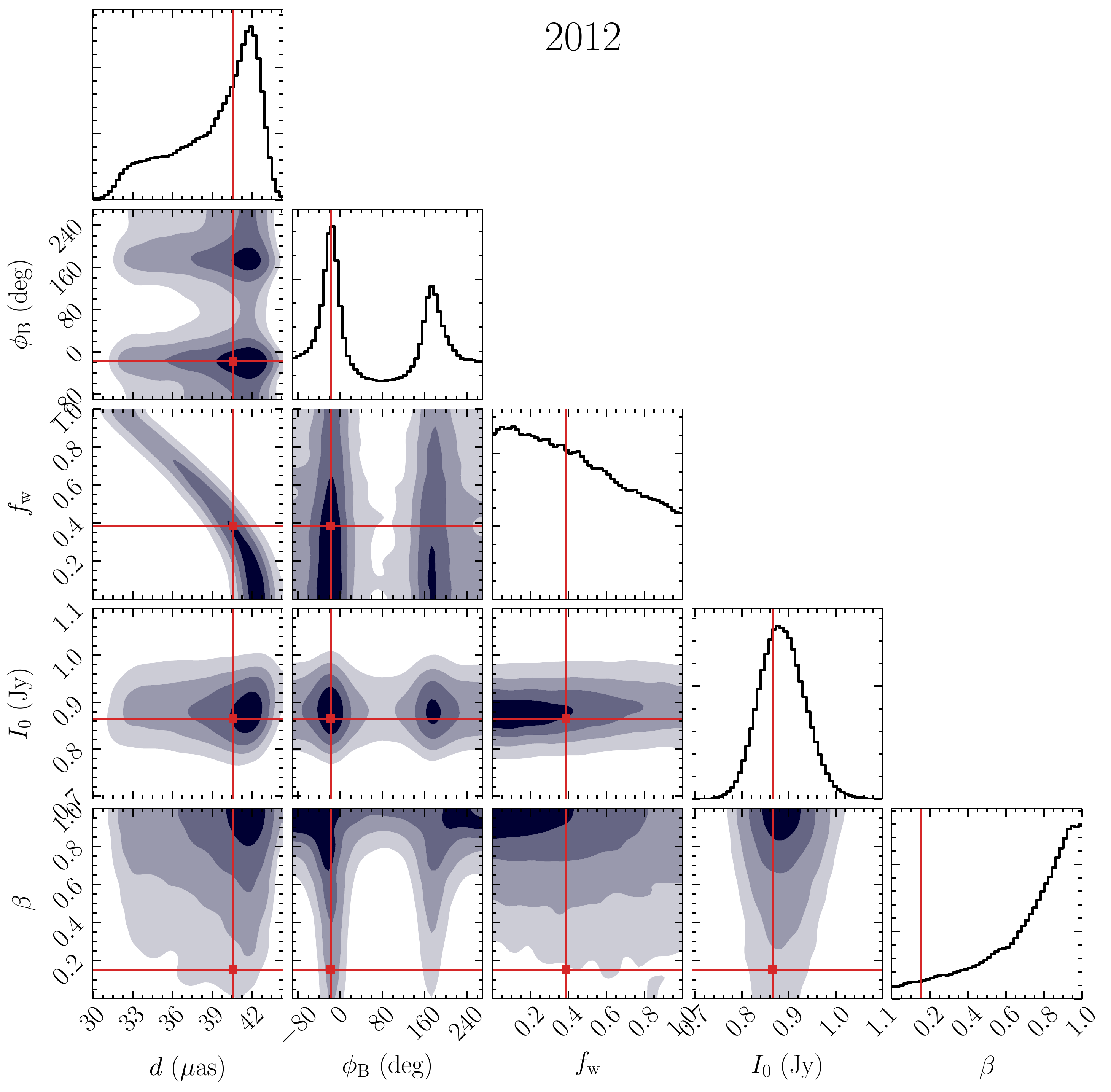}
\includegraphics[width=0.495\columnwidth]{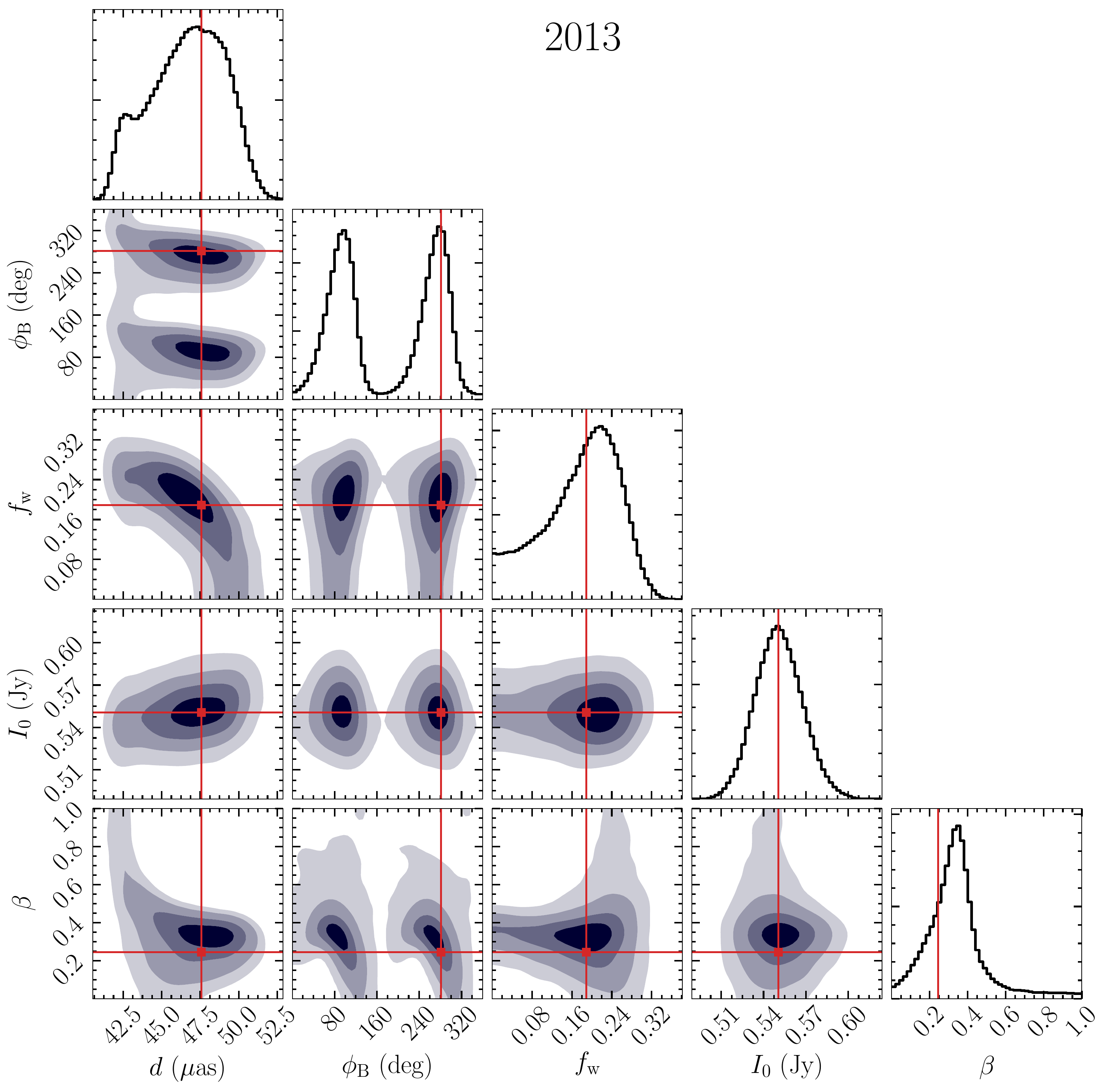}
\caption{Posterior distributions of RT model parameters resulting from fitting to \m87 2009--2013 data, obtained using \themis. The maximum likelihood estimate is indicated with red lines. Contours indicate 0.2, 0.5, 0.8 and 0.95 of the posterior volume.}
\label{fig:cornerRT2009-2013}
\end{figure*}

\begin{figure*}[h!]
\centering
\includegraphics[width=0.495\columnwidth]{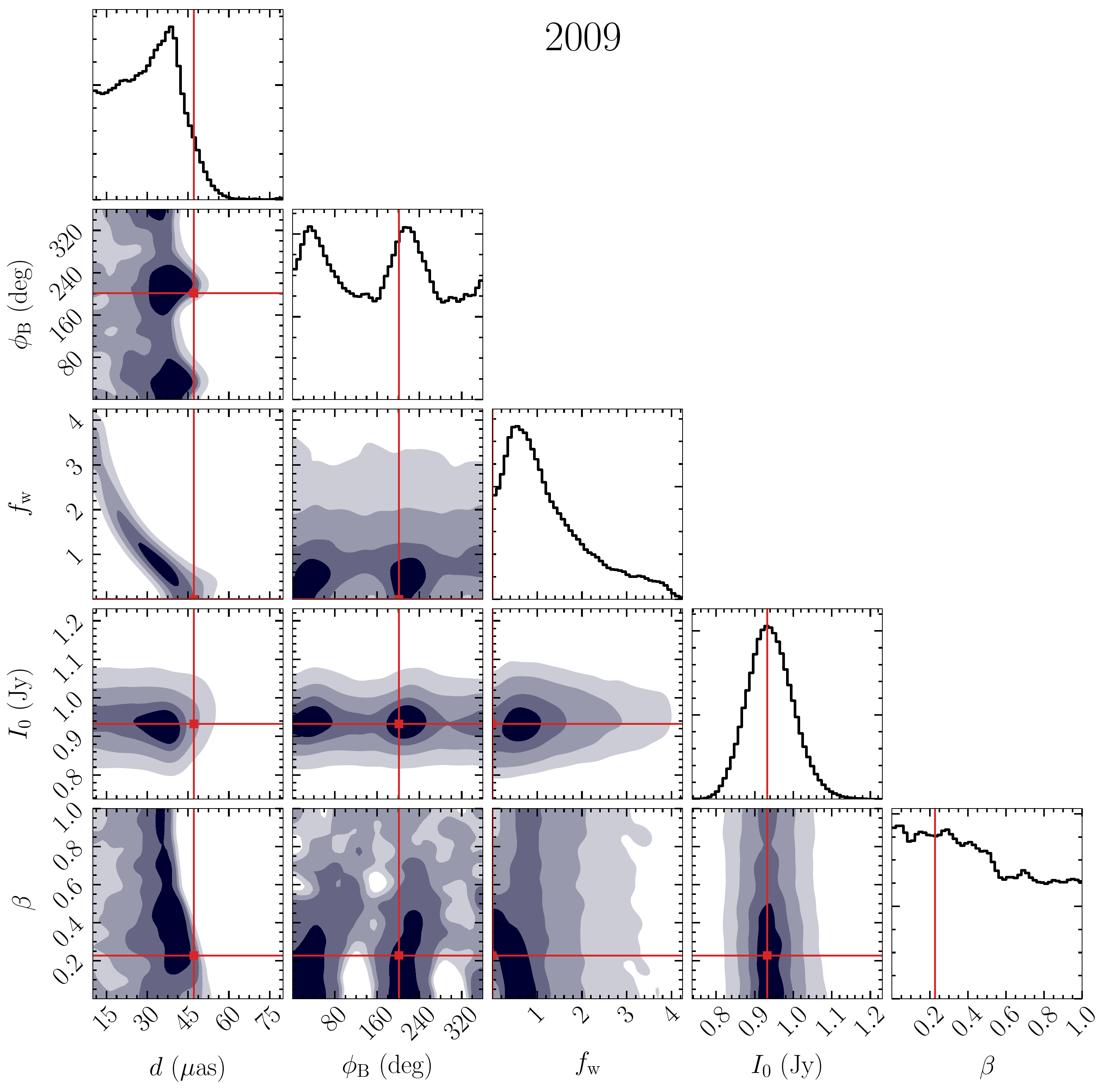}
\includegraphics[width=0.495\columnwidth]{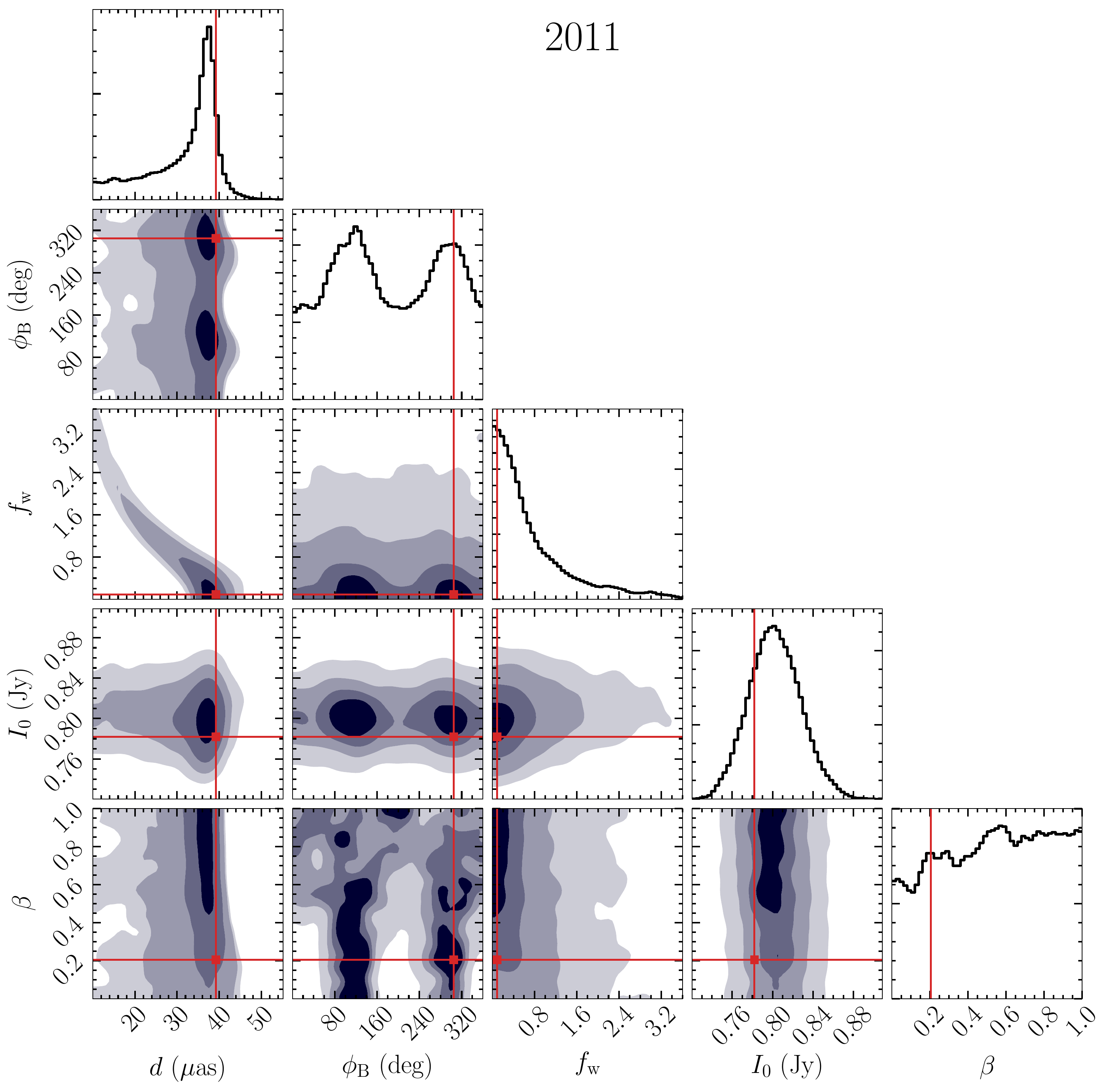}
\includegraphics[width=0.495\columnwidth]{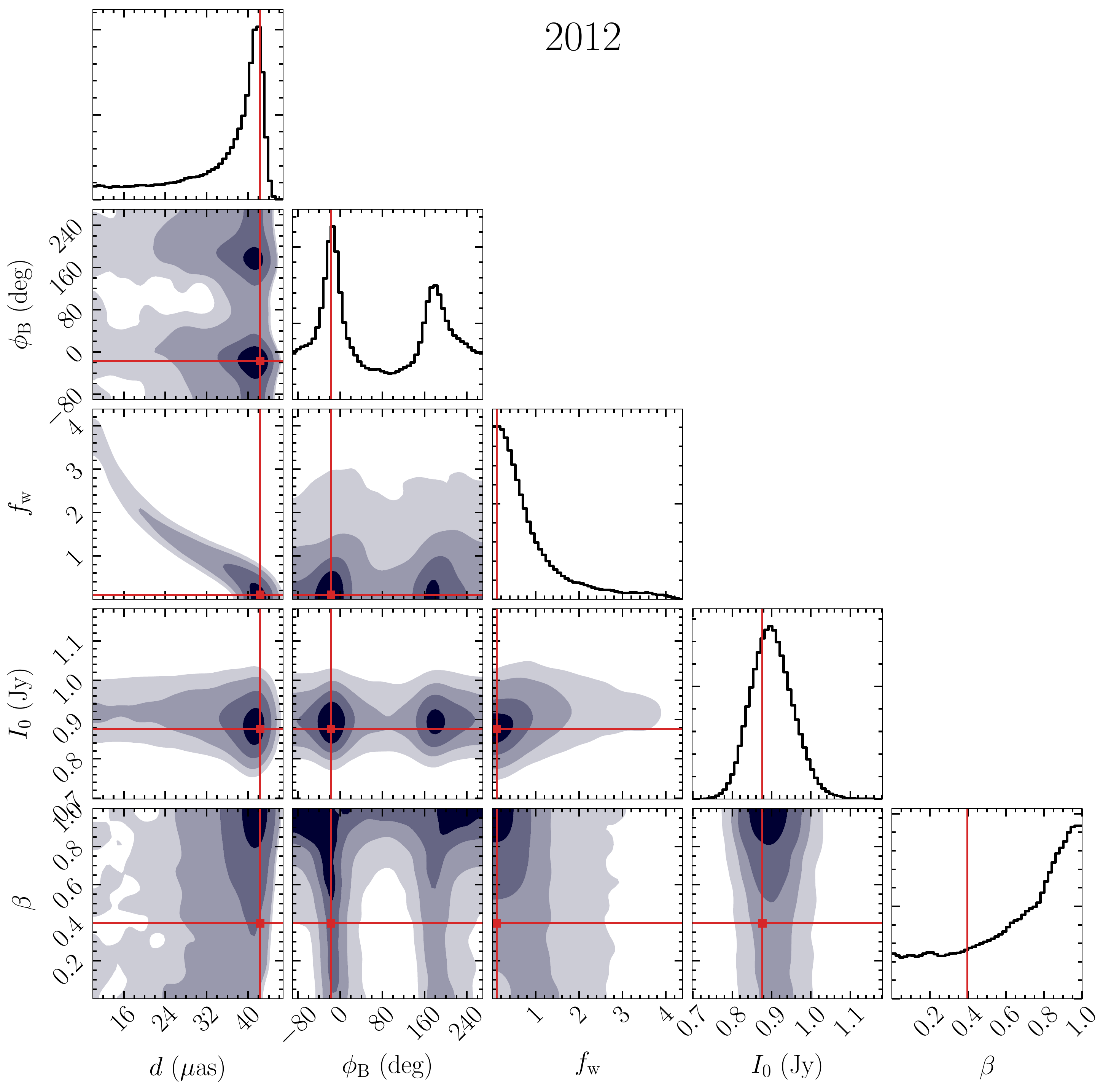}
\includegraphics[width=0.495\columnwidth]{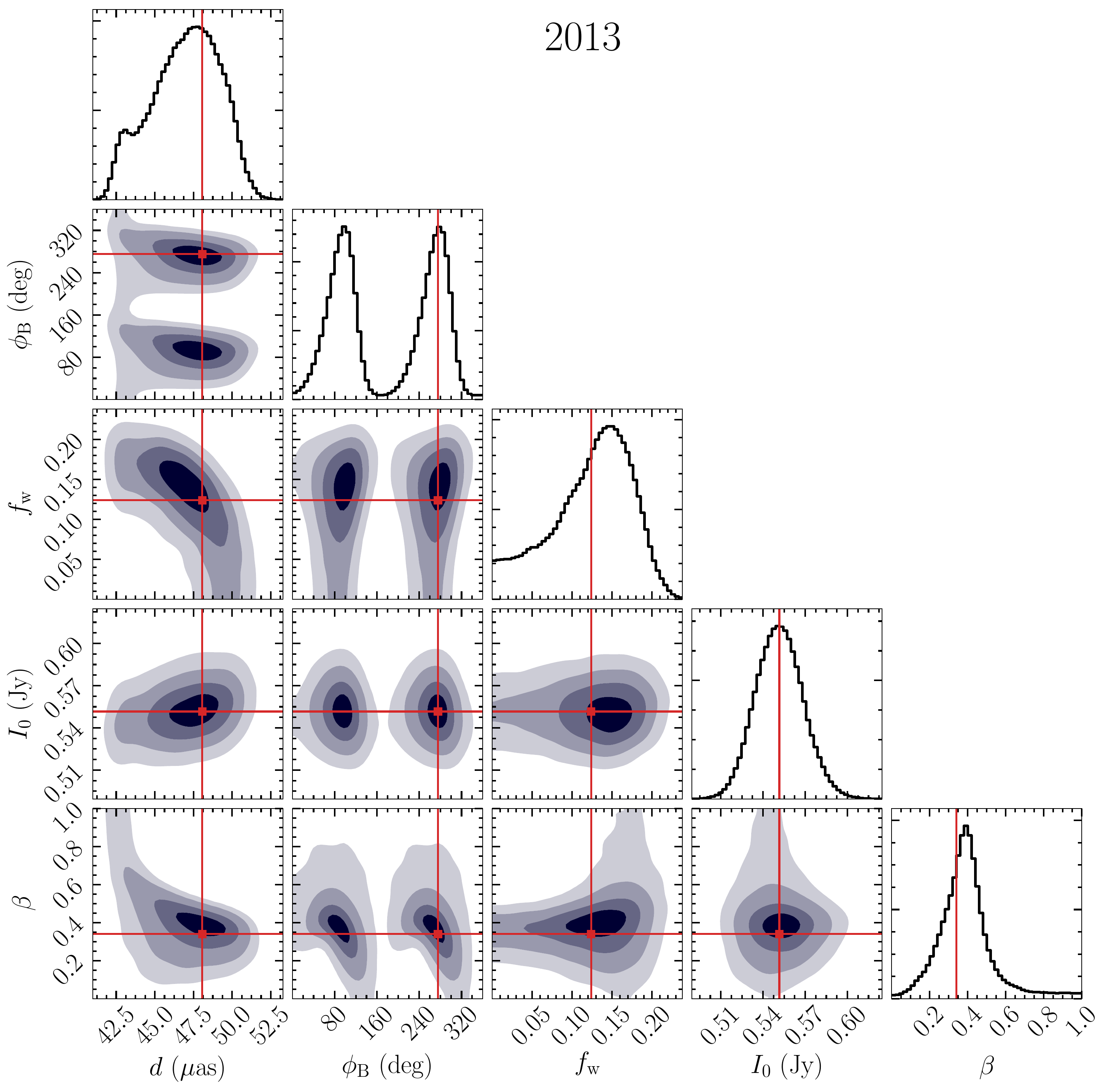}
\caption{Same as \autoref{fig:cornerRT2009-2013}, but for the RG model.}
\label{fig:cornerRG2009-2013}
\end{figure*}

\begin{figure*}
\centering
\includegraphics[width=0.495\columnwidth]{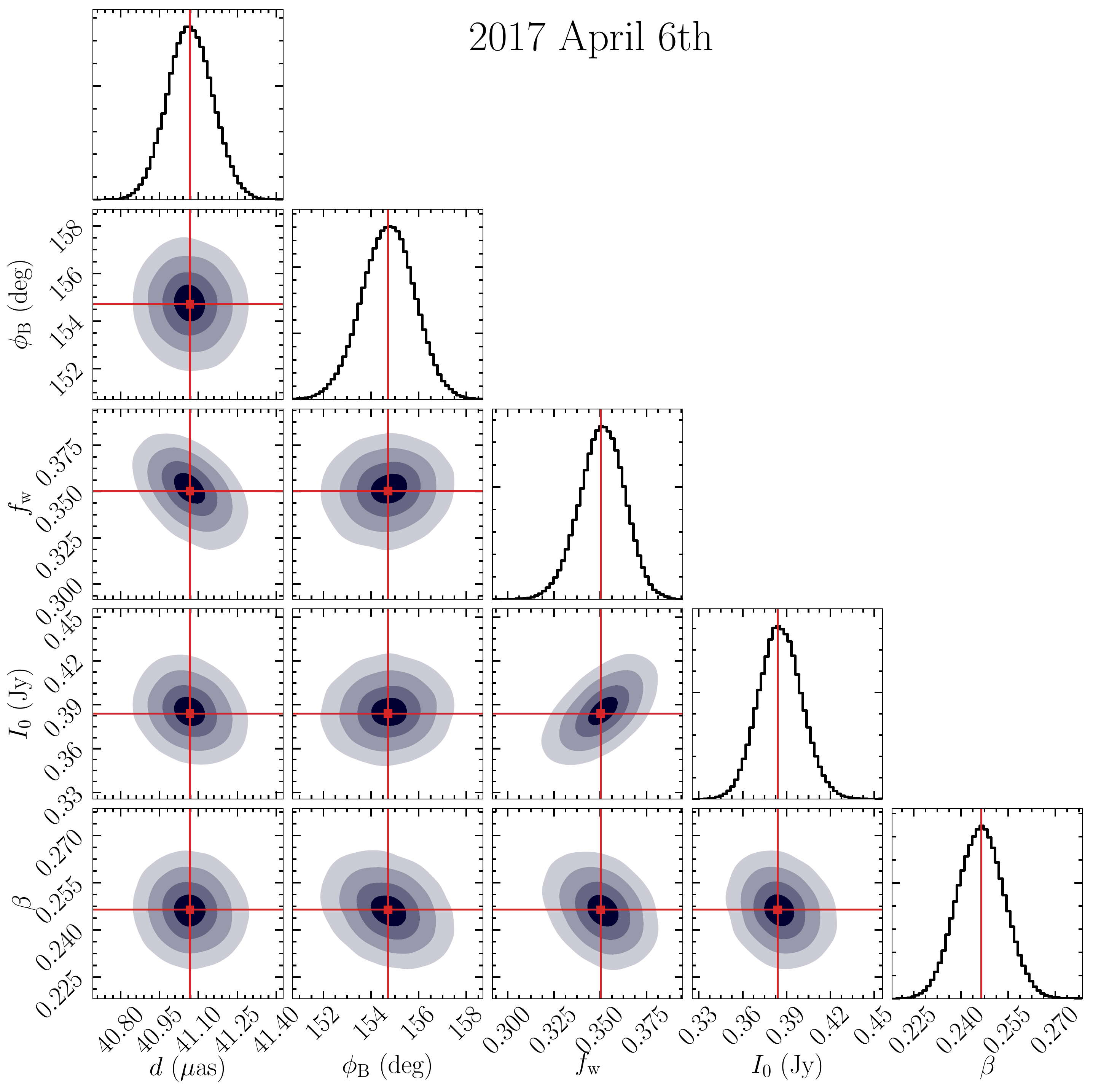}
\includegraphics[width=0.495\columnwidth]{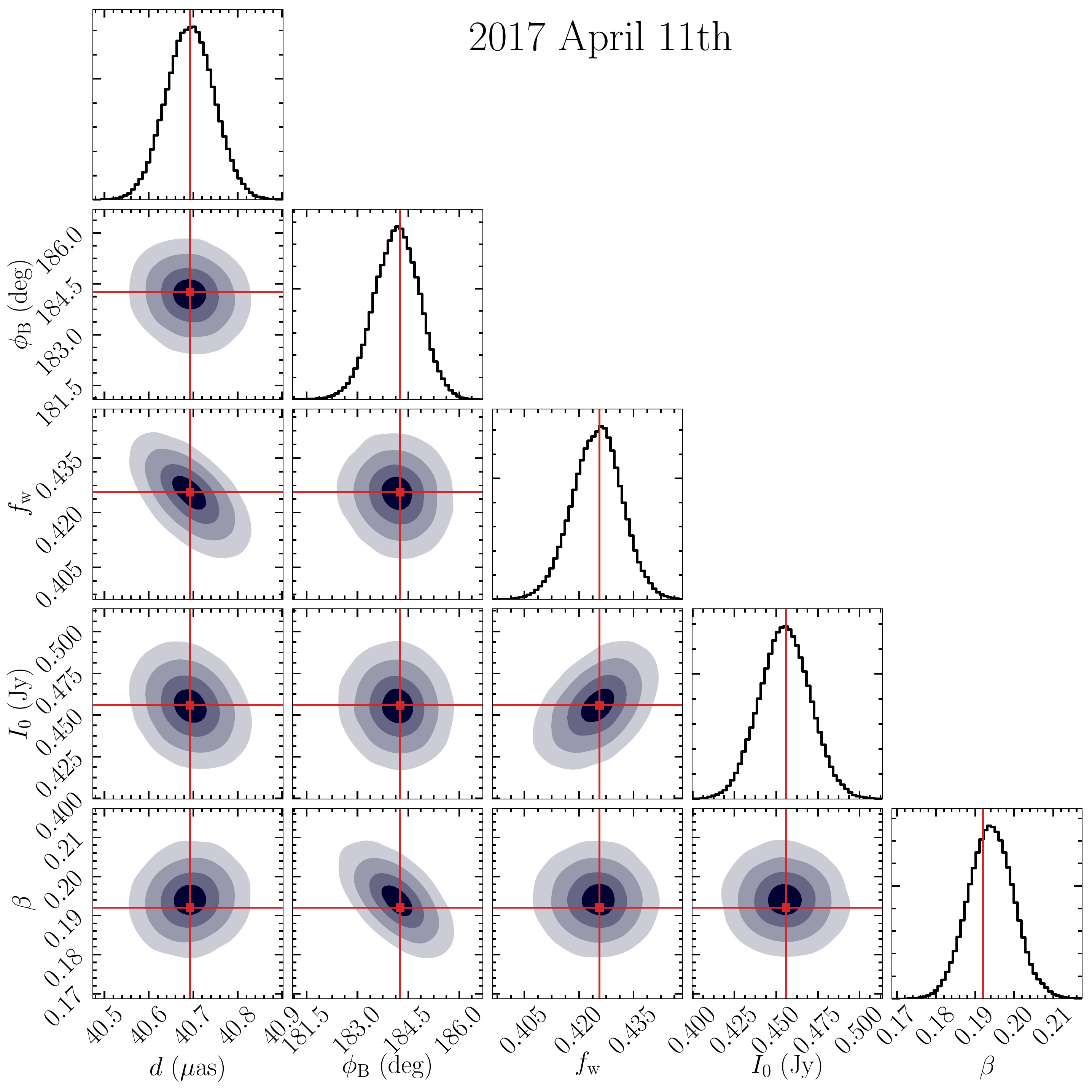}
\caption{Same as \autoref{fig:cornerRT2009-2013}, but for the \m87 observations on 2017 April 6th and April 11th fitted with the RT model.}
\label{fig:corner2017}
\end{figure*}

\begin{figure*}
\centering
\includegraphics[width=0.495\columnwidth]{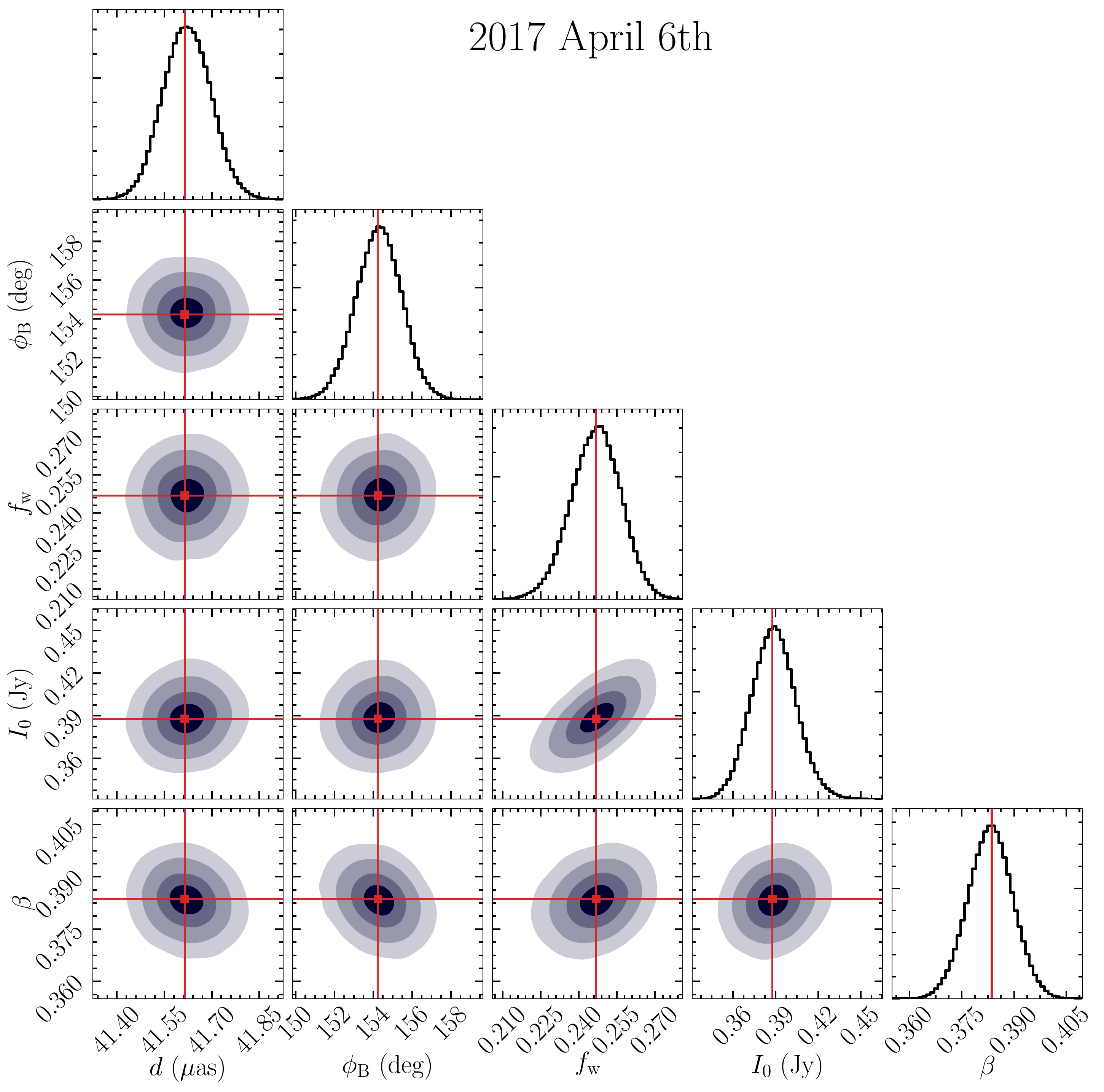}
\includegraphics[width=0.495\columnwidth]{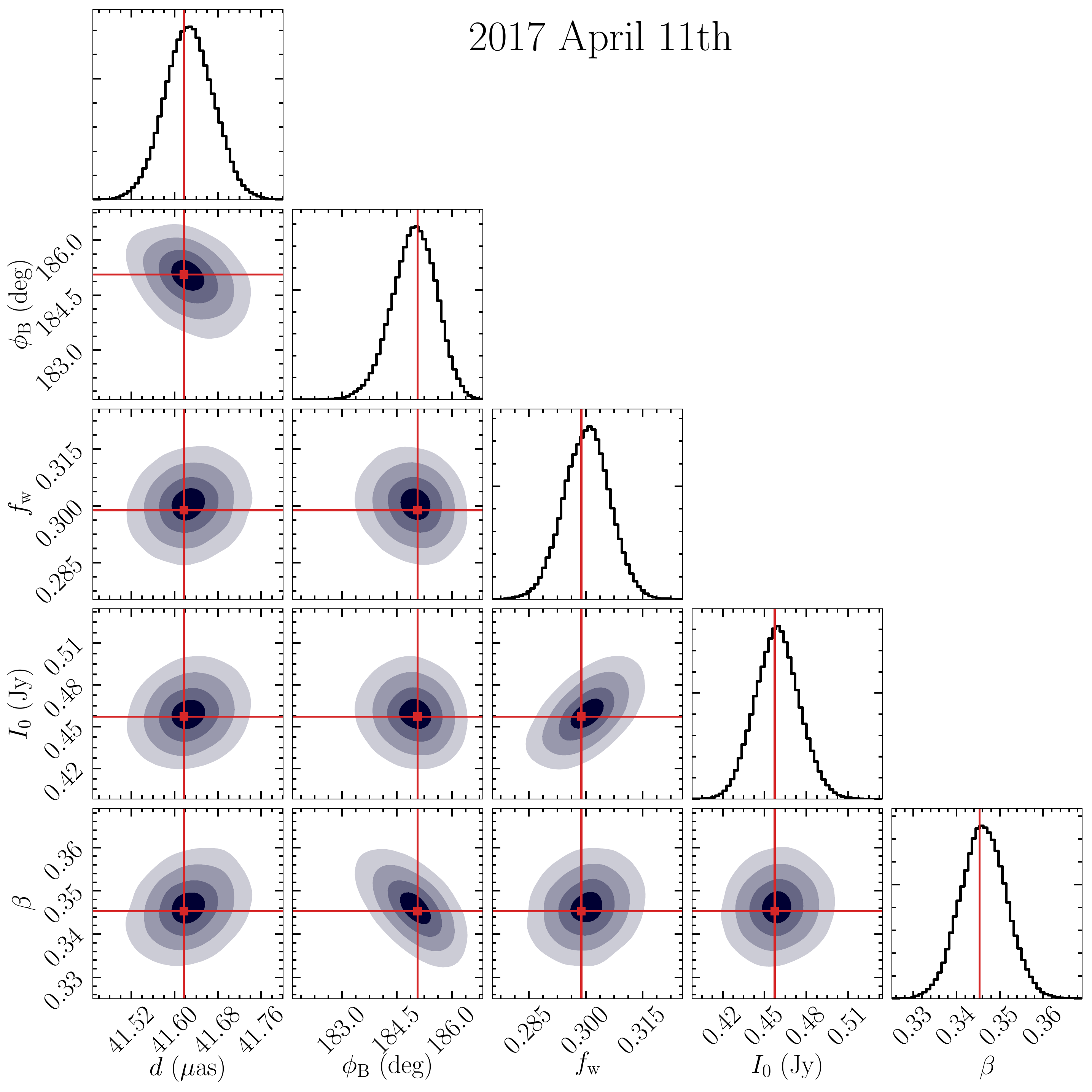}
\caption{Same as \autoref{fig:cornerRT2009-2013}, but for the \m87 observations on 2017 April 6th and April 11th fitted with the RG model.}
\label{fig:corner2017RG}
\end{figure*}

We present the posterior probability distributions corresponding to fitting the 2009--2013 data sets with the RT model, Figure \ref{fig:cornerRT2009-2013}, and with the RG model, Figure \ref{fig:cornerRG2009-2013}. Similarly, for the 2017 data sets, we show the posterior distributions obtained for the RT model, Figure \ref{fig:corner2017}, and for the RG model, Figure \ref{fig:corner2017RG}.

\end{document}